\documentclass[newtx, twocolumn]{aastex7}

\usepackage{enumitem}  %
\usepackage{amsmath}
\usepackage{newtxtext}
\usepackage{newtxmath}
\newcommand{\hst}{HST}
\newcommand{\jwst}{JWST}

\begin{document}

\title{Breaking the AGB Color Degeneracy with Variability and H$_2$O: A Consistent Picture of AGB Chemistry, Dust, and the J-Region Across Environments}
\correspondingauthor{Caroline D. Huang}
\email{caroline.huang@cfa.harvard.edu}

\author[orcid=0000-0001-6169-8586,gname='Caroline',sname='Huang']{Caroline D. Huang}
\altaffiliation{NSF Astronomy \& Astrophysics Postdoctoral Fellow}
\affiliation{Center for Astrophysics | Harvard \& Smithsonian, 60 Garden St., Cambridge, MA, 02138, USA}
\email{caroline.huang@cfa.harvard.edu}

\author[orcid=0000-0001-9910-9230,gname='Massimo', sname='Marengo']{Massimo Marengo} 
\affiliation{Department of Physics, Florida State University, 77 Chieftain Way, Tallahassee, FL 32306, USA}
\email{mmarengo@fsu.edu}

\author[orcid=0009-0008-9156-4640,gname='Antonio', sname='Capodagli']{Antonio Capodagli}
\affiliation{Department of Physics, Florida State University, 77 Chieftain Way, Tallahassee, FL 32306, USA}
\email{acapodagli@fsu.edu}  

\author[orcid=0000-0002-6124-1196,gname='Adam',sname='Riess']{Adam G. Riess}
\affiliation{Department of Physics \& Astronomy, The Johns Hopkins University, 3701 San Martin Drive, Baltimore, MD 21218, USA}
\affiliation{Space Telescope Science Institute, 3700 San Martin Drive, Baltimore, MD 21218, USA}
\email{ariess@stsci.edu}

\author[0000-0003-3889-7709, gname='Louise', sname='Breuval']{Louise Breuval}
\affiliation{European Space Agency (ESA), ESA Office, Space Telescope Science Institute, 3700 San Martin Drive, Baltimore, MD 21218, USA}
\email{lbreuval@stsci.edu}

\author{Stefano Casertano}
\affiliation{Space Telescope Science Institute, 3700 San Martin Drive, Baltimore, MD 21218, USA}
\email{stefano@stsci.edu}

\author[0000-0002-4678-4432,gname='Patricia',sname='Whitelock']{Patricia A. Whitelock}
\affiliation{South African Astronomical Observatory, P.O. Box 9, 7935 Observatory, South Africa}
\affiliation{Department of Astronomy, University of Cape Town, 7701 Rondebosch, South Africa}
\email{paw@salt.ac.za}

\author[0000-0002-5259-2314, gname='Gagandeep', sname='Anand']{Gagandeep S. Anand}
\affiliation{Space Telescope Science Institute, 3700 San Martin Drive, Baltimore, MD 21218, USA}
\email{ganand@stsci.edu}

\author[0000-0003-4284-4167, gname='Randall', sname='Smith']{Randall K. Smith}
\affiliation{Center for Astrophysics | Harvard \& Smithsonian, 60 Garden St., Cambridge, MA, 02138, USA}
\email{rsmith@cfa.harvard.edu}

\author[0000-0001-9420-6525,gname='Wenlong', sname='Yuan']{Wenlong Yuan}
\affiliation{Department of Physics \& Astronomy, The Johns Hopkins University, 3701 San Martin Drive, Baltimore, MD 21218, USA}
\email{wyuan10@jhu.edu}
 
\begin{abstract}

Resolved asymptotic giant branch (AGB) populations present many observational puzzles. We consider three usually treated separately: photometric C/O classification boundaries must be redrawn empirically in new environments; hydrostatic model grids are too blue in bands sampling H$_2$O absorption; and J-region (JAGB) luminosities can differ between fields of a single galaxy. Pulsation connects them into a coherent picture. Using SPHEREx spectroscopy of AGB stars in the Large Magellanic Cloud (LMC), with JWST/NIRCam photometry and HST time-series observations of the SN Ia host M101, we develop an F182M index sensitive to 1.9$\micron$ H$_2$O absorption that chemically classifies variable AGB stars and reveals a degeneracy between oxygen-rich (O-rich) and carbon-rich (C-rich) stars in broadband colors. Because H$_2$O absorption arises in extended atmospheric layers, hydrostatic model grids have difficulty reproducing it. Nearly all stars with substantial infrared excess are variable, while many variable stars show little dust. This supports mass loss initiated by pulsation and enhanced by dust formation. Variable O-rich AGB stars in M101 are up to $\sim$0.7 mag redder than their LMC counterparts in dust-sensitive colors, consistent with more efficient silicate-dust production at higher metallicity, while C-rich populations show weaker environmental dependence. Color windows that select C-rich stars cleanly in the LMC are therefore 33--68\% O-rich in metal-rich M101. This environmental dependence challenges the J-region's universality as a standard candle while evading standard self-consistency checks. We recommend combining variability, or the infrared excess that implies it, with molecular-feature photometry to improve AGB classification at 0.75--5.0$\micron$.

\end{abstract}

\keywords{\uat{Asymptotic giant branch stars}{2100} --- \uat{Carbon stars}{199} --- \uat{Long period variable stars}{935} --- \uat{Stellar classification}{1589} --- \uat{Circumstellar dust}{236} --- \uat{Distance indicators}{394} --- \uat{Large Magellanic Cloud}{903}}

\section{Introduction}\label{sec:intro}

The asymptotic giant branch (AGB) phase is the final stage of a star's nuclear-burning life. Nearly all stars with initial masses ranging from $\sim 0.5-8.0M_{\odot}$ will pass through it. At the start of the AGB phase, stars will be oxygen-rich (or O-rich) AGB stars, defined as having surface chemistry C/O $< 1$. However, during the AGB phase, a process called the third dredge-up can bring carbon from the core, changing the surface chemistry to C/O $> 1$ and creating carbon-rich (C-rich) AGB stars. Surface chemistry directly affects the molecular features and dust present in the spectra of these stars. When both C and O are present on a stellar surface, the highly stable CO diatomic will form abundantly. The remaining unpaired C or O atoms will then be free to make other molecules \citep{Hofner_Olofsson_2018}. These molecules will eventually become the precursors to dust production and mass loss from these highly evolved stars, allowing them to enrich the interstellar medium \citep{Ferrarotti_Gail_2006, Riebel_2012, Hofner_Olofsson_2018}.  

In this work, we focus on the behavior of AGB stars in the near-infrared ($\lambda = 0.75 - 5.0\mu$m) regime.  The typical spectrum of an AGB star in these wavelengths is composed of a cool stellar photosphere (peaking at $\lambda \sim1-2\mu$m), abundant molecular absorption features, and a dust continuum that can dominate at the longer-wavelength ($\lambda \sim 2-5\mu$m) end of this range. Decades of ground- and space-based near-infrared surveys, from the Milky Way and the Local Group to the crowded disks of nearby spirals \citep[e.g.,][]{Skrutskie_2006, Riebel_2012, Dalcanton_2012, Boyer_2015, Boyer_2019}, share a recurring result: broadband near-infrared photometry alone cannot separate the carbon- and oxygen-rich classes of AGB stars. C- and O-rich AGB stars overlap heavily in wide-band colors, particularly once circumstellar reddening becomes substantial \citep{Dalcanton_2012, Boyer_2013}. We refer to this overlap as the AGB color degeneracy.

Given the difficulty of obtaining spectra, medium-band photometry targeting molecular features has become the gold standard for separating C- and O-rich AGB stars using HST WFC3/IR bandpasses. \cite{Boyer_2013} introduced a color-color diagram based on the F127M/F139M/F153M WFC3/IR filters as a method to chemically separate C- and O-AGB by their molecular features. These three bands bracket the 1.4$\mu$m H$_2$O band of oxygen-rich stars and the 1.5-1.6$\mu$m C$_2+$CN absorption of carbon stars, demonstrating clean separation in the high-metallicity inner disk of M31 as well as six metal-poor dwarf galaxies \citep{Boyer_2017}. Ground-based photometric catalogs, such as the OGLE-III Long-Period Variable catalog, employ optical--near-infrared color-color diagrams and planes of the reddening-free Wesenheit index \citep{Madore_1982} for the same purpose, both of which show two distinct AGB populations \citep{Soszynski_2009}. Ground-based spectra classify AGB stars by the stable molecules formed in or near the photosphere --- TiO, CN, C$_2$, and the CO overtones --- because telluric windows block many of the other molecules formed in AGB stars, such as H$_2$O \citep{Verro_2022, Nidever_2026}. 

The James Webb Space Telescope \citep[JWST;][]{2006SSRv..123..485G} and the Spectro-Photometer for the History of the Universe, Epoch of Reionization, and ices Explorer \citep[SPHEREx;][]{Crill_2020} have changed the landscape of possibilities for understanding AGB stars in the near-infrared. JWST is the first single observatory to span the entire $0.75 - 5 \mu$m regime either photometrically or spectroscopically, and SPHEREx is the first to do so spectroscopically at a survey scale. These space-based observations remove the telluric constraints, allowing us to target the full molecular inventory present in dynamic AGB atmospheres. This expanded wavelength range also highlights the potential limitations of hydrostatic model grids in interpreting JWST NIRCam observations.

In an extensive search of JWST/NIRCam filter combinations based on population-synthesis simulations built on hydrostatic model atmospheres, \cite{Boyer_2024} were unable to find a combination that isolates AGB spectral types independently of metallicity, though they note that their models predict easier separation at higher metallicity. The Early Release Science (ERS) study was based on PARSEC-COLIBRI \citep{Marigo_2017, Pastorelli_2020} evolutionary tracks which use AGB spectra derived from hydrostatic COMARCS atmospheres \citep{Aringer_2009, Aringer_2016} with circumstellar dust included in post-processing. Hydrostatic model grids are highly successful at predicting the behavior of stable, photospheric molecules like TiO, VO, C$_2$, and CN. However, as the authors of the models noted, hydrostatic models cannot be expected to fully reproduce the spectra of pulsating AGB atmospheres. More fragile molecular species such as H$_2$O in oxygen-rich stars, and C$_2$H$_2$ and HCN in carbon-rich stars, require a cooler, denser gas than a hydrostatic photosphere can support at these effective temperatures. 

Nevertheless, two decades of infrared spectroscopy and interferometry have shown that these more fragile molecular species \textit{are} abundant in the near-infrared spectra of these stars, and can be even more prominent than the sturdy diatomic molecules. They are expected to form in extended molecular layers levitated by pulsation \citep{Lancon_and_Wood_2000, Hofner_Olofsson_2018} --- coined the ``MOLsphere'' by \cite{Tsuji_2000, Tsuji_2001}. The photometric evidence also indicates this: when defining their WFC3/IR medium-band classification, \citet{Boyer_2017} found the hydrostatic model colors of both C-rich and O-rich stars to be systematically bluer than observed stars in the bands sampling the 1.4~$\mu$m water feature. They suggested the adopted water opacity and departures from hydrostatic equilibrium among the possible causes of this discrepancy, and ultimately drew their O-rich selection boundary empirically from the data rather than from the models. More recently, JWST spectroscopy of an oxygen-rich AGB star in the very metal-poor dwarf galaxy Sextans A ($\sim$1--7\% $Z_\odot$) confirmed strong water vapor absorption at 6.5~$\mu$m, with NIRCam photometry indicating corresponding absorption in the F277W and F300M bands \citep{Boyer_2025}. Water absorption in O-rich stars can therefore remain strong, and photometrically detectable, even at very low metallicity. The most powerful chemical diagnostics may thus be exactly the features hydrostatic grids cannot reproduce --- and their strength should correlate with pulsation, a prediction this paper tests directly.

The Large Magellanic Cloud, which has a well-studied AGB population \citep{Soszynski_2009, Riebel_2015} at a geometrically anchored distance \citep{Pietrzynski_2019}, has been integral in building our local understanding of AGB stars. However, JWST observations of this galaxy cannot be made directly. AGB stars will saturate NIRCam in most observing modes, and the spatial distribution of the population spans degrees while the NIRCam field of view spans only arcminutes. Currently, only SPHEREx is able to bridge the gap between the LMC and JWST observations in other galaxies through its all-sky spectrophotometry. Using near-infrared SPHEREx spectra, we can directly investigate the LMC AGB populations through synthetic photometry of JWST NIRCam filters. 

Developing photometric frameworks for the classification of AGB stars is important for both AGB physics and the recent application of AGB stars in the cosmic distance ladder. Physically, carbon dust and silicate (oxygen-based) dust have different dust-formation mechanisms \citep{Ventura_2012, Hofner_Olofsson_2018} --- incorrectly identifying the chemistry of AGB stars obscures the underlying physics. Likewise, chemistry determines the composition of the material returned to the interstellar medium by mass-loss, not just the amount. At the population level, the C/M ratio can also be used as a diagnostic of metallicity, initial mass, and dredge-up efficiency \citep[e.g.,][]{Battinelli_and_Demers_2005, Cioni_2009}.  Recently, AGB stars have also gained renewed interest as distance indicators. Both of the most commonly used applications require reliable chemical separation: Mira period-luminosity relations (PLRs) used for Hubble Constant (H$_0$) measurements are calibrated on oxygen-rich stars \citep[e.g.,][]{Huang_2018, 2024ApJ...963...8H, Bhardwaj_2025} while the J-region AGB (JAGB) method assumes its color-selected sample is purely carbon-rich \citep[e.g.,][]{Madore_2022, Lee_2023, Li_2025}. Uncorrected chemical contamination propagates into distance systematics --- and, because the classification errors have a directional effect on photometric measurements, they bias the distances measured rather than simply introduce scatter. 

Distance-ladder-based studies have thus far not employed the robust, medium-band classification developed by \citet{Boyer_2017} because the narrower throughput of the WFC3/IR medium bands makes them prohibitively expensive at the distance of SN Ia host galaxies. Unlike HST, JWST has the necessary sensitivity to utilize medium-band photometry at much larger distances. In addition to JWST and SPHEREx, with the impending flood of data from Roman and Rubin, we will frequently identify stars as variable, even when we do not have enough epochs to derive formal periods. Here, we will use knowledge of variability as a prior while ultimately focusing on the population-level statistics of C- and O-AGB stars in the near-infrared color-magnitude diagrams (CMDs). We leave the full time-series analysis (as in PLRs) to a companion paper. 

The J-region was first identified by \cite{Nikolaev_and_Weinberg_2000} as a region of the 2MASS $J-K_s$ color-magnitude diagram that appeared to consist almost entirely of carbon-rich AGB stars. In recent years, it has gained renewed attention as a carbon-star standard candle \citep[e.g.,][]{Ripoche_2020, Madore_and_Freedman_2020}. Unlike Cepheid or Mira variables (Miras), the JAGB method typically relies on a single-epoch CMD. Stars are selected in a (typically broadband) color window in which only the cooler carbon-rich AGBs are believed to reside. A statistic of their luminosity function (LF) --- most often the mode, the peak of the smoothed luminosity function --- is taken as the fiducial JAGB luminosity. Gaussianity of the luminosity function is typically taken to indicate that the J-region is chemically pure, and agreement between modal, mean, and median magnitudes is often used as a proxy for the stability of the JAGB luminosity. 

The ease of use and relatively short observing time requirements have led to the JAGB method being widely adopted by both the SH0ES and CCHP groups for the calibration of the fiducial Type Ia Supernova (SN Ia) peak luminosity \citep[e.g.,][]{Freedman_and_Madore_2020, Madore_2022, Lee_2025_JWSTJAGB, Li_2025}. In addition, many other groups have also produced JAGB distances to galaxies within the Local Group \citep[e.g.,][]{Zgirski_2021, Parada_2023}. While JAGB and the Tip of the Red Giant Branch (TRGB) measurements made by the same groups have typically confirmed the consistency between the methods, cross-program comparisons have recently yielded larger disagreements \citep{Anand_2025_RNAAS}.

Fewer comparisons have been made at metallicities higher than that of the LMC. This higher-metallicity regime is also where most of the SN Ia hosts used for distance calibration reside. Galactic calibrations from Gaia EDR3 parallaxes of field carbon stars \citep{Lee_2021} and from open clusters \citep{Madore_2022_OC} agree with the Magellanic Cloud calibrations, which has been interpreted as evidence that the fiducial JAGB zeropoint is independent of metallicity up to solar values. However, \citet{Magnus_2024} re-derived the Galactic calibration using AGB stars chemically classified with Gaia RP spectra and found J-region magnitudes systematically fainter than the Magellanic Clouds by 0.2--0.4 mag, in line with the fainter Galactic median found earlier by \citet{Ripoche_2020}. \citet{Magnus_2024} conclude that a reliable calibration at near-solar metallicity must await improved parallaxes or be carried out in another environment.

Physically, the JAGB method rests on two main assumptions. The first is a compositional assumption --- that the color-selected J-region is composed of C-rich AGB stars. The second is a population-level assumption --- that the luminosity function of the C-rich AGB stars inhabiting the J-region is the same in every environment and constant over a specific color range. This work addresses the first assumption only. The second requires independent geometric distances or pulsation periods to serve as mass proxies and is thus deferred to future work. 

We show that simply identifying the presence of pulsation (a variability prior) is sufficient to break the AGB color degeneracy in both the LMC and M101. We use this knowledge to reconcile several findings that have often been treated as separate puzzles in relatively disconnected literatures. While the microphysics of any one AGB star remains difficult to predict, with pulsation as an additional axis of information, the behavior of a resolved AGB population can be described consistently across environments: 
\begin{enumerate}
    \item Medium-band classification boundaries have had to be redrawn empirically in new environments \citep[e.g.,][]{Boyer_2017, Boyer_2019}. We show that metallicity inevitably shifts the color boundaries between C- and O-AGB stars and assert that classifiers whose boundaries can be drawn \textit{in situ} within the science sample are preferable because disagreements between models and observations then constrain the models rather than propagating into the classification.
    \item Hydrostatic model grids depart from observations of AGB stars in bands that sample water absorption \citep{Aringer_2016, Boyer_2017}. Extended molecular layers have been detected spectroscopically and interferometrically for decades \citep{Tsuji_2000, Hofner_Olofsson_2018}. We show that the departures from model grids are strongest in the molecules formed above the photosphere and that their size correlates with pulsation, implicating these pulsation-levitated molecular layers as their cause. 
    \item Fiducial J-region magnitudes differ between fields of a single galaxy \citep[NGC 4258;][]{Li_2025} and between inner and outer disks \citep{Lee_2025_JWSTJAGB}, and the Galactic calibration is fainter than the Magellanic Clouds \citep[e.g.,][]{Ripoche_2020, Magnus_2024}, even where standard purity and robustness checks \citep[e.g.,][]{Freedman_and_Madore_2020, Madore_2022} are met. We show that oxygen-rich contamination of the J-region grows with metallicity and dust content without registering in these checks. 
    \item Dust production among AGB stars is known to switch on with pulsation \citep{McDonald_Zijlstra_2016, McDonald_and_Trabucchi_2019}, and this onset has been interpreted as pulsation-driven. Correlation alone, however, cannot establish which condition is necessary for the other. By measuring both conditional probabilities, $P({\rm Variable}\,|\,{\rm Dusty})$ and $P({\rm Dusty}\,|\,{\rm Variable})$, we test this ordering directly and show that the link is asymmetric in both the LMC and M101. Essentially all stars with substantial infrared excess are variable, while many variables show little dust. Infrared excess can therefore serve as a proxy for variability, which allows the F182M-based index that we develop to be applied to single-epoch photometry where the broadband colors are most degenerate (Section \ref{sec:appendix_classification}). 

\end{enumerate}

Because the paper connects several subfields, readers might prefer to approach it grouped by topic, which we outline below. Section \ref{sec:data} describes the observations, Section \ref{sec:methods} the construction of the samples, Section \ref{sec:results} the measurements, Section \ref{sec:discussion} their interpretation, Section \ref{sec:application} the resulting recommendations and the M101 catalog, and Section \ref{sec:conclusions} our conclusions. Tables \ref{tab:datasets} and \ref{tab:analysis_samples} serve as a key to the datasets and sample names used throughout. 
\begin{itemize}
\item \textit{Pulsation, dust, and environment:} Sections \ref{sec:var_vs_cmd}, \ref{sec:variability_dust}, \ref{sec:pulsation_mass_loss}, and \ref{sec:color_environment}; Figures \ref{fig:var_vs_nonvar} and \ref{fig:capstone_figure}; Tables \ref{tab:color_offset_bracket} and \ref{tab:var_amp_2x2}.
\item \textit{AGB stars as distance indicators:} Sections \ref{sec:JAGB_contamination}, 
\ref{sec:jagb_contamination_discussion}, and \ref{sec:application_JAGB}; Figure \ref{fig:jagb_cmds}; Tables \ref{tab:jagb_contamination}, \ref{tab:jagb_mean_mode}, and \ref{tab:jagb_environments}.
\item \textit{Classifying AGB stars with NIRCam, including how to use pulsation without time-series data (color as a proxy for pulsation), and the M101 catalog:} Sections \ref{sec:m101_variable_selection} and \ref{sec:application_classification}--\ref{sec:application_scope}; Appendix \ref{sec:appendix_classification}; Table \ref{tab:catalog_columns}.
\item \textit{SPHEREx spectra of LMC AGB stars and their synthetic photometry:} Section \ref{sec:spherex_analysis}, Appendix \ref{sec:appendix_synth_mag}; Figure \ref{fig:moleculesandatoms}; Table \ref{tab:obs_vs_synth}.
\end{itemize}
Throughout the paper we use Vega magnitudes. Amplitude is defined as the difference in peak-to-trough magnitude.

\begin{deluxetable*}{llllllc}
\tabletypesize{\footnotesize}
\tablewidth{0pt}
\tablecaption{Datasets used in this work.\label{tab:datasets}}
\tablehead{
\colhead{Dataset} & \colhead{Kind} & \colhead{$\lambda$/bands} &
\colhead{Cadence} & \colhead{Years} & \colhead{Provides} & \colhead{Ref.}
}
\startdata
\noalign{\vskip 2pt}\hline\noalign{\vskip 3pt}
\multicolumn{7}{c}{LMC (calibration \& anchor)\tablenotemark{a}} \\
\noalign{\vskip 3pt}\hline\noalign{\vskip 4pt}
\textbf{OGLE-III} & ground optical & $V$, $I$ & time series & 2001--2009 & $P$, $A_1$, variability & 1 \\
\textbf{LPV catalog} & photometry & (0.55, 0.8\,$\mu$m) & ($\approx 500$ epochs) & & types; phot.\ C/O types & \\[4pt]
\textbf{2MASS} & ground NIR & $JHK_s$ & single epoch & $\sim$2000 & observed NIR colors & 2 \\
 & photometry & (1.2--2.2\,$\mu$m) & & & & \\[4pt]
\textbf{WISE} & space MIR & $W1$/$W2$ & multi-epoch mean & 2010--2018 & flux-averaged & 3 \\
\textbf{(CatWISE2020)} & photometry & (3.4/4.6\,$\mu$m) & & & observed colors & \\[4pt]
\textbf{Spitzer IRAC} & space MIR & [3.6]/[4.5]/[5.8]/ & 1-2 epochs & $\sim$2005 & observed MIR colors & 4, 5 \\
\textbf{(SAGE)} & photometry & [8.0]\,$\mu$m & & & & \\[4pt]
\textbf{SPHEREx} & space & 0.75--5.0\,$\mu$m & multi-epoch & 2025--2026\tablenotemark{b} & mean spectra; spec.\ & 6, 7 \\
 & spectroscopy & $R \approx 35$--130 & (median span & & C/O types; synthetic & \\
 & & & $\approx$282\,d) & & phot.\ in any bandpass & \\[4pt]
\noalign{\vskip 2pt}\hline\noalign{\vskip 3pt}
\multicolumn{7}{c}{M101 (application \& SN Ia host)\tablenotemark{c}} \\
\noalign{\vskip 3pt}\hline\noalign{\vskip 4pt}
\textbf{HST WFC3/IR} & space & F160W (+F110W) & time series & 2014--2024 & LPV detection, periods & 8, this \\
 & photometry & (1.1--1.6\,$\mu$m) & 13-15 epochs & & $\rightarrow$ variables & work \\[4pt]
\textbf{JWST NIRCam} & space & F115W/F150W/F182M/ & single epoch & 2024 & $I_{\rm F182M}$, CMDs, & this \\
 & photometry & F277W/F356W/F444W & & & CMD-based selection & work \\
 & & (1.15--4.44\,$\mu$m) & & & & \\
\enddata
\tablenotetext{a}{Much of the NIR characterization and classification of AGB stars was first calibrated in the LMC and later applied to more distant galaxies; this work follows that structure. The LMC's geometric distance error of $\sim$1\% \citep{Pietrzynski_2019} also makes it an anchor galaxy for many extragalactic distance ladders.}
\tablenotetext{b}{The SPHEREx mission is ongoing and will continue
collecting data; the years quoted are those of the observations used
in this work.}
\tablenotetext{c}{Distant galaxies such as M101 apply physics and
color information calibrated in the LMC to study their AGB stars; as
a SN Ia host, M101 enables a measurement of $H_0$ via the calibration of the SN Ia peak magnitude.}
\tablecomments{A unified OGLE-III, 2MASS, WISE, and IRAC photometry catalog will be described in detail and published in Capodagli et al., in preparation. Details of the creation are outlined in Section \ref{sec:data} (3\arcsec\ matches via IRSA; $W1$/$W2$ are
CatWISE2020 multi-epoch flux-averaged magnitudes).}
\tablerefs{(1) \citet{Soszynski_2009}; (2) \citet{Skrutskie_2006};
(3) \citet{Wright_2010}, \citet{Eisenhardt_2020};
(4) \citet{Fazio_2004}; (5) \citet{Meixner_2006}; (6) \citet{Bock_2026};
(7) \citet{Akeson_2025}; (8) \citet{2024ApJ...963...8H}.}
\end{deluxetable*}

\begin{deluxetable*}{lllcc}
\tabletypesize{\footnotesize}
\tablewidth{0pt}
\tablecaption{Analysis samples.\label{tab:analysis_samples}}
\tablehead{
\colhead{Sample} & \colhead{Data Source} & \colhead{Selection Criteria} &
\colhead{$N$(O)/$N$(C)} & \colhead{Used in}
}
\startdata
\noalign{\vskip 2pt}\hline\noalign{\vskip 3pt}
\multicolumn{5}{c}{LMC (calibration \& anchor)} \\
\noalign{\vskip 3pt}\hline\noalign{\vskip 4pt}
\textbf{LMC CMD-selected} & OGLE-III + 2MASS & $K_s$ vs.\ $J-K_s$ CMD selection & 11,330/8,016 & \S\ref{sec:var_vs_cmd}, \S\ref{sec:variability_dust},
 \\ 
\textbf{AGBs} & (unified catalog) & (TRGB bounded + sloped blue edge, & & and \S\ref{sec:color_environment} \\
 & & \S\ref{sec:lmc_cmd_agb} -- mirrors \S\ref{sec:m101_cmd_agb}); C/O from & & \\
 & & OGLE catalog labels & & \\[4pt]
\textbf{LMC SRVs + Miras} & OGLE-III + 2MASS & same CMD selection + OGLE LPV & 3,980/5,781 & \S\ref{sec:var_vs_cmd} and \\ %
 & (unified catalog) & SRV/Mira types only & &  \S\ref{sec:variability_dust}\\
 & & (excl.\ OSARGs) & & \\[4pt]
\textbf{LMC SPHEREx AGBs} & SPHEREx + OGLE & random SRV+Mira retrievals, & 59/34 & \S\ref{sec:lmc_m101_color_comparison}, \S\ref{sec:co_color_sep}, \\
\textbf{(representative sample)} & & spectroscopically reclassified & & and \S\ref{sec:JAGB_contamination} \\
 & & (excl.\ OSARG / JAGB-table / & & \\
 & & Mira-only additions) & & \\[4pt]
\textbf{LMC SPHEREx AGBs} & SPHEREx + OGLE & random SRV+Mira retrievals + & 73/73\tablenotemark{a} & \S\ref{sec:spectral_classification} and \\
\textbf{(full retrieved sample)} & & additional OSARG, Mira, and & & \S\ref{sec:variable_classification}\\
 & & J-region retrievals; & &\\
 & & spectroscopically reclassified & & \\[4pt]
\noalign{\vskip 2pt}\hline\noalign{\vskip 3pt}
\multicolumn{5}{c}{M101 (application \& SN Ia host)} \\
\noalign{\vskip 3pt}\hline\noalign{\vskip 4pt}
\textbf{M101 variability-} & HST + JWST & LPV detection + water & 818/196 & \S\ref{sec:lmc_m101_color_comparison},  \S\ref{sec:var_vs_cmd}, \\
\textbf{detected} & & index at $I_{\rm F182M} = 0.40$ mag & & and \S\ref{sec:variability_dust}\\[4pt]
\textbf{M101 CMD-only} & JWST & CMD selection (AGB region $\cup$ dusty & 13,686/1,440\tablenotemark{b} & \S\ref{sec:JAGB_contamination}, \S\ref{sec:variability_dust}, \\
 & & tail $-$ supergiant strip), & & and \S\ref{sec:application_catalog} \\
 & & F150W$-$F182M $= 0.64$ & & \\
\enddata
\tablenotetext{a}{The full retrieved sample comprises \emph{all}
retrieved SPHEREx mean-spectra stars regardless of OGLE class (146
stars; 73 O / 73 C after spectroscopic reclassification). It is used
only in targeted checks of boundary conditions --- e.g., the quality
of the photometric OSARG classifications (\S\ref{sec:variable_classification}) and the spectral features of evolved stars --- and does not enter the calibrations, which use the representative sample.}
\tablenotetext{b}{O-like/C-like. Throughout, ``O-like/C-like''
denotes the CMD-based color classification and ``O-rich/C-rich''
candidates the water-index classification of variables; spectroscopic
classes (LMC only) are stated as such.}
\tablecomments{Sample names are used verbatim in all figure legends
and text. }
\end{deluxetable*}

\section{Observations, Data Reduction, and Photometry}\label{sec:data}
We use HST and JWST observations of SN Ia host galaxy M101 and 2MASS, OGLE, CatWISE, Spitzer IRAC, and SPHEREx observations of the Large Magellanic Cloud (LMC). Table \ref{tab:datasets} contains a summary of the individual datasets used in this paper and the information used from each one. Below, we expand on the details of the newer SPHEREx, HST, and JWST observations. 

\subsection{M101 Observations}
\label{sec:observations_m101}
\begin{figure}[]
\includegraphics[width=0.47\textwidth]{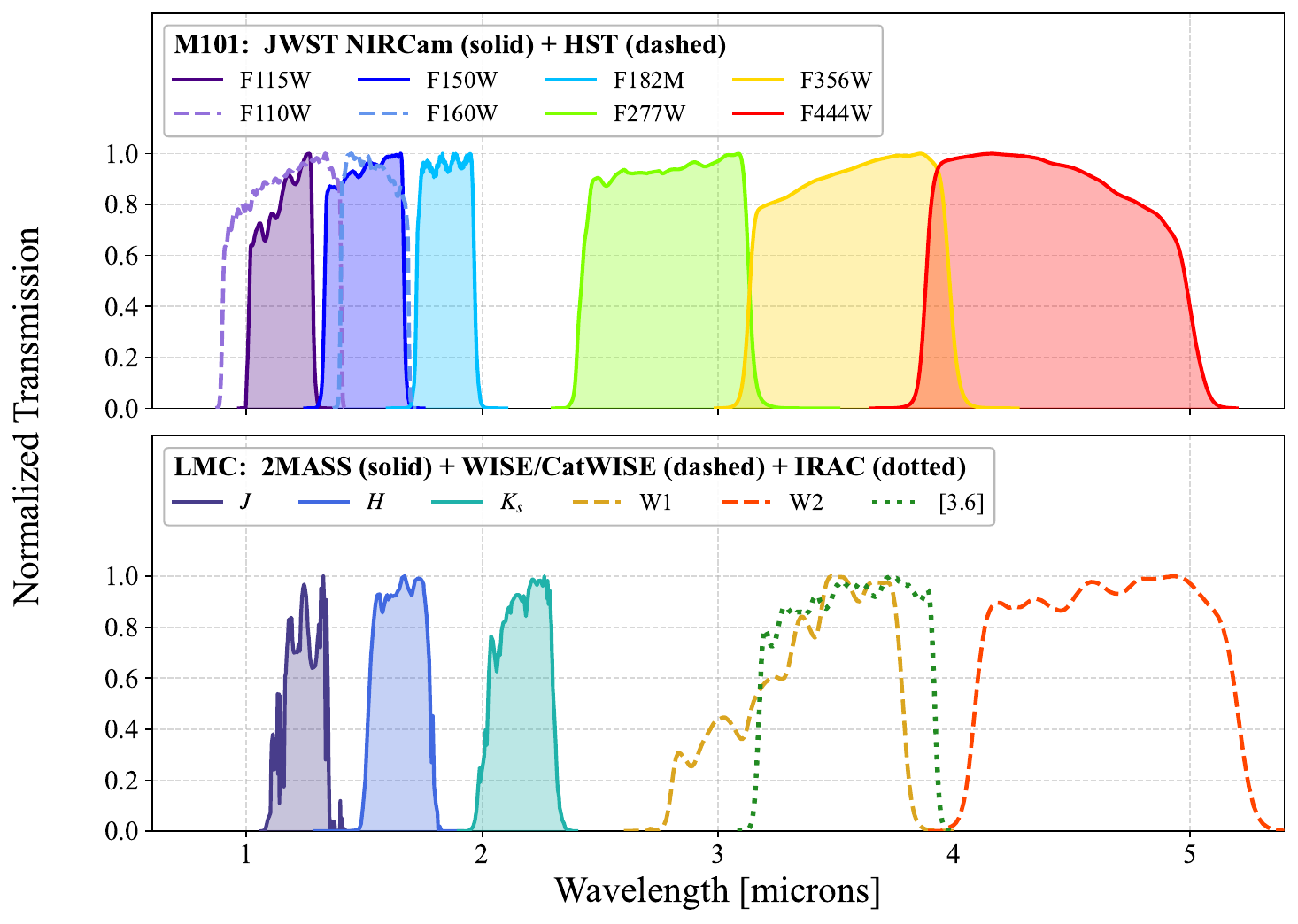}
\caption{Normalized transmission curves of the bandpasses used in this work. \textit{Top:} the M101 filters, the six JWST NIRCam bands (solid: F115W, F150W, F182M, F277W, F356W, F444W) and the two HST WFC3/IR bands of the time-series data (dashed: F110W, F160W). \textit{Bottom:} The archival LMC filters, 2MASS $J$, $H$, and $K_s$ (solid), WISE $W1$, $W2$ (dashed; photometry from CatWISE2020), and Spitzer IRAC [3.6] (dotted). 
\label{fig:filters}}
\end{figure}

JWST NIRCam observations of M101 were obtained in the JWST Cycle 2 General Observer Program 4087 (PI Huang) on February 23, 2024 using F115W, F150W, and F182M filters in the short wavelength (SW) channel and F277W, F356W, and F444W filters in the long wavelength (LW) channel (transmission curves shown in Figure \ref{fig:filters} alongside the other bands used in this work). The exposure setup consisted of one integration of eight {\tt SHALLOW4} groups, taken with four subpixel dithers using the {\tt SMALL-GRID-DITHER} pattern. The total exposure time was 1675s per pair of SW and LW filters. NIRCam Module B covered the field of SN-2011fe ($\alpha= 14^{\rm h}03^{\rm m}05.43^{\rm s}$, $\delta = +54^\circ16'47.6''$, Equinox: J2000) where the Miras from \cite{2024ApJ...963...8H} are located while NIRCam Module A covered one of the Cepheid fields ($\alpha = 14^{\rm h} 02^{\rm m} 57.73$, $\delta = +54^\circ 19'29.4''$, Equinox: J2000) analyzed in several previous works \citep[e.g.,][]{Shappee_and_Stanek_2011, Mager_2013, Riess_2022, Riess_2024}. A summary of the JWST/NIRCam observations used in this paper can be found in Table \ref{tab:jwst_observations}.

F182M was included in the SW channel as a chemical-classification diagnostic. NIRCam's medium bands offer the best available compromise between molecular-feature selectivity and observing depth. We chose F182M in particular because it falls entirely within the 1.9$\mu$m H$_2$O complex, which is expected to be the deepest of the near-infrared water bands in O-rich AGB spectra.  Because the observing program was primarily designed around cross-checking the systematics of the O-rich Mira distance to this galaxy (the subject of a companion paper currently in preparation), we targeted the water bands rather than the carbon-based ones. The corresponding carbon-side option would have been F300M, which covers the 3.1$\mu$m C$_2$H$_2$ + HCN feature. Model spectra of K and M giants \citep{Aringer_2016} and carbon stars \citep{Aringer_2009} were used to locate molecular features in wavelength space, with the understanding that hydrostatic grids might fail to reliably predict their observed strength in pulsating stars. The selection was therefore based on the location of the molecular features rather than on model-predicted classification performance. Earlier WFC3/IR medium-band surveys had already hinted in this direction: the observed colors of O-rich AGB stars were systematically offset from hydrostatic-model expectations, requiring their classification boundaries to be drawn empirically \citep{Boyer_2017} -- a sign that water absorption may have been more prominent in these stars than the models suggested.

In addition to these observations, we also obtained simultaneous HST observations from GO-17312 (PI Huang) which provide time-series information (2-3 new epochs) for the sources in both fields. These observations bracket the JWST observations in order to constrain the phase of the JWST observations and decrease the scatter of the PLRs created with the single-epoch JWST data. These were also reduced in a single photometric run with archival HST observations, which provide a full time-series baseline of $\sim$3,600 days. A summary of all the HST observations used in this paper can be found in the Appendix, Table \ref{tab:hst_observations}. In \citet{2024ApJ...963...8H}, we adopted a metallicity of approximately half-solar for this field, based on the abundance gradient of \citet{Mager_2013}. In Section \ref{sec:methods_metallicity} we refine this estimate and test its internal consistency with the observed C/M ratio. 

\begin{figure*}[]
\includegraphics[width=\textwidth]{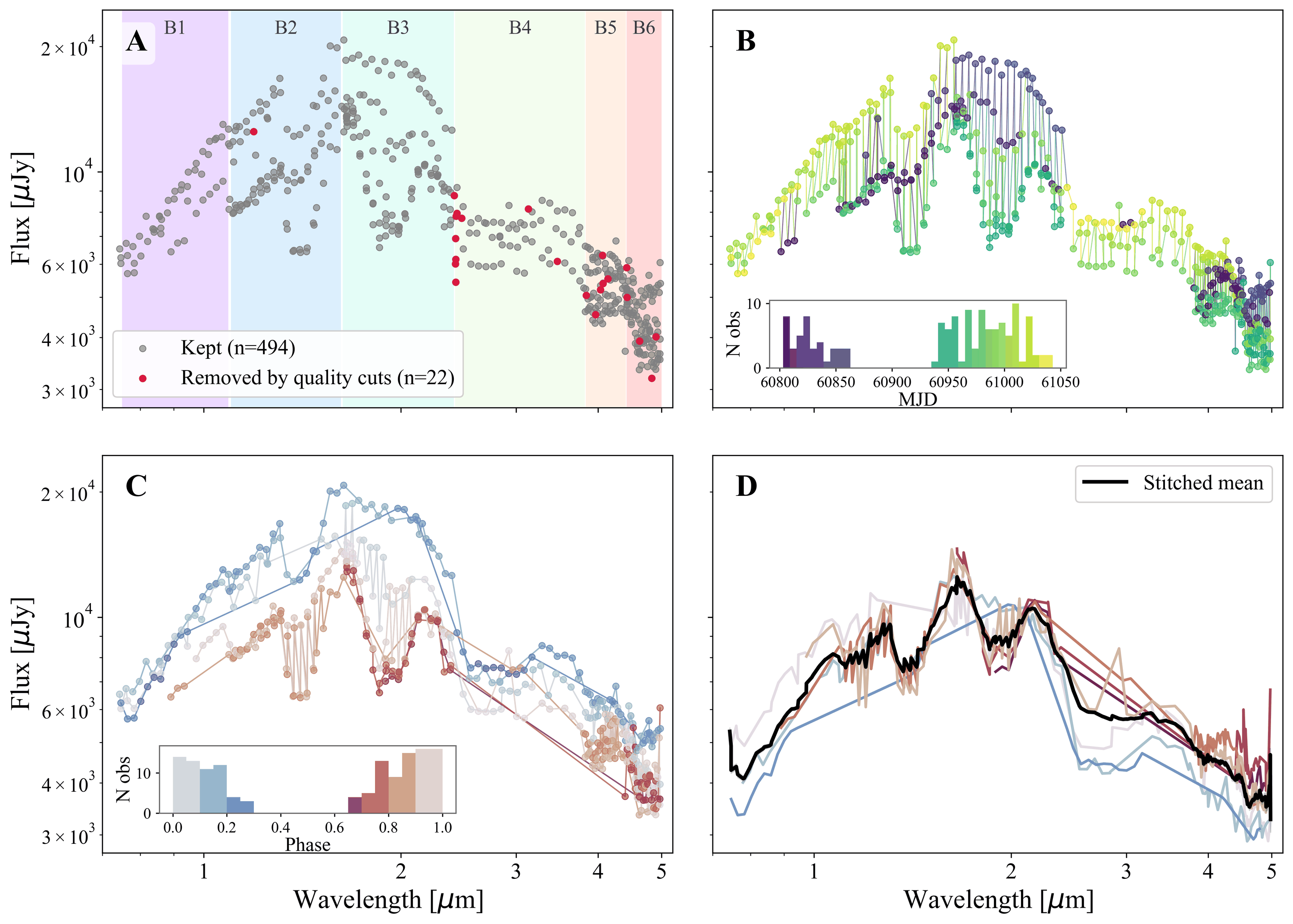}
\caption{Plot depicting our custom data-reduction pipeline for SPHEREx spectra of variable stars. Panel A shows the raw observations (gray points) and the points removed by quality cuts (red points). Panel B shows the raw observations colored by date of observation, along with the histogram (inset, lower left) showing the number of observations at each epoch. Points are connected in wavelength space to show the spread in flux measured at each wavelength. Panel C shows the spectra created by grouping points by phase. Colors are shared with the shifted phased spectra shown in Panel D, and the final mean spectrum in black. See text for full description of mean spectrum creation process.
\label{fig:mean_spectrum_creation}}
\end{figure*}

\subsection{M101 Data Reduction}
We retrieve all WFC3/IR images from the Mikulski Archive for Space Telescopes (MAST) as {\tt flt} files, which are individual exposures that have been pipeline-processed, calibrated, and flat-fielded. A detailed description of WFC3/IR observations of the field and the {\tt DAOPHOT/ALLSTAR/ALLFRAME} \citep{Stetson_1987, Stetson_1994}-based time-series photometry process is presented in Section 2.3 of \citet{2024ApJ...963...8H}. 

For the JWST data, we retrieved the Stage 2 calibrated exposures and the Stage 3 mosaics from MAST. We performed PSF photometry with \textsc{Dolphot} NIRCam module \citep{Dolphin_2016, Weisz_2024} on the individual calibrated exposures (four dithers per filter), using the F150W mosaic as the astrometric reference frame and the parameter values recommended by the JWST Resolved Stellar Populations Early Release Science program \citep{Weisz_2024}. Because the PSF and degree of crowding differ between detectors, each of the four short-wavelength (SW) chips was photometered independently (fitting the F115W, F150W, and F182M exposures simultaneously), with the long-wavelength (LW) detector treated as a single additional run (F277W, F356W, F444W). 

For this work, we do not perform artificial-star tests. The previous HST WFC3/IR-based time-series study from \citet{2024ApJ...963...8H} found crowding of the order of $\approx 0.05$ mag per star. This is therefore the upper bound expected for corrections for NIRCam photometry, which has a sharper PSF at short wavelengths and is comparable to that of WFC3/IR even at F444W, particularly given our stringent crowding and sharpness cuts. Because this work does not attempt to measure a distance modulus, the classification boundaries are measured within the M101 sample itself, and the LMC--M101 color offsets we report are several times larger than this bound, any residual systematic at this level -- including a potential color-dependent term from the wavelength dependence of the PSF -- would not alter our conclusions. 

All of the HST and JWST data presented in this article were obtained from the Mikulski Archive for Space Telescopes (MAST) at the Space Telescope Science Institute. The specific observations analyzed can be accessed via \dataset[doi:10.17909/168e-1683]{https://doi.org/10.17909/168e-1683}.

\subsection{HST and JWST Crossmatching}
\label{sec:hst_jwst_crossmatch}

The variable sources identified using \hst\, time series are then cross-matched with our single-epoch multi-band JWST NIRCam observations. First, we visually confirmed cross matches between \hst\, and \jwst\, for bright, relatively isolated stars previously used in the \hst\, DAOPHOT photometric calibration. Using these confirmed matches as an input catalog, we fit a single affine transformation (consisting of translation, rotation, and scaling) at the pixel level for this source list, described by,
\begin{equation}
\begin{aligned}
x_{\mathrm{JWST}} &= k (x_{\mathrm{HST}} \cos \theta - y_{\mathrm{HST}}  \sin\theta) + \Delta x \\
y_{\mathrm{JWST}} &= k (x_{\mathrm{HST}} \sin \theta + y_{\mathrm{HST}} \cos\theta) + \Delta y
\end{aligned}
\end{equation}
where $\Delta x$ and  $\Delta y$ describe the translation between \hst\, and \jwst, $\theta$ is rotation angle, and $k$ is the pixel-scale ratio between WFC3/IR and NIRCam. In our case, the \hst\, master image had a pixel scale of 0.12$''$/pix while the \jwst\, image had a pixel scale of 0.031$''$/pix. The scale factor was treated as a free parameter and converged within 0.1\% of the expected pixel-scale ratio, consistent with no significant residual scale mismatch between the two instruments' astrometric solutions over the field. 

To crossmatch the rest of the \hst\, catalog, we first transformed the WFC3/IR pixel coordinates into NIRCam pixel coordinates and then used a one-to-one nearest neighbor crossmatch within a radius of two NIRCam pixels. Within NIRCam short- and long-wavelength observations, we used a tolerance of 0.4 NIRCam pixels.

\subsection{LMC Observations}\label{sec:observations_lmc}
Our Large Magellanic Cloud observations include SPHEREx spectrophotometry and 2MASS, OGLE, Spitzer, and CatWISE photometry. The catalog photometry is drawn from a unified OGLE-III, 2MASS, CatWISE2020 and Spitzer/SAGE catalog, constructed via 3\arcsec\ crossmatches through IRSA, with $W1$/$W2$ taken as multi-epoch, flux-averaged CatWISE2020 magnitudes. The catalog construction and validation will be described in detail in Capodagli et al.\ (in preparation).  We chose the LMC in part for its extensive coverage thus far during the SPHEREx mission, which allows us to derive synthetic photometry for the \jwst\,  NIRCam filters that correspond to those observed in M101. The LPV population in the LMC is also one of the best characterized, having been extensively studied by many groups \citep[e.g.,][]{Fraser_2005, Soszynski_2009, Spano_2011, Riebel_2015, Hey_2025}. Furthermore, the LMC has a geometric distance from detached eclipsing binaries which is precise to 1\% \citep{Pietrzynski_2019}. As a result, it is one of the most commonly used anchor galaxies in the extragalactic distance scale. Thus, understanding the effect of environmental differences on its stellar populations is particularly relevant for examining the consistency of stellar populations in anchor and SN Ia host galaxies in stellar-distance ladders. Although we do not make comparisons to the Small Magellanic Cloud (SMC) here, its AGB populations have historically been studied alongside those of the LMC to probe the effects of lower metallicity. 

The results of spectral surveys by \citet{Blanco_McCarthy_Blanco_1980} produced the first C- and M-star censuses in the Magellanic Cloud fields based on the optical regime. Later, this was expanded into the near-infrared by \citet{Cohen_1981}, establishing the NIR colors and bolometric luminosity functions of C-stars. After that, NIR surveys split AGB stars by CMD locations \citep{Nikolaev_and_Weinberg_2000}, identified C/M ratios as a tracer of metallicity \citep{Cioni_2006}, and built our current picture of AGB populations through Spitzer \citep{Boyer_2011} and Gaia \citep{Lebzelter_2018}. Similarly, from the variability and Period-Luminosity (PL) relations side, the discovery of the Mira PL relation \citep{Glass_Lloyd_Evans_1981}, the first PL sequences A-E (\citealt{Wood_1999} later refined by \citealt{Ita_2004}), and the current OGLE-III LPV Catalogs \citep{Soszynski_2009} were also focused on studies of the variable populations of the LMC. 

These findings have had an impact beyond the LMC because they were also exported to studies in more distant galaxies. On the stellar population side, the C/M tracer was expanded to M33 \citep{Cioni_2009}, NGC 6822 \citep{Sibbons_2012}, and M31 \citep{Boyer_2013, Boyer_2019}. Mira PL relations have also been applied to Local Group dwarfs \citep{Whitelock_2013}, M33 \citep{Yuan_2018, Konchady_2024}, and to galaxies in the cosmic distance ladder via NGC 4258 \citep{Huang_2018}, NGC 1559 \citep{Huang_2020}, and M101 \citep{2024ApJ...963...8H}. On the JAGB side, two independent studies originated in the LMC \citep{Madore_and_Freedman_2020, Ripoche_2020} and expanded to other nearby galaxies \citep[e.g.,][]{Freedman_and_Madore_2020, Zgirski_2021, Parada_2023} and into SN Ia hosts with JWST \citep{Lee_2025_JWSTJAGB, Li_2025}. 

While Galactic AGBs may be more similar to the M101 AGB population, most of these AGBs are too bright for the SPHEREx flux limits or do not yet have sufficient wavelength coverage for producing robust synthetic photometry. Galactic AGBs also suffer from larger distance uncertainties, which make population-level direct comparisons more difficult \citep{Andriantsaralaza_2022}. 

\subsection{SPHEREx Observations}

SPHEREx is the first all-sky spectral survey covering $0.75-5\mu$m in 102 spectral channels \citep{Crill_2020, Bock_2026}. Launched on March 11, 2025, its nominal science operations began on May 1, 2025. SPHEREx will eventually survey the entire sky four times over the course of its 25-month mission. SPHEREx scans great circles perpendicular to its polar orbit, and thus its all-sky coverage peaks towards the ecliptic poles, where its two Deep Fields are located. These Deep Fields are intended for intensity mapping studies and thus the LMC is excluded from the deep mosaic. However, the LMC lies only $\sim 4-5^\circ$ from the Southern Ecliptic Pole and is closer to the pole center than the center of the Southern Deep Field itself. As a result of its favorable location relative to the survey geometry, the LMC still accumulates hundreds of individual spectral samples per source per year. This paper makes use of the first year of SPHEREx observations.

\subsubsection{Target Selection}

We select AGB targets from the OGLE Catalog of Variable stars \citep{Soszynski_2009}.  While all AGBs are expected to be variable to some degree, not all LPVs are AGBs --- red supergiant stars, binaries, OGLE Small Amplitude Red Giants (OSARGs), and Long Secondary Periods (LSPs) --- also exhibit long period variations. To minimize contamination from non-AGB sources, we limit catalog members to include only semi-regular variables (SRVs) and Miras. The OSARGs are a mixture of pulsating small amplitude red giants and asymptotic giant branch stars. 

Then, we query the SPHEREx Spectral Image search to confirm that each target has at least 100 total observations and has been observed in all six SPHEREx bandpasses. For sources that passed these criteria (virtually all LMC sources we scanned), we then retrieve the spectra from the SPHEREx Data Explorer using the Spectrophotometry Tool \citep{Akeson_2025}. The retrieved spectra cover observations from April 24, 2025 to March 22, 2026; a small fraction of epochs ($\sim0.5\%$) were obtained during the final week of commissioning, before nominal operations began on May 1, 2025. Observations spanned between $238-321$ days per star, with a median baseline of 282 days and with a median number of $\sim 900$ good observations per star. We retrieved a total of 59 O-rich spectra and 34 C-rich spectra. 

\subsubsection{Spectral Data Cleaning}

We use bit-wise pixel quality flags (bitflags) provided by the SPHEREx Spectrophotometry Tool to apply cuts to the retrieved spectra \citep{Akeson_2025}. We removed any points with the {\tt NONLINEAR}, {\tt NONFUNC}, {\tt MISSING$\_$DATA}, {\tt FIT$\_$ERROR}, {\tt REFERENCE}, {\tt DICHROIC}, {\tt PHANMISS}, and {\tt SUR$\_$MISMATCH} flags. We consider these to be serious or fatal errors and likely to make any flagged observations unusable or unreliable. We also remove any negative flux measurements or measurements marked with the {\tt CONTAINS$\_$BAD$\_$PIXEL} product flag.

However, we keep points with bitflags that we deem moderate or minor in severity, under the assumption that the SPHEREx pipeline has at least partially corrected them: {\tt OVERFLOW}, {\tt HOT}, {\tt COLD}, {\tt PERSIST}, {\tt PERSIST$\_$UNK}, {\tt VAR$\_$UNDER}, {\tt CROSSTALK}, {\tt PHANTOM}. Finally, we considered {\tt TRANSIENT}, {\tt OUTLIER}, {\tt SUR$\_$ERROR} to be minor flags and also kept any pixels with these flags. As noted by \cite{Rustamkulov_2026}, data releases before December 2025 under-flagged bad pixels. This has now been corrected, but the flagging update was not applied retroactively. 

A comparison of epochs before and after December 2025 shows a $3.9\%$ rate of serious and fatal flags before and a $7.01\%$ rate after the fix was implemented. This suggests that on average $\sim 3\%$ of the earlier measurements are under-flagged. However, we do note that the exact rate appears band-dependent. For Band 3, there was no detected change in the percentage of flagged pixels, while in Band 4, the change was nearly 4\%. For the remaining three bands, the increase in flags ranged from $1.3-2.3$\%. Panel (a) of Figure \ref{fig:mean_spectrum_creation} shows the raw points (in gray) along with the bitflag-rejected points in red for one O-rich Mira in the sample. 

Because each SPHEREx observation samples only a narrow wavelength range, wavelength and pulsational phase can become correlated for a variable star that is well-detected during only part of its cycle. Naively averaging such a spectrum would then create biases of different amounts at different wavelengths. We screened all of the retrieved stars for this effect and discovered only a single star (OGLE-LMC-LPV-05800) flagged for incomplete cycle coverage. We rebuilt its mean spectrum instead by excluding its two sparsely sampled phase bins. 

\section{Methodology}
\label{sec:methods}

All AGB stars are variable, and the majority can be classified using the Long Period Variable (LPV) sequences on a Period-Luminosity Diagram. These sequences were first identified by \cite{Wood_1999} from MACHO Project data. While AGB stars are typically identified primarily via their brighter-than-the-TRGB CMD location, given that we also have time-series observations of both galaxies, we were able to employ two methods of selection: 1) a more traditional, CMD-based classification using the approximate TRGB location to mark AGB status and F182M excess to determine O- or C-like surface chemistry 2) a variability-based selection using long-period variability to identify AGB membership and a `water-index' (anticipated to be more effective on pulsating stars) for chemical classification. Throughout the paper, we compare the findings from both selection methods. Here, we use only the approximate amplitude of variability and not period. 

\begin{figure}[]
\includegraphics[width=0.47\textwidth]{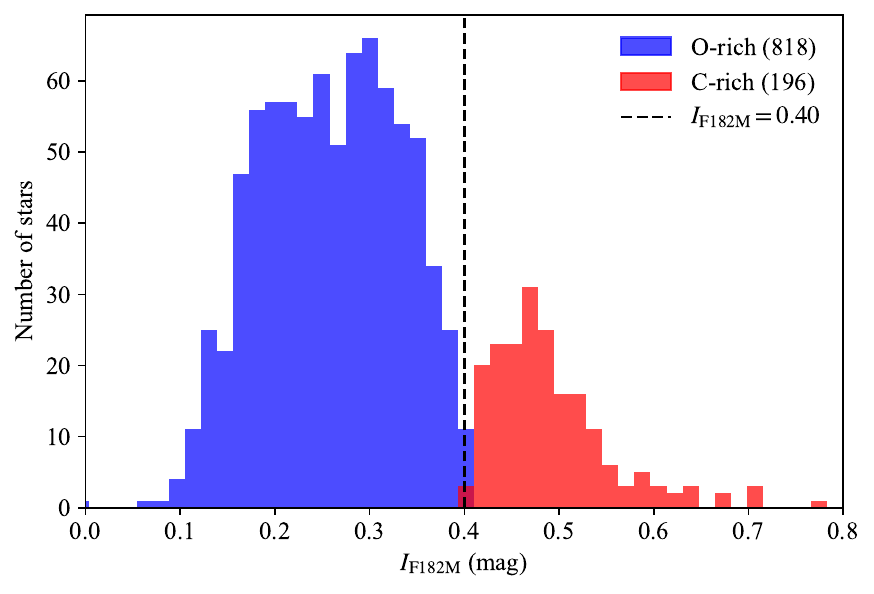}
\caption{The distribution of $I_{\rm F182M}$ values for variability-selected AGB stars in M101, with the empirically chosen $0.40$ value used to define the C- and O-AGB classifications. A small number ($N=5$) of stars had negative $I_{\rm F182M}$ values and were cut off from the plot to better show the distribution, but were still included in the analysis as O-rich stars. 
\label{fig:classification}}
\end{figure}

For the LMC population, we limit our OGLE-LPV population to only semi-regular and Mira variable stars for our `variable' selection. OSARGs are included only in the CMD-based selection (Section \ref{sec:lmc_cmd_agb}). While OSARGs also contain low-amplitude and early-AGB stars, they will not be detected as variable at the distance of an SN Ia host such as M101, and thus this split more closely reproduces the selection in M101. 

\subsection{M101 Variability-based AGB Selection}
\label{sec:m101_variable_selection}

\begin{figure*}[]
\includegraphics[width=\textwidth]{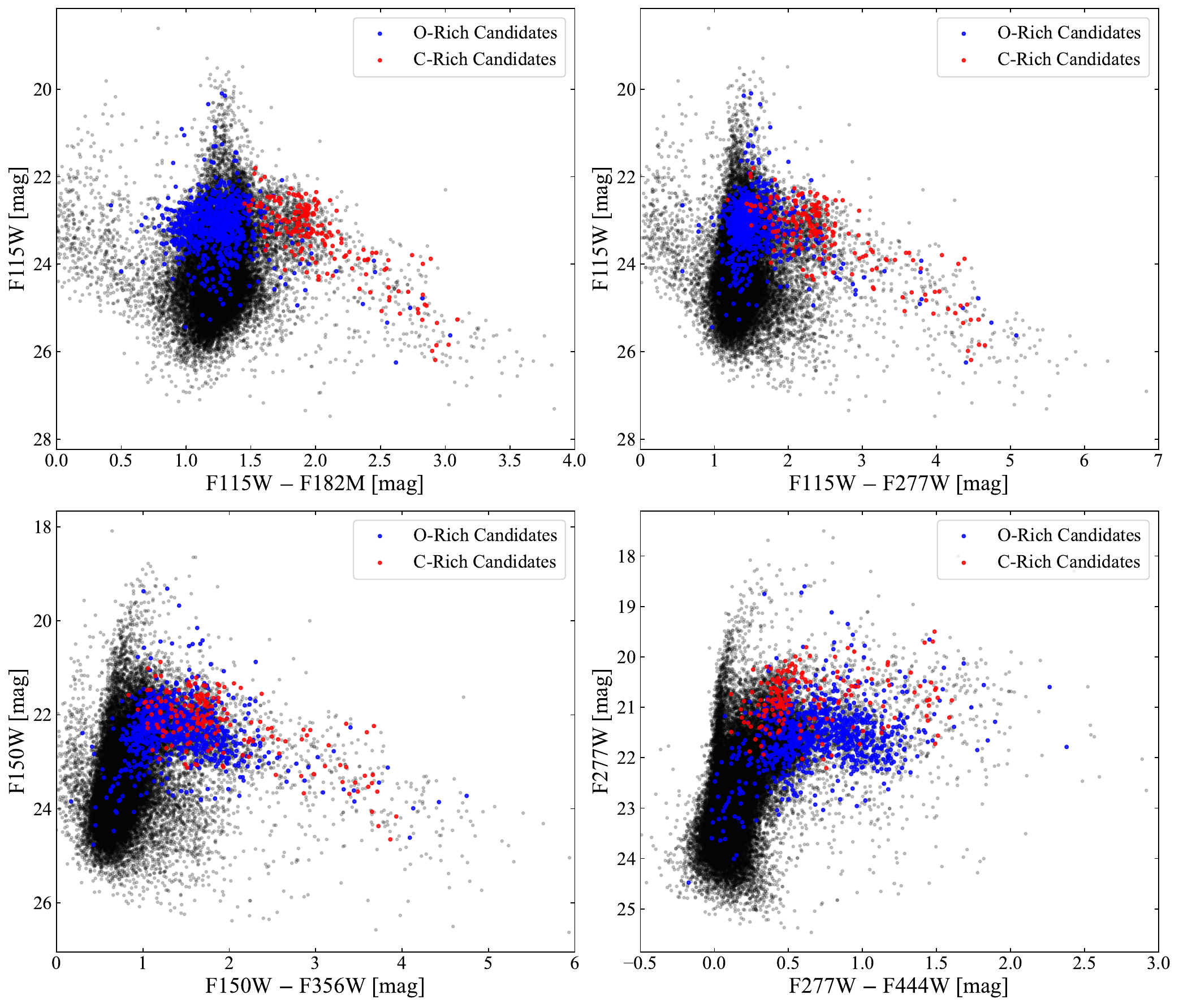}
\caption{NIRCam color magnitude diagrams of M101 with variability-selected O- (blue) and C- (red) AGBs highlighted. Classification was made following the selection criteria shown in Figure \ref{fig:classification} and explained in the text in Section \ref{sec:m101_variable_selection}.  
\label{fig:variability_cmds}}
\end{figure*}

To select long-period variables in M101, we obtained initial HST WFC3/IR F110W and F160W light curves and determined the Welch-Stetson variability index $L$ using the {\tt TRIALP} code provided by Peter Stetson \citep{Stetson_1996}. We consider only stars with at least 13 epochs of observation and $L > 0.75$, leaving us with a total of 2106 sources. While this is an aggressive selection that likely removes many true variables, our intention was to keep the sample as free of non-variable contaminants as possible. 

Next, we remove any potentially blended objects. This threshold is defined as any source that is within a 2.5 pixel radius of an equally bright or brighter source in the full catalog, leaving us with 1832 sources. We then do a preliminary fit of the periods in this remaining sample using $\chi^2$ minimization on a grid search of periods ranging from 40 to 1500 days. While fitting the periods, each light curve is modeled as a sine function and we leave only the amplitude, mean magnitude, and phase of the sine as free parameters. We also compared these results with Lomb-Scargle fits to the period and found agreement in $>$90\% of cases. Finally, we tested that a sinusoidal fit was favored over a straight-line fit for each source to confirm periodic behavior, resulting in a final catalog of 1811 long-period variables. We then applied a minimum amplitude cut of $0.2$ mag in WFC3/IR F160W. The period information is used only to limit our sample to long-period variables. Analysis of the period distribution will be discussed in a follow up work. 

For the C/O classification of the variables, we use a combination of three NIRCam filters: F150W, F182M, and F277W. For this analysis, we treat the two broadband F150W and F277W as pseudo-continuum filters. While they do have overlap with molecular features in both C- and O-AGB, the filter width also spans beyond the absorption bands of individual molecules. The three bandpasses do not overlap (Figure \ref{fig:filters}). The half-max widths of F150W and F277W are 1.33 -- 1.67 and 2.42 -- 3.13 $\mu$m, respectively. F182M (1.72--1.97 $\mu$m) falls in the gap between them and sits entirely within the $\sim 1.9\mu$m water band that is present only in O-AGB stars. Thus, it is used to separate between C- and O-rich surface chemistries.  We use the F150W and F277W filters to perform a linear interpolation in pivot wavelength between the two anchor magnitudes, evaluated at the F182M pivot wavelength. We call the pseudo-continuum expected magnitude at F182M $m_{\rm F182M}^{\rm cont}$. Its value is defined as
\begin{equation}
m_{\rm F182M}^{\rm cont} = m_{\rm F150W} + \frac{\lambda_{\rm F182M}-\lambda_{\rm F150W}}{\lambda_{\rm F277W}-\lambda_{\rm F150W}}\left(m_{\rm F277W}-m_{\rm F150W}\right).
\end{equation}
The pivot wavelengths for each of these filters are $\lambda_{\rm F150W} = 1.501\mu$m, $\lambda_{\rm F182M} = 1.845\mu$m, and $\lambda_{\rm F277W} = 2.776\mu$m. We then calculate a `water index', $I_{\rm F182M}$, defined as 
\begin{equation}\label{eq:IF182M}
I_{\rm F182M} = m_{\rm F182M}^{\rm cont}  - m_{\rm F182M}
\end{equation}
which measures the difference between the observed and linearly interpolated magnitude. We interpolate linearly in wavelength; interpolating in $\log \lambda$ shifts the weight towards the more separated F277W anchor and also folds in more of the dust-sensitive F150W - F277W color combination into the index, which degrades the separation of the two classes, erasing the bimodal distribution. More details of attempted variants are discussed in Appendix section \ref{sec:appendix_classification}. Fully expanded and with the pivot wavelengths substituted, this becomes, 
\begin{equation}
I_{\rm F182M} = 0.7302\,m_{\rm F150W} + 0.2698\, m_{\rm F277W} - m_{\rm F182M}.
\end{equation}
Figure \ref{fig:classification} shows the distribution of $I_{\rm F182M}$ values for all of the variability-selected AGBs. The index value is positive for nearly all stars because the linear interpolation between the F150W and F277W magnitudes is not a measurement of the physical continuum, but rather a comparison baseline. A cool star's SED introduces curvature at these wavelengths, which places its F182M flux above the interpolated line even when water absorption is present. Therefore, we use only the relative, rather than absolute, values of the $I_{\rm F182M}$ to determine if a star is O- or C-rich. Because water absorption is expected to depress the F182M magnitude and lower the $I_{\rm F182M}$ value, O-rich stars are expected to have systematically smaller $I_{\rm F182M}$ values than C-rich stars.  As a result, we see two distinct peaks in the $I_{\rm F182M}$ histogram: one we attribute to O-rich stars at median $\approx 0.23$ mag and one we attribute to C-rich stars, with median $\approx 0.47$ mag. To locate the boundary between the two groups without referencing the classifications themselves, we identify the local minimum of a Gaussian kernel density estimate (KDE) of the $I_{\rm F182M}$ distribution, which lies at $I_{\rm F182M} \approx 0.42$ mag. We then adopt $I_{\rm F182M} > 0.40$ mag for the C-rich classification (the difference between the adopted and measured values shifts the classification of only 20 of 1,014 variables). 

The $I_{\rm F182M}$ and F150W - F182M bimodalities are unlikely to reflect circumstellar dust rather than water absorption in the extended atmosphere. Firstly, $I_{\rm F182M}$ has been constructed to be differential compared to the local pseudo-continuum, so the broadband dust slope is expected to be largely divided out by construction. Secondly, we show in Sections \ref{sec:color_environment} and \ref{sec:variable_classification} that for the variability-detected AGB stars, the wide-band colors sensitive to dust (e.g., F150W - F444W) are $0.5-0.7$ mag redder in M101 than in the LMC, while the F150W - F182M width is consistent to $\approx0.05$ mag in both the LMC and M101, showing that its behavior is decoupled from general dust reddening. We therefore treat the bimodality in $I_{\rm F182M}$ and F150W - F182M as primarily tracing surface chemistry. 

We then plot the distribution of the variability-identified C- and O-AGBs on several combinations of NIRCam filters in Figure \ref{fig:variability_cmds}. We note that it is very likely that nearly all of the black sources in the same general region of our identified AGBs will also be variable. However, our strict variability criteria preclude many real variables from being included in the sample in order to minimize any spurious sources. 

\begin{figure*}[]
\includegraphics[width=\textwidth]{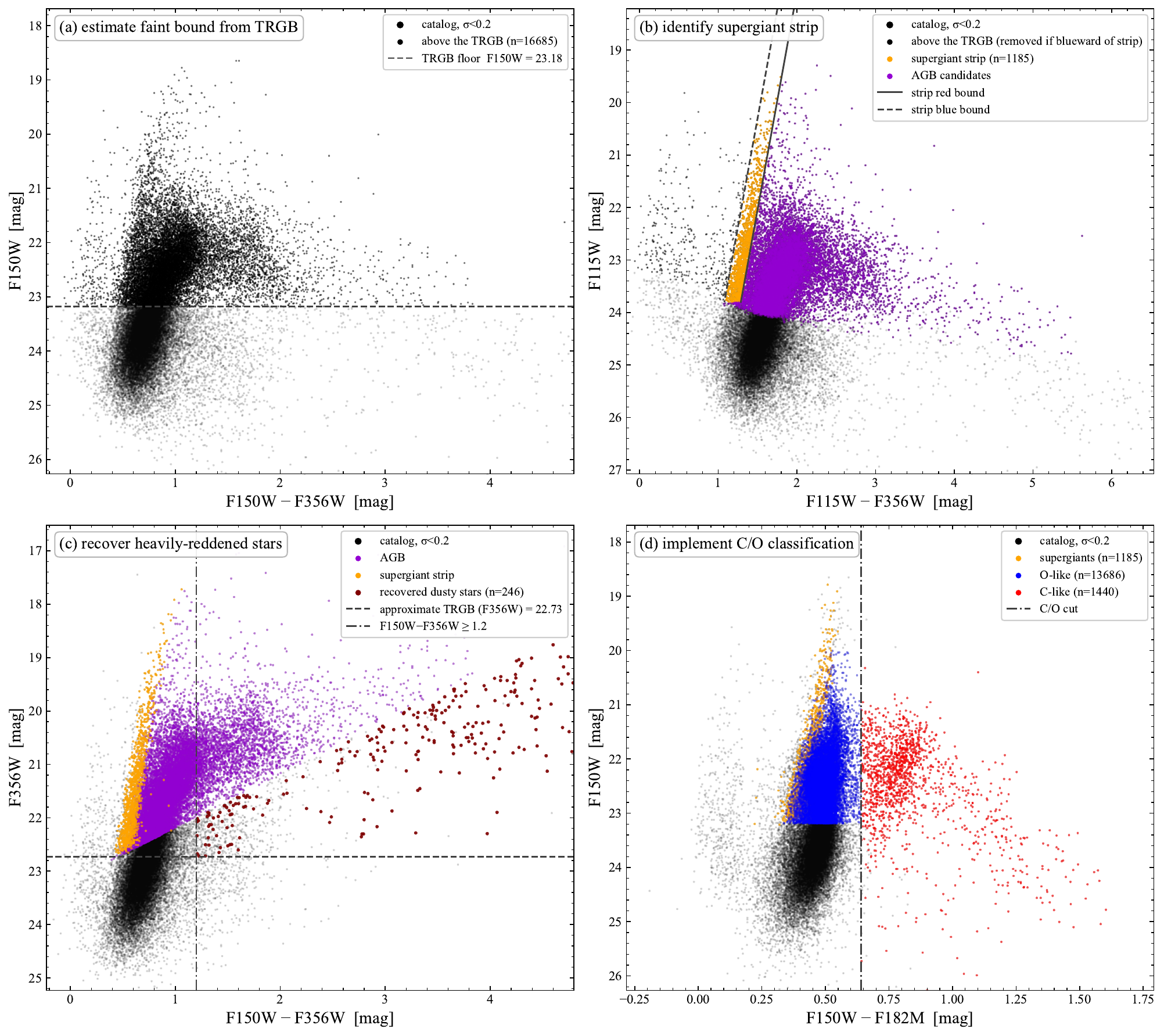}
\caption{CMD-based selection from M101 showing the different steps: a) TRGB faint magnitude bound b) identification of the supergiant strip c) recovery of reddened dusty stars d) C/O classification and additional dusty star recovery. 
\label{fig:m101_CMD_selection_process}}
\end{figure*}

\subsection{M101 CMD-based AGB Selection}
\label{sec:m101_cmd_agb}

\begin{figure*}[]
\includegraphics[width=\textwidth]{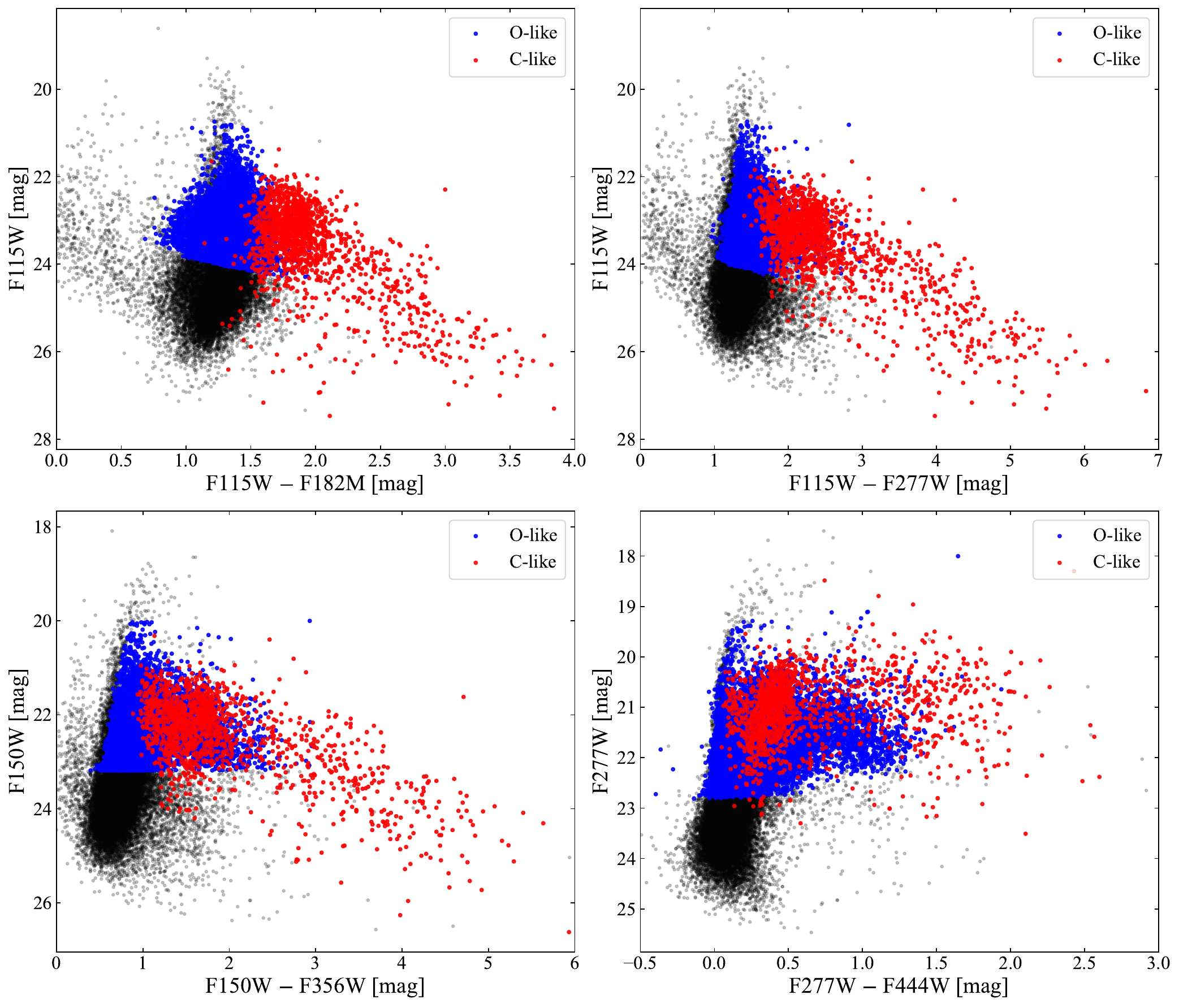}
\caption{NIRCam color-magnitude diagrams with M101 AGBs selected using their CMD characteristics as explained in Section \ref{sec:m101_cmd_agb}, colored by their presumed chemical classification (blue for O-like, red for C-like). The individual panels have the same extent and colors as shown in Figure \ref{fig:variability_cmds}.  
\label{fig:static_cmds}}
\end{figure*}

In addition to the AGB stars selected based on variability criteria, described in Section \ref{sec:m101_variable_selection}, we also select AGB stars based on their location on the CMD as a control. For this sample, we first create a cleaned CMD using strict quality cuts to remove background galaxies and other contaminants. These are based on the following thresholds combining both the recommendations of \cite{Warfield_2023} and the criteria used in \cite{Anand_2025} to separate stars and background galaxies:  
\begin{itemize}
    \item[] \texttt{CROWDING < 0.5,}
    \item[] \texttt{SNR > 5,}
    \item[] \texttt{FILTER\_SHARP\^{}2 <= 0.01,}
    \item[] \texttt{FILTER\_CROWD <= 0.5,}
    \item[] \texttt{FILTER\_FLAG <= 2, and}
    \item[] \texttt{OBJECT\_TYPE <= 2}
\end{itemize}
These are similar to the prescriptions used by \cite{Anand_2025}, but we apply these criteria to all of the filters rather than to any single pair of filters. This allows us to create one AGB sample that is stable in membership across all of the NIRCam band CMDs at the cost of removing more stars.  We also apply a cut requiring photometric error to be $< 0.2$ mag in all six bands. 

Next, we isolate the AGB-region of the CMD. First, we place a faint bound at F150W = 23.18 mag, an approximate TRGB location for the adopted $\mu=29.18$ mag and M$_{\rm TRGB, F150W} = -6.0$ mag (close to the tip magnitude at the red end of the color range calibrated by \citealt{Newman_2024}). The near-infrared TRGB is known to be sloped, brightening towards redder colors \citep[e.g.,][]{Dalcanton_2012, Newman_2024}. We do not model or attempt to fit this slope because the floor will only affect stars within $\pm0.1$ mag of it and most of our subsequent analysis uses medians and 16th-84th percentile ranges, which are insensitive to both the brightest and faintest stars in the distribution. The exact choice of bound therefore has little effect on our conclusions; in the LMC, moving the tip by $\pm 0.1$ mag changes our color comparisons by at most 0.004 mag (see Section \ref{sec:lmc_cmd_agb}). 

Because the floor is applied as a horizontal cut in F150W, the same stars show a diagonal lower envelope in CMDs created from other magnitudes (Panels (b) and (c) of Figure \ref{fig:m101_CMD_selection_process}); this envelope reflects the cut, not the TRGB slope. We also employ a hard luminosity cutoff at m$_{\rm F150W} = 20.0$ mag ($M_{\rm F150W} \approx -9.2$ mag), which removes stars brighter than nearly all AGB stars. In the LMC, only 0.2\% of OGLE-III SRVs and Miras are brighter than the equivalent absolute magnitude in the 2MASS H band ($M_H \approx -9.2$ mag). Such stars may be of interest in their own right, but they are outside the scope of this work. The faint bound is shown in Panel (a) of Figure \ref{fig:m101_CMD_selection_process}. Next, we identify the supergiant strip (in yellow) by eye in the F115W$-$F356W CMD, removing both it and everything blueward of the strip. Removing the strip, which contains only O-like stars and no variables, changes the C$-$O color separations by at most 0.03 mag and leaves the J-region samples unchanged. The remaining stars are left as candidate AGBs (purple points in Panel (b) of Figure \ref{fig:m101_CMD_selection_process}). 

Next, we expect that the dusty reddened tail of the CMD contains real AGB stars whose circumstellar dust prevents them from being detected above the F150W TRGB. These we recover by employing two criteria to select heavily reddened stars while minimizing the number of truly fainter RGB stars that we include. The first is the F150W - F356W $\geq 1.2$~mag cut, shown in Panel (c), which admits stars that are above where the blue edge of the TRGB appears to start in F150W - F356W and had significant F356W excess (maroon points). These stars must also be brighter than the F356W TRGB measured from this catalog (F356W $< 22.73$ mag; Panel (c)). The second are the stars redward of F150W$-$F182M = 0.64 mag, which we also use to classify stars as either C- or O-rich. Faint, red point sources can potentially include compact background galaxies that survive the quality cuts. However, 26 of 246 of these dusty sources are detected variables (Section \ref{sec:m101_variable_selection}), nearly twice the rate in the main CMD-selected AGB sample, and their spatial distribution is indistinguishable from those of AGB stars of similar F150W magnitude. We note that the AGB stars in this field are distributed nearly uniformly, as are the Miras in NGC 4258 \citep[see Figures 18 and 19 of][]{Huang_2018}. 

The threshold value of 0.64 mag is the midpoint between the median colors of the O- and C-star classes from the variability-detected sample (Section~\ref{sec:m101_variable_selection}). As a check that the boundary is meaningful for this photometrically selected population, we also apply the same Gaussian-KDE minimum detection to the CMD-selected sample's color distribution, with no prior knowledge of classification. It independently shows a valley at $0.663$ mag. Adopting either value reclassifies only 80 of 15,126 stars. We include stars redder than this threshold regardless of their F150W magnitudes because these are likely to be C-rich stars, which will primarily be AGB stars by construction (at the distance of M101, extrinsic carbon stars are expected to be difficult to detect). The O-AGB stars are shown as blue points, C-AGB as red points, and supergiants as yellow points in Panel (d) of Figure \ref{fig:m101_CMD_selection_process}. Finally, we also show the distribution of these AGB stars across the same representative CMDs as used for the variability-classified AGB sample in Figure \ref{fig:static_cmds}.
\\

\subsection{LMC CMD-based AGB Selection}
\label{sec:lmc_cmd_agb}

For the LMC, we primarily use the full OGLE-III LPV catalog \citep{Soszynski_2009} crossmatched with 2MASS, WISE, and Spitzer IRAC [3.6] photometry described in Section~\ref{sec:data} (detailed description to appear in Capodagli et al., in preparation).  In order to create an LMC counterpart for the variability-blind, CMD-only selection used for M101 (Section~\ref{sec:m101_cmd_agb}), we also select the LMC AGB region directly via the $K_s$ versus $J-K_s$ CMD without using the OGLE variability classes. This forms the ``LMC CMD-selected'' sample presented in Table~\ref{tab:analysis_samples}. This allows us to retain the OSARGs that OGLE classifies as AGB stars (OSARGs include both RGB and AGB stars), which more closely mirror the stars in the AGB region of the M101 CMD but would have no detected variability at the distance of M101 (Section \ref{sec:var_vs_cmd_setup}). 

For the faint bound, we use the literature TRGB value of $K_{s, \rm TRGB} = 11.94$ mag \citep{Cioni_2000}. While their value is dereddened and our catalog magnitudes are not, varying the tip value by $\pm 0.1$ mag does not affect any color comparisons we derive in Section \ref{sec:var_vs_cmd} by more than 0.004 mag. We use a bright bound of $K_s = 8.76$, in order to match the range of the CMD region used to select AGB stars in M101, described in Section \ref{sec:m101_cmd_agb}. In order to fit the sloped blue edge, we use a straight-line fit to the blue (first-percentile) envelope of the SRV and Mira colors in five $K_s$ bins. This gives us a blue edge of $J-K_s \ge 2.089 - 0.0898\, K_s$. We do not use a red edge. These selection criteria result in a total of 19,346 LPVs consisting of 9,585 OSARGs, 8,416 SRVs, and 1,345 Miras. 

Because the LPV catalog is variability-selected, we also verify that it is an unbiased representation of the CMD, so that selecting from it is equivalent to using a variability-blind selection. To do this, we retrieved a subset of the 2MASS Point Source Catalog (2$^{\circ}$ cone search on the bar, $J<16.5$, AAA quality) --- a selection with no variability information --- and applied the same CMD cuts as described above. This yields a smaller catalog of 12,758 sources (the smaller count reflects the smaller footprint, and the requirement to have A quality in all three bands). Of these, we found that 97.8\% (12,474) have an OGLE-III LPV counterpart within 1$''$ (6,401 SRV + Mira, 6,073 OSARGs); the remaining 284 have no OGLE-III LPV counterpart. Considering that even the AAA catalog likely includes some foreground contamination, effectively \textit{all} the stars in this region of the CMD are variable. This is also in agreement with theoretical predictions that all AGB stars are variable to some degree. The fraction of non-variable stars in the AGB region is the number most affected by the choice of TRGB tip, increasing from 2.1 to 2.7\% from brighter to fainter tip values, likely because a lower TRGB threshold allows more non-variable RGB stars. The OSARGs also reproduce the median colors of the 284 non-LPV catalog members to $\le 0.01$ mag, so treating them as equivalent to the CMD-only ``no detected variability" population introduces no measurable color bias.

\subsection{Analysis of SPHEREx Spectra}\label{sec:spherex_analysis}

In order to compare the SPHEREx spectra with M101 photometry, we first created mean spectra from the individual SPHEREx observations. We then use this as input for synthetic photometry.

\subsubsection{Mean Spectrum Creation}
\label{sec:mean_spec}

Given that AGB stars are variable by nature, their total flux and spectral energy distributions change between epochs. In addition, the strength of molecular absorption features in AGB spectra are known to be phase-dependent \citep[e.g.,][]{Lebzelter_2001, Wittkowski_2007}. SPHEREx will visit each star many times over the survey window, so the final spectrum will consist of observations taken at different pulsation phases, potentially across multiple pulsation cycles (shown in Panel B of Figure \ref{fig:mean_spectrum_creation}). In most epochs, only a few of the six SPHEREx bands will be observed, thus leading to inconsistent wavelength coverage from epoch-to-epoch. Naively averaging these spectra in flux would weight each phase by how often it happened to be observed and by its brightness, favoring the brightest phases, which near the SED peak and on the Wien side are generally the warmest. Because each epoch covers only some of the bands, different bands would also be averaged over different sets of phases. Together, these effects can smear the phase-dependent structure together and bias the colors, potentially toward a featureless continuum. 

For each star, we instead group the observations by pulsational phase as predicted by the OGLE periods (see Panel C of Figure \ref{fig:mean_spectrum_creation}), average over pulsational cycle-to-cycle variations, and then combine all of the phased means into one representative spectrum, which we call the final `mean spectrum.' This preserves the features of the spectrum at each phase in order to more accurately reproduce mean colors at the cost of losing absolute flux information. It is important to note that the mean spectrum does not represent a specific phase of the star's pulsational cycle since it averages over every phase observed.

\begin{figure*}[]
\includegraphics[width=\textwidth]{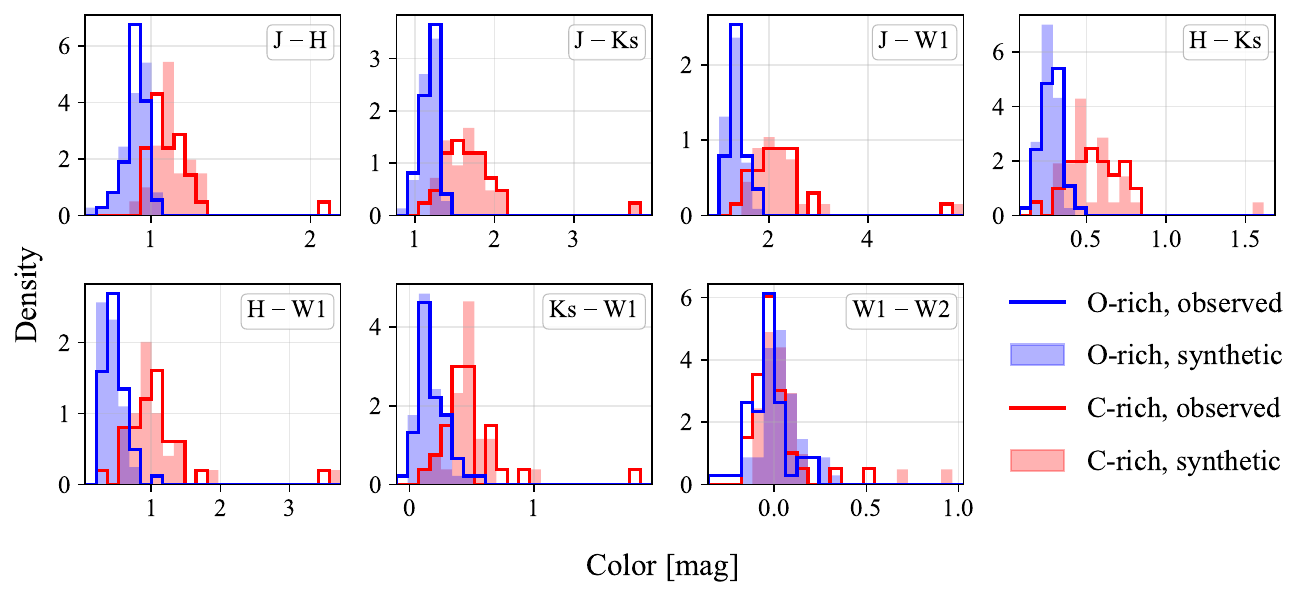}
\caption{Comparison of the observed 2MASS and CatWISE colors (unfilled histograms) of O- and C-rich AGB stars in the LMC with the synthetic colors of the same stars derived from their SPHEREx spectra (filled histograms), after the $J$-band correction described in Section \ref{sec:obs_synth_comparison}.  \label{fig:obs_synth_comparison}}
\end{figure*}

We fold each star's observations using its OGLE period and sort them into 10 phase bins. For stars with periods shorter than the SPHEREx baseline, a phase bin can contain observations from several pulsation cycles. Within a phase bin, we assume that the \textit{shape} of the spectrum is the same from cycle to cycle but that its \textit{flux} may differ, and that the flux difference depends on wavelength (it is larger at shorter wavelengths). We therefore take the cycle with the most observations in each phase bin as the reference, and scale the observations from each other cycle, separately in each of the six SPHEREx bands, to minimize the string length of the combined spectrum. The scaled observations are then combined with inverse-variance weights into one spectrum per phase bin. 

In order to combine the phase bins, we fit a single wavelength-independent scale factor for each phase-bin spectrum over $1.2-2.5\mu$m that matches it to the mean of all phase bins, iterating three times, and set the overall flux level to that of the median phase bin. The final mean spectrum is the inverse-variance-weighted mean of these aligned phase spectra. Because each phase is scaled to a common flux level before the phases are combined, the brightest phases do not dominate the mean, as they would in a straight average in flux. A representative mean spectrum is shown in Panel D of Figure \ref{fig:mean_spectrum_creation}. We then use transmission functions obtained from the Spanish Virtual Observatory (SVO) Filter Profile Service\footnote{\url{https://svo2.cab.inta-csic.es/theory/fps/}}\citep{Rodrigo_2012, Rodrigo_2020, Rodrigo_2024} to compute synthetic Vega magnitudes for each star. Details regarding the computation of synthetic magnitudes can be found in Appendix \ref{sec:appendix_synth_mag}. 

\subsubsection{Comparison with Observed Magnitudes}
\label{sec:obs_synth_comparison}

\begin{deluxetable}{lccccccc}
\tablecaption{Observed vs.\ synthetic photometry of the LMC sample\label{tab:obs_vs_synth}}
\tablewidth{0pt}
\tablehead{\multicolumn{8}{c}{Individual Bands}}
\startdata
Band & $n_{\mathrm{O}}$ & $n_{\mathrm{C}}$ & $\Delta_{\mathrm{O}}$ (mag) & $\Delta_{\mathrm{C}}$ (mag) & $\sigma_{\mathrm{O}}$ (mag) & $\sigma_{\mathrm{C}}$ (mag) & \\
\hline
$J$ & 53 & 30 & $+0.154$ & $+0.169$ & 0.151 & 0.207 & \\
$H$ & 53 & 31 & $+0.246$ & $+0.242$ & 0.162 & 0.224 & \\
$K_s$ & 53 & 30 & $+0.271$ & $+0.308$ & 0.158 & 0.193 & \\
W1 & 57 & 34 & $+0.301$ & $+0.290$ & 0.176 & 0.232 & \\
W2 & 57 & 34 & $+0.257$ & $+0.277$ & 0.164 & 0.222 & \\
\hline
\noalign{\vskip 1.5ex}
\hline
\multicolumn{8}{c}{Colors} \\
\hline
Color & $n_{\mathrm{O}}$ & $n_{\mathrm{C}}$ & $\Delta_{\mathrm{O}}$ (mag) & $\Delta_{\mathrm{C}}$ (mag) & shift (mag) & KS$_{\mathrm{O}}$ ($\sigma$) & KS$_{\mathrm{C}}$ ($\sigma$) \\
\hline
J$-$H & 53 & 30 & $-0.090$ & $-0.056$ & $-0.080$ & $>5$ & 1.8 \\
J$-$Ks & 53 & 30 & $-0.115$ & $-0.101$ & $-0.114$ & $>5$ & 0.9 \\
J$-$W1 & 51 & 30 & $-0.140$ & $-0.067$ & $-0.098$ & 3.2 & 0.2 \\
J$-$W2 & 51 & 30 & $-0.069$ & $-0.054$ & $-0.059$ & 2.5 & 0.1 \\
H$-$Ks & 53 & 30 & $-0.020$ & $-0.051$ & $-0.025$ & 1.3 & 0.5 \\
H$-$W1 & 51 & 31 & $-0.040$ & $-0.017$ & $-0.031$ & 1.8 & 0.2 \\
H$-$W2 & 51 & 31 & $+0.004$ & $+0.021$ & $+0.016$ & 0.3 & 0.2 \\
Ks$-$W1 & 51 & 30 & $-0.021$ & $+0.041$ & $+0.011$ & 1.1 & 0.9 \\
Ks$-$W2 & 51 & 30 & $+0.030$ & $+0.056$ & $+0.048$ & 1.1 & 0.9 \\
W1$-$W2 & 57 & 34 & $+0.059$ & $+0.041$ & $+0.056$ & 3.4 & 1.3 \\
\enddata
\tablecomments{ $n$ is the number of stars with observed photometry available from each class; $\Delta$ is the median (across all stars) of the synthetic $-$ observed residuals for C and O stars; $\sigma$ is the standard deviation of the residuals; ``shift'' is the median offset of the full (C+O) set subtracted from their synthetic colors; KS gives the Gaussian-equivalent significance of a two-sample KS difference between the uncorrected synthetic and observed distributions. See Section \ref{sec:obs_synth_comparison} for full details.}
\end{deluxetable}

We compare the phase-averaged synthetic magnitudes computed from the SPHEREx mean spectra (Section~\ref{sec:mean_spec}) with the observed catalog photometry of the same stars described in Section \ref{sec:observations_lmc}: single-epoch, random-phase 2MASS $J$, $H$, and $K_s$, and multi-epoch, flux-averaged CatWISE2020 $W1$ and $W2$. The synthetic magnitudes are $\sim 0.15-0.30$ mag fainter than the observed magnitudes. However, the range is largely driven by the difference in the $J$ band observed vs. synthetic magnitudes. For the four other filters, the zeropoint offset is nearly uniform (spanning $\sim 0.22 - 0.29$ mag). Table \ref{tab:obs_vs_synth} shows the difference in the median observed vs. synthetic data in each bandpass as well as the approximate Gaussian-equivalent significance of the differences. We use the median rather than the mean to be less sensitive to the long tails of heavily reddened stars. %

More relevant to our analysis, however, is not the magnitude offset in any individual band, which is set by the normalization of the mean spectra (Section~\ref{sec:mean_spec}), but rather the zeropoint offset in color space for each pair of filters. Because the zeropoint offset between observed and synthetic magnitudes in each filter is nearly the same, the difference between observed and synthetic \textit{colors} is less than $\sim 0.05$ mag for any combinations that do not use the 2MASS $J$ band. As shown in Table \ref{tab:obs_vs_synth}, only five out of 20 distributions are not consistent to within $\sim2\sigma$, four of which involve the 2MASS $J$ band. Of the LMC observations we can compare to, the 2MASS $J$ band is the only one primarily using SPHEREx Band 2. 

The last remaining outlier, $W1-W2$, has a small color offset between O- and C-rich subtypes. Both bands sample the Rayleigh-Jeans-like tail of the SED, so the color carries almost no temperature information. Thus, a small offset in color will result in a significant detection for a color that is nearly degenerate with zero. The $W2$ filter also extends beyond the red limit of the SPHEREx coverage, leaving 17.5\% of the band uncovered, which causes a bright bias in the synthetic flux because we use a mean flux based on the covered part of the $W2$ band. 

Because the synthetic and observed colors are measured for the same stars, the two-sample Kolmogorov-Smirnov (KS) \citep{Kolmogorov_1933, Smirnov_1948} test — which assumes independent samples — is not strictly applicable; we therefore treat its $p$-values (and their Gaussian equivalents) only as a qualitative check that the color distributions agree in shape. We base all quantitative statements on the median offset of each star rather than making any claims about the shape of the overall distributions. However, it is expected that the observed and synthetic photometry will have scatter since the 2MASS observations are randomly phased, and the WISE and SPHEREx observations are mean-phase. 

Relative to the other synthetic bands we have calculated, the $J$ band appears to have a $\sim 0.1$ mag excess. The larger number of O-rich star spectra makes this effect more significantly detectable in the O-rich star sample, but this value is consistent between both C and O classes (only a $\sim 0.015$ mag differential between the two). Thus, it does not appear to be easily explained by telluric absorption or chemical class. Furthermore, we verify that this excess is present even if we use SPHEREx raw spectral points for synthetic photometry rather than the mean spectra, so it is not an artifact of the mean-spectrum creation process. 

If we correct for this excess, we find that, as shown in Figure \ref{fig:obs_synth_comparison}, the observed and synthetic photometry are fully consistent to within the uncertainties for every band and chemical type combination. However, the rest of our analysis does not make use of this empirical correction because we are primarily concerned with the differences in color between C and O subtypes (where the $J$-band shift largely cancels) and because we do not understand the origin of the offset. 

\subsection{Classification using Spectra}
\label{sec:spectral_classification}

We classify each star as carbon- or oxygen-rich using visual inspection of its mean SPHEREx spectrum. While all OGLE LPVs already have an OGLE photometric classification of its chemical type, this allows us to cross-check these chemical classifications with our mean spectra for consistency. AGB spectra in the near-infrared regime are made up of three main components: 1) the stellar continuum, which peaks at $1-2\mu$m depending on temperature, 2) chemistry-dependent molecular absorption features, which can be found throughout the $0.75 - 5.0 \mu$m regime studied here, and 3) infrared dust excess, which typically begins at around $2\mu$m for many stars and increases towards longer wavelengths. Broadband color classifications often attempt to measure effective temperature or dust excess in an attempt to classify C- and O-AGB stars \citep[e.g.,][]{Nikolaev_and_Weinberg_2000, Cioni_2006, Boyer_2011}, but these can be misleading and are often quite sensitive to the exact bandpasses chosen \citep[e.g.,][]{Dalcanton_2012, Boyer_2013, Boyer_2024}. 

Literature classifications based on spectra have most commonly focused on using $<2.5 \mu$m molecular absorption bands to discriminate between the carbon- and oxygen-rich chemical subtypes \citep[e.g.,][]{Lancon_and_Wood_2000, LiL_2024, Li_2026, Nidever_2026}. In O-rich stars, there are TiO and VO bands; in C-rich stars, there are CN and C$_2$ bands. However, the choice of wavelength range is the result of necessity --- ground-based spectra have telluric lines that overlap with some of the most prominent molecular features in AGB spectra starting at 1.4 microns and longer wavelengths. 

\begin{figure*}[]
\centering
\includegraphics[width=\textwidth]{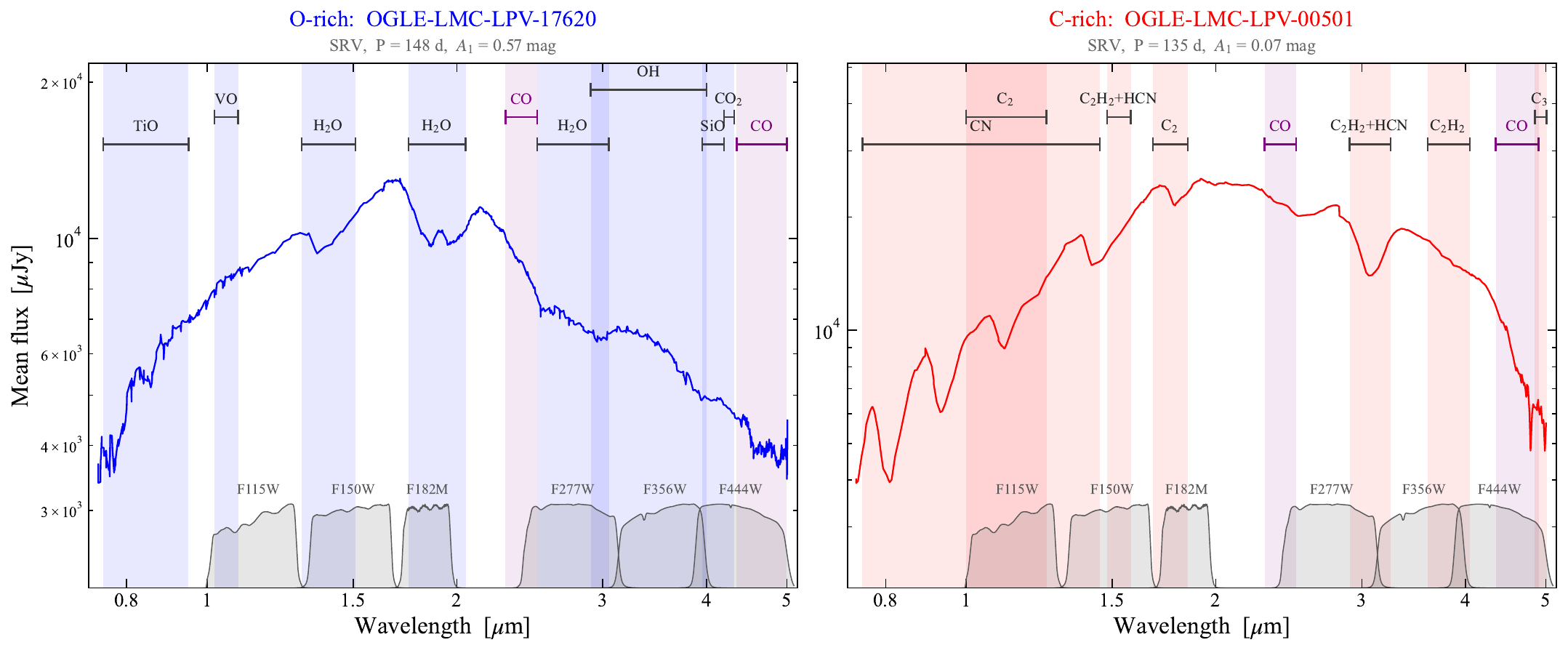}
\caption{SPHEREx spectra of C- and O-rich AGB stars of comparable period. The locations of known molecular bands in both types are overlaid in pale blue (corresponding to O-based molecules), pale red (C-based molecules), and purple (present in both types of stars). The gray filled curves on the bottom of each panel show the wavelength location of the transmission curves of the six NIRCam bands used in this work drawn on an arbitrary vertical scale. 
\label{fig:moleculesandatoms}}
\end{figure*}

Our analysis of the entire $0.75 - 5 \mu$m range suggests that the most visually distinctive features, when present, are the C$_2$H$_2 +$HCN absorption band at $3.05\mu$m in C-rich stars and the coherent H$_2$O band pattern at 1.4, 1.9, and 2.7$\mu$m in O-rich stars (see Figure \ref{fig:moleculesandatoms}). These are the features that we use for classification. The CN bands below 1.5$\mu$m are also present in most C-rich spectra (Figure \ref{fig:moleculesandatoms}), but they fall in the same wavelength range as the narrower TiO and VO bands of O-rich stars and are occasionally difficult to tell apart by eye without marking band positions. When the H$_2$O and C$_2$H$_2 +$HCN bands are not present, we instead rely on the TiO + VO for O-rich classification and CN + C$_2$ for C-rich classification. Finally, if the spectra showed no identifiable molecular absorption features at all and was within the appropriate flux range to be an LMC AGB star, we classified it as O-rich. 

Using our classification scheme, only one SRV originally classified photometrically by \cite{Soszynski_2009} as O-rich (OGLE-LMC-LPV-13034) was found to instead be a C-rich SRV based on the presence of distinctive 3$\mu$m C$_2$H$_2 +$HCN bands. While our sample consists of only 93 stars, preliminary independent verification with members of the SPHEREx team confirmed similar rates. Out of 932 Miras and SRVs identified as C-rich via Gaussian process regression classification of SPHEREx spectra, 926 of them were also identified as C-AGBs based on OGLE photometric classifications. The remaining six were originally classified as O-AGB stars by OGLE but showed the same 3$\mu$m C$_2$H$_2 +$HCN bands. 

\subsection{SPHEREx Sample Representativeness}
\label{sec:subsample_rep}

The SPHEREx Spectrophotometry pipeline is computationally expensive. Thus, we are only able to retrieve a relatively small sample (93 spectra) of spectra for LMC stars. To verify that the stars with retrieved SPHEREx spectra are indeed an unbiased and representative subsample of the parent population, we compare the observed 2MASS and CatWISE colors of the SPHEREx sample with the full OGLE-III SRV and Mira Catalog (12,795 stars: 6,445 O-rich, 6,350 C-rich). 

\begin{figure*}[]
\includegraphics[width=\textwidth]{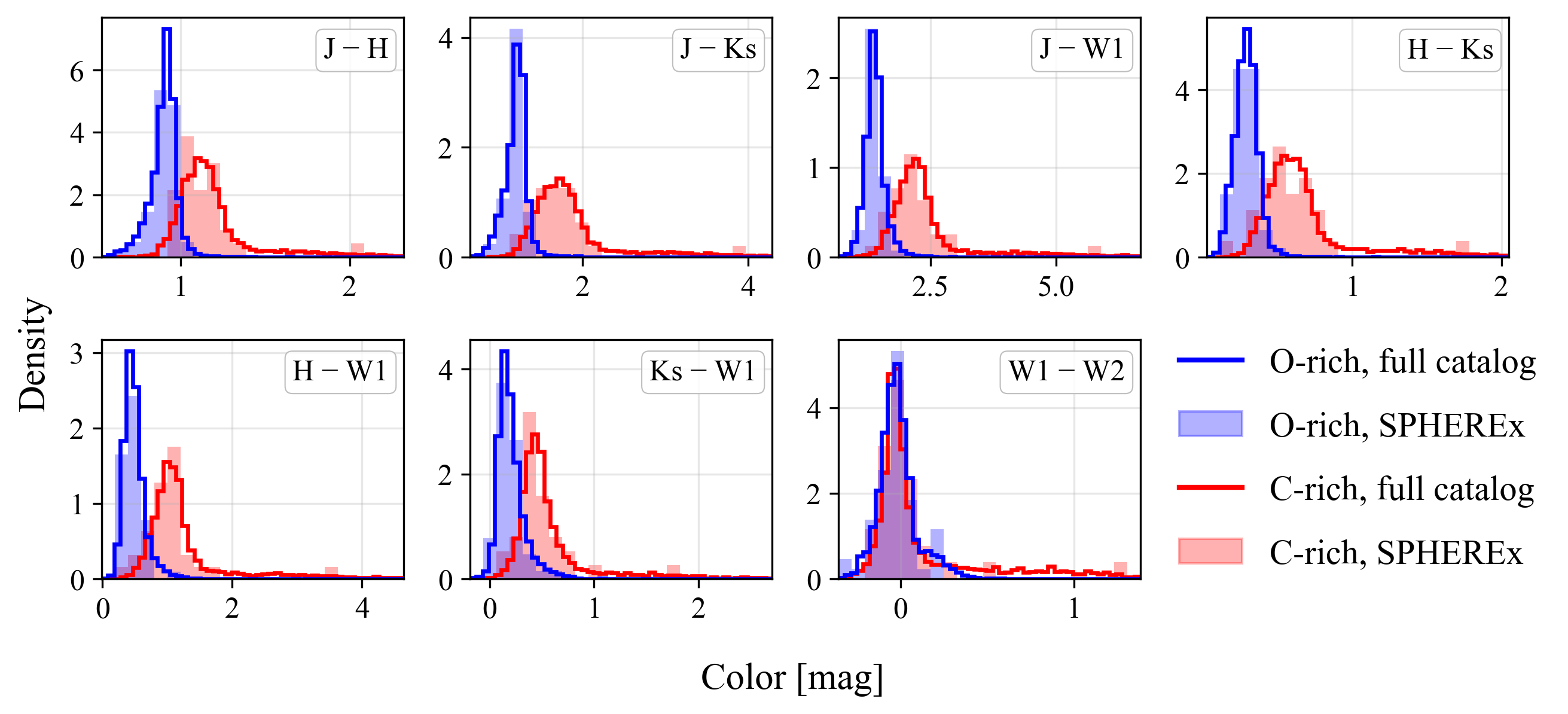}
\caption{Comparison between the density-normalized observed color distribution of the subset of AGB stars with retrieved SPHEREx spectra (shaded, filled histograms) with the observed color distribution of the full OGLE SRV and Mira parent sample (unfilled step histograms). 
\label{fig:synth_vs_observed}}
\end{figure*}

Figure \ref{fig:synth_vs_observed} shows the density-normalized distribution of observed colors for the stars in the SPHEREx sample (filled histograms) overlaid on the distribution of the observed colors of the broader OGLE parent sample (unfilled histograms). The parent sample consists of SRVs and Miras in the OGLE LMC LPV catalog. While Figure \ref{fig:synth_vs_observed} shows only seven of the 2MASS and CatWISE colors, we tested all 10 color combinations. 

We employ a two-sample KS test for both C and O stars in all 10 possible colors using $J, H, K_s, W1,$ and $W2$. We find that a KS test does not reject a common parent distribution in any of the 20 cases. We also use an Anderson-Darling $k$-sample test \citep{Scholz_1987} --- which is more sensitive to the tails of the distribution --- which also does not reject any color distributions. The small size ($\mathcal{O}(100)$) of the SPHEREx sample is the main limit on the statistical power of these tests. Finally, we also compare the medians directly, and find that the O-rich sample matches the catalog to within $0.04$ mag in every color, and the C-rich matches the sample to within $0.1$ mag in every color. Both are consistent to within their uncertainties. 

Thus we conclude that the SPHEREx sample is representative of the full OGLE catalog in the location, width, and skew of the colors for both C- and O-AGB classes, separately. We note however that the ratio of C-to-O stars of the SPHEREx sample is not representative of the overall distribution of C and O stars in the LMC --- this discrepancy is a result of the successful retrieval fractions from the SPHEREx Spectrophotometry tool.

\subsection{Field Metallicities}
\label{sec:methods_metallicity}

Neither the LMC nor our M101 field has a precisely known absolute metallicity for its intermediate-age AGB stars. In \citet{2024ApJ...963...8H}, we adopted a metallicity of approximately half solar based on the abundance gradient of \citet{Mager_2013}, evaluated at the galactocentric radius of our field. However, this estimate assumes a smooth gradient and is based on a nebular abundance scale from \citet{Kennicutt_2003}. The LMC is far better studied, but the value relevant to its intermediate-age AGB population still depends on the metallicity tracer used. Calcium-triplet spectroscopy of red giants in the bar gives a median [Fe/H] = $-0.40$ dex \citep{Cole_2005}, photometric metallicity maps of the red giant branch give a galaxy-wide mean of $-0.42$ dex with a shallow radial gradient \citep{Choudhury_2021}, and intermediate-age star clusters --- which are coeval with the progenitors of current AGB stars and are thus the most directly applicable tracer --- give a mean of -0.48 dex with an rms of only 0.09 dex \citep{Grocholski_2006}. We therefore assume [Fe/H] $\approx -0.5 \pm 0.1$ ($\approx 1/3 Z_{\odot}$) as the literature value most appropriate for the mean metallicity of the LMC's AGB population. These are iron abundances calibrated from the calcium triplet \citep{Cole_2005, Grocholski_2006}. Intermediate-age LMC stars are mildly $\alpha$-poor ([$\alpha$/Fe] $\approx -0.1$ to $-0.2$ at these metallicities, as adopted by \citealt{Cole_2005}), so their total metallicity is [M/H] $\approx$ [Fe/H] $-0.1$ \citep{Salaris_1993}, which would only widen the difference from our M101 field. 

Given the differences in methodology for obtaining these metallicity measurements, it is difficult to directly compare the metallicities of the LMC and M101. However, we can estimate the metallicity difference between the two galaxies directly from our data using the number ratio of C-rich to O-rich AGB stars (C/M), which is inversely correlated with metallicity \citep{Battinelli_and_Demers_2005, Cioni_2009}. While absolute metallicities are difficult to derive from the C/M ratio due to differences in how the O-rich sample is defined between studies \citep[metallicity values are expected to vary by a factor of 1.5--4.0,][]{Boyer_2019}, we can make a differential comparison by applying the same classification to both galaxies. We find that C/M $\approx 0.7$ in the LMC and C/M $\approx 0.1$ in M101 using shared criteria (see Table \ref{tab:analysis_samples}), a difference of roughly a factor of 7, which exceeds any expected definitional uncertainty. This difference indicates that our M101 field is substantially more metal-rich than the LMC and is independent of any adopted absolute value for either galaxy. 

The absolute value of the C/M ratio can also be used to constrain the field's metallicity through the C/M-metallicity relation. \citet{Boyer_2019} found that the C/M--metallicity relation appears to have an inflection point, steepening sharply somewhere between the metallicity of the LMC and that of M31's disk. Among more metal-poor galaxies, $\Delta \log$(C/M)/$\Delta$[M/H] $\approx -1.1$, while across M31's disk,  $\Delta \log$(C/M)/$\Delta$[M/H] $\approx -6.5$. Our measured C/M $\approx 0.1$ puts our field in the steep regime and corresponds to their outermost M31 fields, which have C/M $\approx 0.1$ at [M/H] $\approx -0.15$ on the CMD-based scale of \citet{Gregersen_2015}.  With a slope of $\approx - 6.5$, even a factor-of-a-few error in the C/M ratio only shifts the inferred metallicity by $<0.1$ dex. We therefore assume a $\sim 0.1$ dex systematic budget since the C/M ratio also depends on the star-formation history of the population, not only its metallicity, and adopt [M/H] $\approx -0.2 \pm 0.1$ ($\approx 2/3 Z_{\odot}$) as our best estimate for the metallicity of our M101 field's AGB population. While simple extrapolation from the LMC using the shallower slope allows for [M/H] as high as $+0.25$, we can rule out a near-solar or super-solar metallicity: the C/M ratio falls below $\sim 0.01$ in M31's inner disk \citep{Boyer_2019}, an order of magnitude less than our measured ratio. 

\section{Results}
\label{sec:results}

Using our synthetic LMC observations as a proxy for the LMC AGB stars, we are able to test how the populations of C- and O-AGB stars measured in the LMC compare to those detected using variability or CMD-only criteria in M101. 

\subsection{Comparison of AGB Colors}
\label{sec:lmc_m101_color_comparison}

\begin{deluxetable}{lcccc}
\tablecaption{M101 vs. LMC Colors of AGB stars
\label{tab:color_offset_bracket}}
\tablehead{ & \multicolumn{2}{c}{O-rich} & \multicolumn{2}{c}{C-rich} \\
\colhead{Color} & \colhead{CMD-only} & \colhead{Variable} &
\colhead{CMD-only} & \colhead{Variable} \\
 & \colhead{(mag)} & \colhead{(mag)} & \colhead{(mag)} & \colhead{(mag)}}
\setlength{\tabcolsep}{4pt}
\startdata
F115W $-$ F150W & $+0.09$ & $+0.02$ & $+0.14$ & $+0.17$ \\
F115W $-$ F182M & $+0.18$ & $+0.08$ & $+0.28$ & $+0.35$ \\
F115W $-$ F277W & $+0.15$ & $+0.20$ & $+0.30$ & $+0.35$ \\
F115W $-$ F356W & $+0.25$ & $+0.55$ & $+0.37$ & $+0.44$ \\
F115W $-$ F444W & $+0.24$ & $+0.70$ & $+0.46$ & $+0.57$ \\
F150W $-$ F182M & $+0.09$ & $+0.05$ & $+0.16$ & $+0.20$ \\
F150W $-$ F277W & $+0.07$ & $+0.22$ & $+0.15$ & $+0.15$ \\
F150W $-$ F356W & $+0.16$ & $+0.58$ & $+0.20$ & $+0.24$ \\
F150W $-$ F444W & $+0.15$ & $+0.74$ & $+0.27$ & $+0.36$ \\
F182M $-$ F277W & $-0.01$ & $+0.19$ & $0.00$ & $-0.03$ \\
F182M $-$ F356W & $+0.07$ & $+0.55$ & $+0.04$ & $+0.04$ \\
F182M $-$ F444W & $+0.06$ & $+0.73$ & $+0.10$ & $+0.16$ \\
F277W $-$ F356W & $+0.09$ & $+0.36$ & $+0.06$ & $+0.10$ \\
F277W $-$ F444W & $+0.09$ & $+0.53$ & $+0.13$ & $+0.20$ \\
F356W $-$ F444W & $-0.04$ & $+0.13$ & $+0.08$ & $+0.12$ \\
\hline
median over colors & $+0.09$ & $+0.36$ & $+0.15$ & $+0.20$ \\
\enddata

\tablecomments{The entries are median$_{\rm color}$(M101)$-$median$_{\rm color}$(LMC). Thus, a positive number indicates a redder population in that color in M101. ``CMD-only'' compares the M101 AGB population selected using the F150W vs.\ F150W$-$F182M CMD alone with LMC SRV+Mira sample, ``Variable'' compares the SRV+Mira sample in M101 with the SRV+Mira sample from the LMC. }
\end{deluxetable}

\begin{figure*}[ht!]
 \centering
 \includegraphics[width=0.49\textwidth]{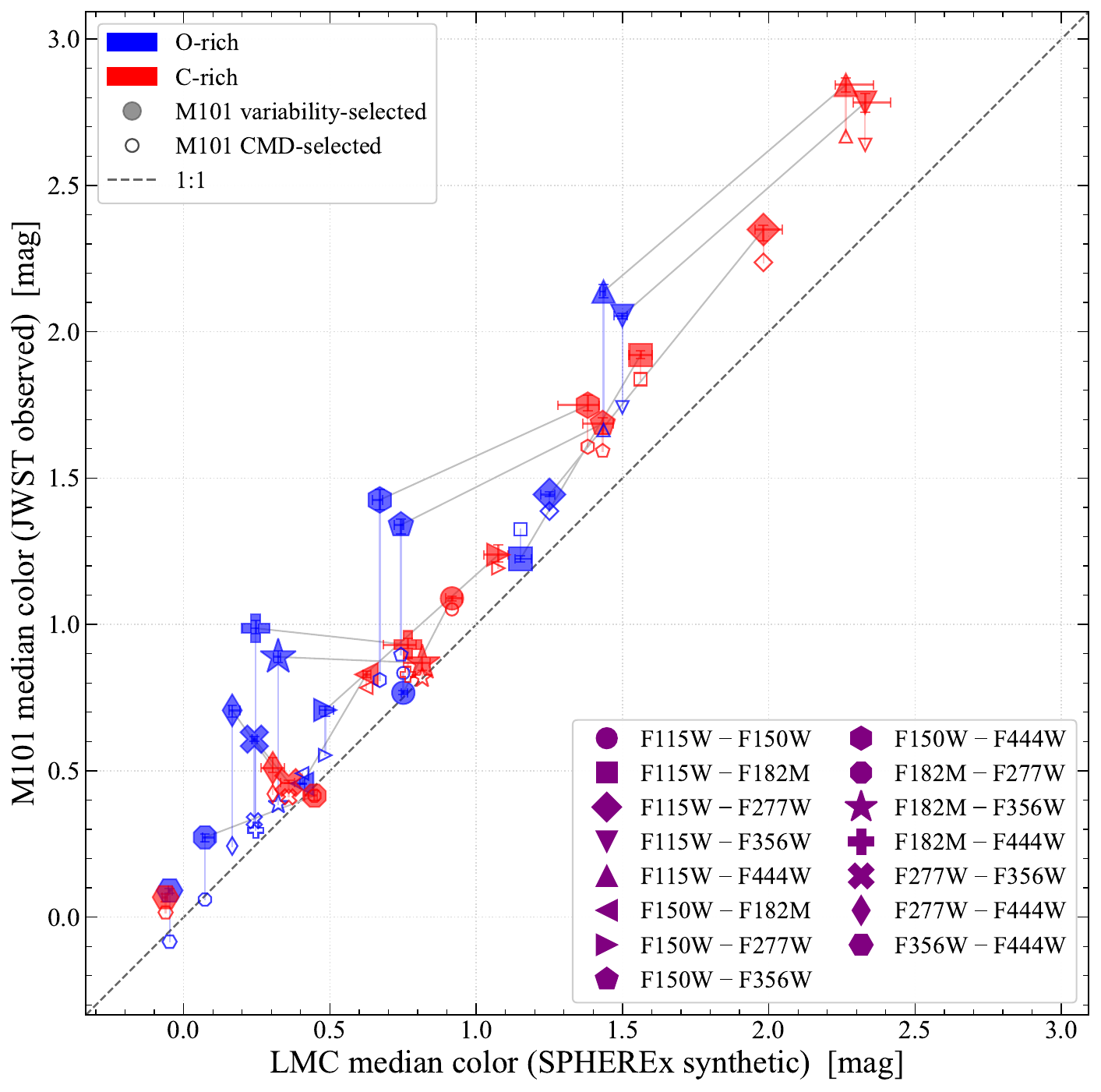}
  \includegraphics[width=0.49\textwidth]{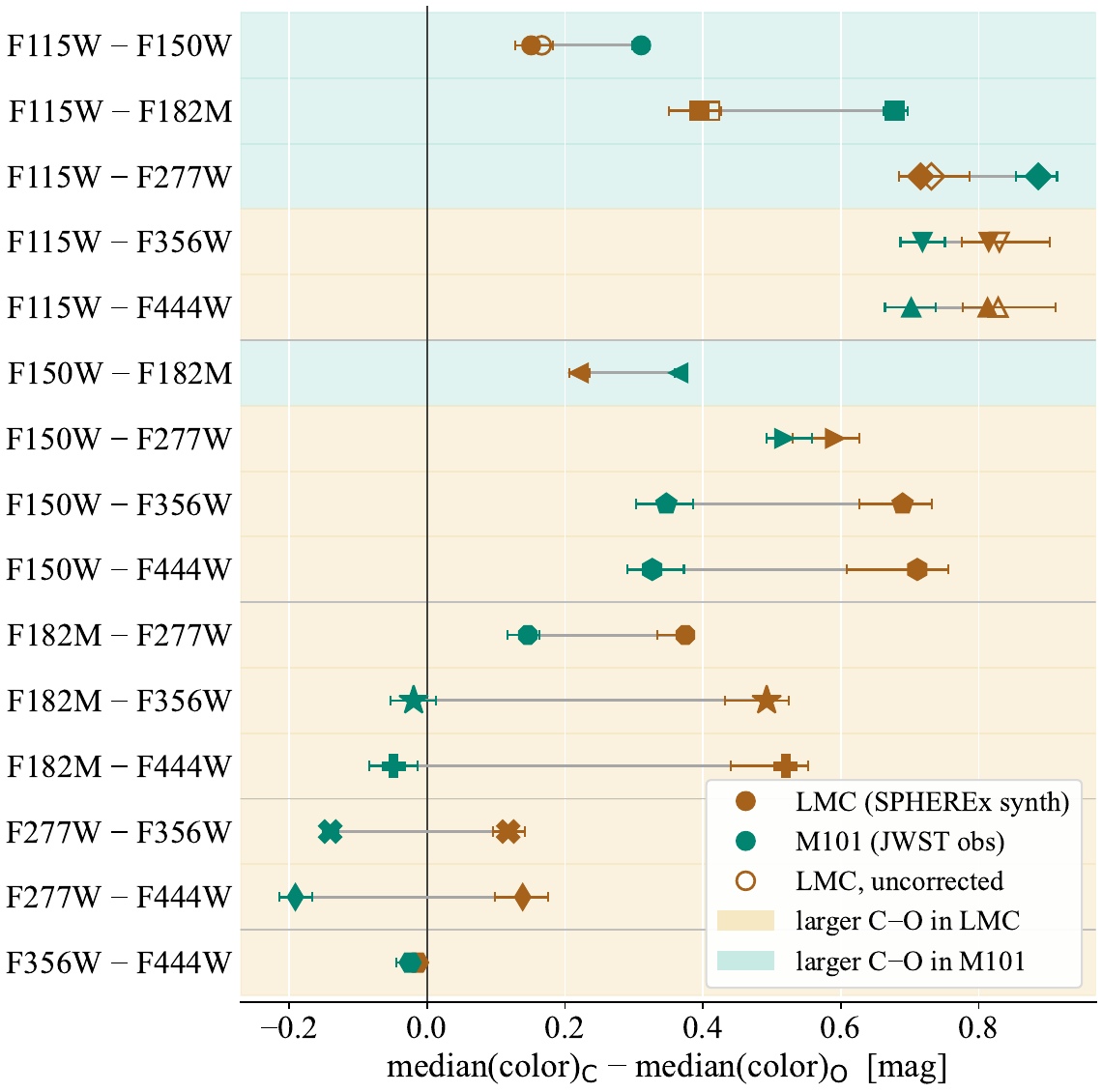}
  \caption{\textit{Left:} Median colors of O- and C-rich AGB stars in M101 against those in the LMC (SPHEREx synthetic photometry), with one symbol per NIRCam color. M101 variability-selected AGBs are shown as filled points and CMD-selected AGBs as smaller hollow points; the LMC medians all use the SRV and Mira classes. Gray lines connect the O- and C-rich medians of the same color (filled points), and the faint vertical lines connect the variability- and CMD-selected M101 medians of the same color. Points above the dashed one-to-one line are redder in M101. 
  \textit{Right:} The separation in median C-AGB and O-AGB color in
  both M101 NIRCam observations and LMC from SPHEREx synthetic photometry. Positive numbers indicate that C-rich stars are redder than O-rich stars in those colors, while negative numbers indicate that C-rich stars are bluer than O-rich stars. Colors where the two groups are considered `easier' to separate in color space in the LMC are shaded tan while colors where the two groups are easier to separate in M101 are shaded teal.
  \label{fig:c_minus_o_colors}}
\end{figure*}

Here we compare the synthetic NIRCam colors of LMC AGBs identified through variability with the observed NIRCam colors of M101 AGBs identified through either variability or CMD location in M101. Since M101's much larger distance ($>6$ Mpc) precludes us from obtaining a complete sample of its SRVs, which are not easily detected as variable, we are unable to match the variability criterion exactly between the LMC and M101. We instead compare LMC SRVs and Miras with M101 variability-selected AGBs and M101 CMD-selected AGBs to approximate the range of uncertainty due to differences in variability. The distribution of colors in the LMC and M101 both have long tails (a result of heavy dust reddening in the most-evolved stars) so we use the median rather than mean color for comparison. We use 16-84th percentile ranges rather than standard Gaussian errors to reflect the non-Gaussian distribution of the underlying population. Table \ref{tab:color_offset_bracket} lists the resulting median color offsets (M101$-$LMC) for each NIRCam color, chemical type, and M101 selection method. The left panel of Figure~\ref{fig:c_minus_o_colors} shows the medians from both galaxies plotted against each other. 

\begin{figure*}[]
\centering
\includegraphics[width=\textwidth]{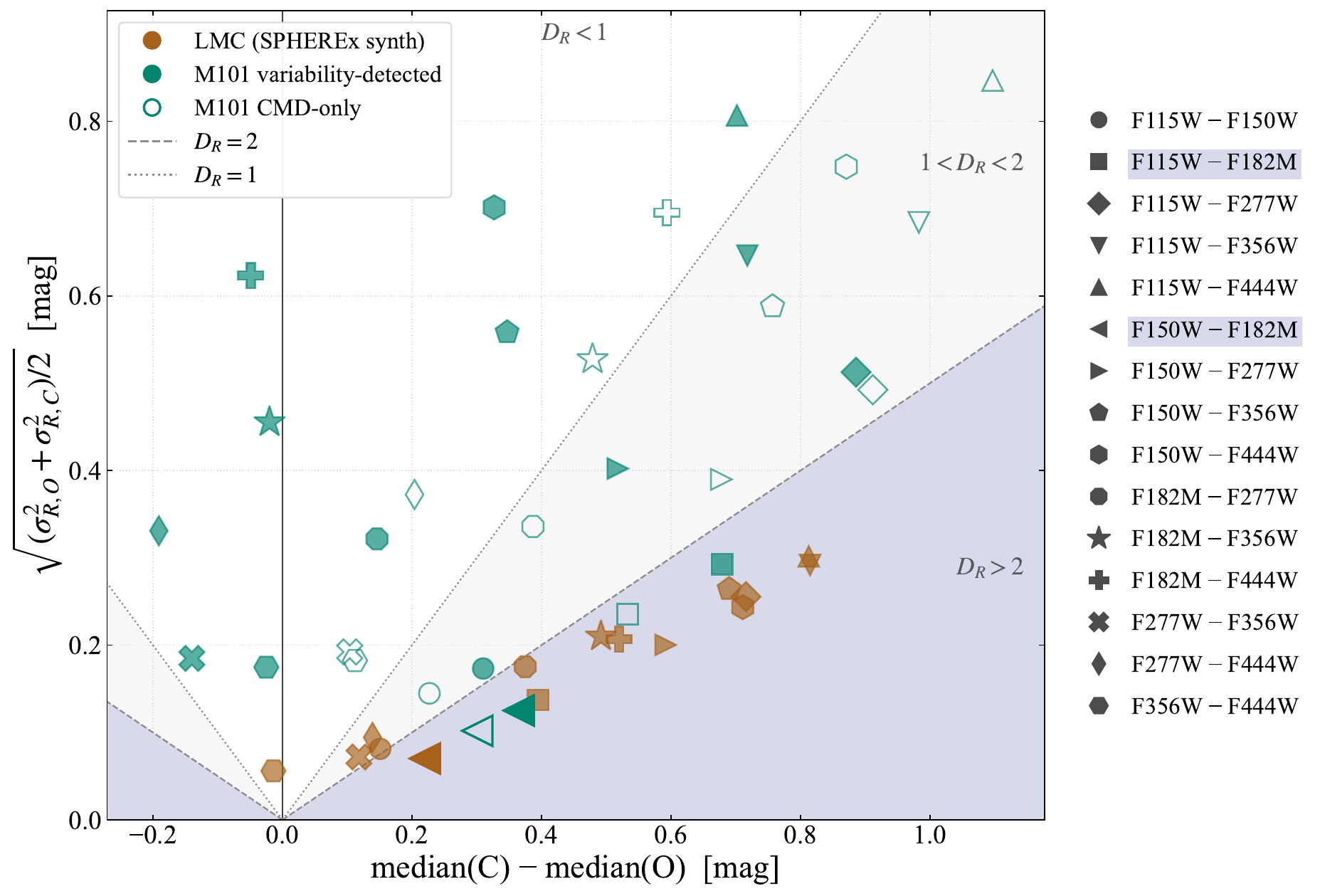}
\caption{Robust separation metric $D_R$ for C- and O-rich AGB stars in both the LMC and M101. The horizontal axis shows the difference between the medians of the C- and O-rich AGB populations and the vertical axis displays the combined robust dispersion derived from the 16th and 84th percentiles of each population (the numerator and denominator of Equation \ref{eq:DR_definition}, respectively). Diagonal lines indicate contours of constant $D_R$, with the shaded region highlighting the region that is `cleanly' separated ($D_R>2$). Brown symbols are used for the synthetic LMC observations, solid teal symbols denote the variability-detected M101 stars, and the open teal symbols show the CMD-only selected M101 stars. The two colors for which all three samples remain in the $D_R > 2$ regime are highlighted in light purple. 
\label{fig:c_sep_plane}}
\end{figure*}

We find that, regardless of the way AGB stars are selected in M101, their colors are redder than those of the AGB stars in the LMC (see Figure~\ref{fig:c_minus_o_colors}, left panel). Because the direction of this effect is not sensitive to the method used to select the M101 AGB population, it appears to be a real effect resulting from environmental differences in the two galaxies. However, the degree to which the M101 AGB stars are redder than the LMC AGB stars does not remain uniform across all wavelengths and colors. For colors using only bands shorter than $\sim 2 \mu$m --- including the F150W$-$F182M color used to separate the two chemical classes in M101 --- the difference in color between the two galaxies is small and nearly independent of the selection method. The offset in median color between the galaxies (see Table~\ref{tab:color_offset_bracket}) ranges from $+0.02$ to $+0.18$ mag for O-rich and $+0.14$ to $+0.35$ mag for C-rich stars, with positive numbers indicating redder colors in M101.

The increase in reddening in M101 is significantly more noticeable when pairing a short-wavelength band with either F356W or F444W. For O-rich stars, the degree of reddening depends on how the M101 sample is selected, ranging from +0.55 to +0.74 mag when comparing variables to at most +0.25 mag when employing a CMD-based selection in M101. For the C-rich stars, their median color offsets change by no more than $\sim 0.1$ mag regardless of the selection criteria used in either galaxy. 

In other words, the colors formed from the shorter-wavelength bands, which are dominated by the stellar photosphere and its molecular bands, appear less strongly affected by the stellar environment, regardless of how the samples are chosen, while the colors sensitive to emission from circumstellar dust may be strongly affected by environmental factors, such as metallicity. This is expected if selecting more variable stars also selects for dustier O-rich stars, an effect which is then layered on top of the smaller reddening that is present in all colors, classes, and selection methods. The relative stability in the C-rich stars' colors is very likely due to C-rich stars already being selected \textit{evolutionarily}: C-rich stars only become C-rich through the third dredge-up process, which brings synthesized carbon from the core to the surface.  Meanwhile, O-rich stars can be at any evolutionary stage along the AGB track, leaving some relatively unobscured by dust and others heavily dust-reddened. O-rich stars also span a wider range of initial masses, so their colors depend on the star-formation history. A younger population contains more massive O-rich stars undergoing hot-bottom burning, which can be among the dustiest AGB stars \citep[e.g.,][]{Nally_2026}, whereas C-rich stars form only over a limited range of masses and are less affected. For a full discussion of the implications see Section \ref{sec:color_environment}. 

\subsection{Color Separation of C- and O-rich Populations}
\label{sec:co_color_sep}

\begin{figure*}[]
\centering
\includegraphics[width=\textwidth]{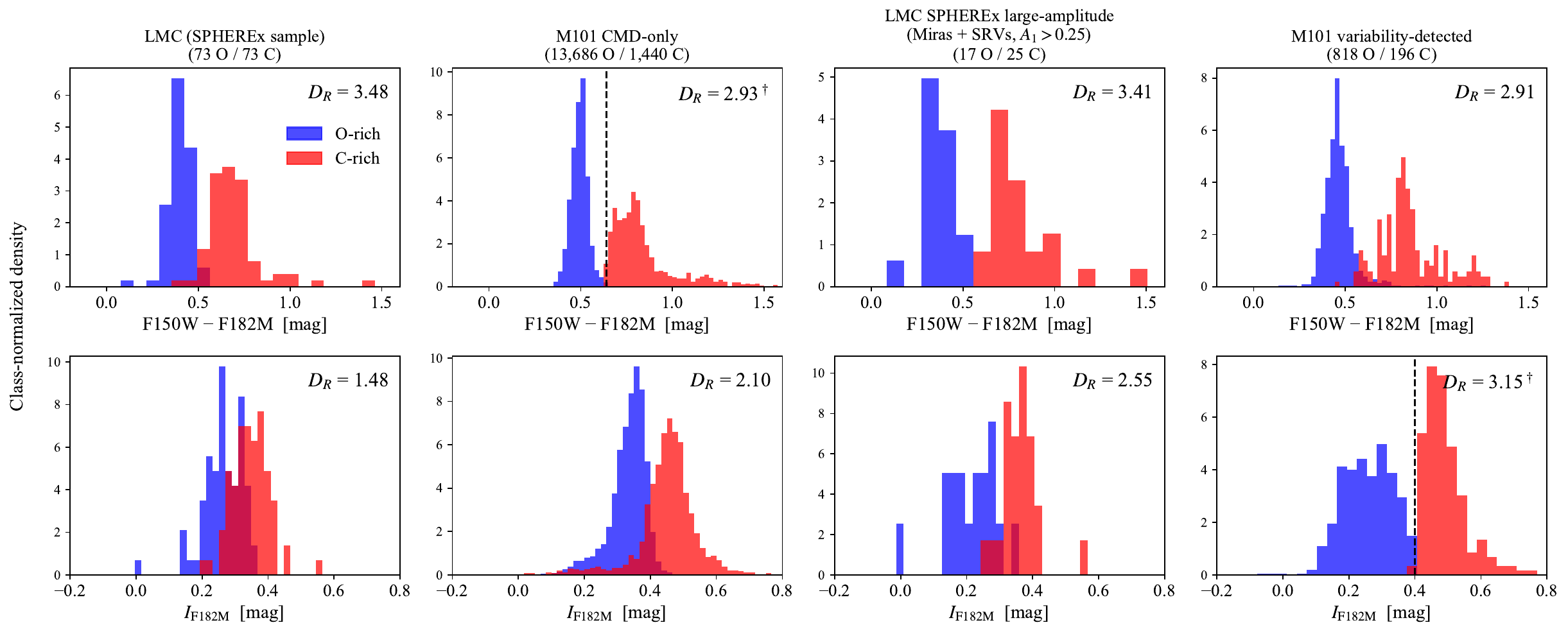}
\caption{Scaled histograms showing the separation of C- and O-rich AGB stars by F150W$-$F182M color (\textit{top row}) and by $I_{\rm F182M}$ (\textit{bottom row}) in four samples. From left to right, the samples are: the spectroscopically classified LMC SPHEREx sample (73 O-rich/ 73 C-rich), the M101 CMD-only sample (13,686 /1,440), the large-amplitude LMC subsample (Miras and SRVs with $A_1 > 0.25$ mag; 17/ 25), and the M101 variability-detected sample (818/196). Each class is normalized to unit area, so the heights do not reflect the relative numbers of O- and C-rich stars (given in the panel titles); O-rich stars are shown in blue and C-rich in red, and each panel lists the calculated $D_R$ of the two groups (Section \ref{sec:co_color_sep}). The dashed lines mark the classification boundaries that we adopted, which are F150W$-$F182M = 0.64 mag for the CMD-only sample and $I_{\rm F182M} = 0.40$ mag for the variability-detected sample. Daggers mark the two panels where the separation is partly by construction because the quantity plotted also defines the classes for that sample. \label{fig:windex_vs_color}}
\end{figure*}

Here we investigate how well different color combinations can effectively separate AGB stars. We hypothesize that a necessary but not sufficient criterion for determining whether AGB stars can be separated in a given color is whether the AGB distribution is (at least) bimodal in that color. We refer to this approach, classifying with the distribution of a single color or index and placing the boundary at its valley, as a \textit{single-color bimodality}, in contrast to boundaries drawn on color-color diagrams. If there is only one detected peak in the color distribution, then it is clear that both C- and O-AGBs must be mixed in color space for that particular filter combination. However, if there is detectable bimodality, then the bimodality can be caused by the distinct chemical compositions of O- and C-AGBs. 

To investigate the degree of AGB bimodality, we first examine which colors produce the largest separation in the median color of C- and O-AGB stars. Generally, C-rich stars are expected to be redder than O-rich stars as a result of their lower photospheric temperature and greater reddening due to dust. The right panel of Figure \ref{fig:c_minus_o_colors} shows the median colors of both the C and O variables for the NIRCam bands we have observed or modeled in both the LMC and M101. 

Figure \ref{fig:c_minus_o_colors} (right panel) shows that while for every band combination that we tested, this trend (C-AGB redder than O-AGB) holds true in the LMC (brown points), it is not generally true for variability-detected AGB stars in M101 (teal points). The increased reddening seen in the evolved O-rich stars in M101 makes their median color redder than those of C-rich stars in colors using NIRCam long-wavelength bands. This is seen most clearly in F277W$-$F356W and F277W$-$F444W, which have the most-negative values ($-0.14$ mag and $-0.19$ mag) in the right panel of Figure \ref{fig:c_minus_o_colors}, indicating that the O-rich stars are \textit{redder} in those colors than the C-rich stars. Colors anchored with the shorter-wavelength F115W and F150W bands keep C-rich stars redder in both galaxies. 

This same reversal (O-rich stars becoming redder than C-rich stars) is also seen in SAGE observations of the LMC. However, in the LMC, this only occurs at $\gtrsim 8~\mu$m, where the warm circumstellar dust emission dominates the photospheric flux \citep{Blum_2006, Srinivasan_2009}. Because the O-rich stars in the LMC show much smaller excesses at $\sim$3.6--4.4~$\mu$m than those in M101 (Table \ref{tab:color_offset_bracket}), their dust emission only overtakes the photospheric flux at much longer wavelengths. As a result, an assumption that C stars are redder than O stars, which works in the LMC up to $\sim 8\micron$, fails at much shorter wavelengths in M101. For all color combinations where separation becomes more difficult in M101 relative to the LMC (either via the relative color flipping between the two galaxies or the separation growing smaller in M101), we have highlighted the background in tan.

However, there are also a few cases, particularly in filter combinations using F115W, F150W, and F182M, where C-AGB stars are redder and the separation between the C and O classes grows in M101 relative to the LMC. We have highlighted these cases in pale teal in Figure \ref{fig:c_minus_o_colors}.

The width of the distributions of C and O stars in a given color also affects how well they can be separated. Because O-rich stars are more common than C-rich stars in nearly every environment (stars enter the AGB as O-rich stars and only those of a relatively narrow mass range evolve into C-rich stars), if even the tail end of the distribution of O-rich stars overlaps significantly with the distribution of C-rich stars, there can be relatively large O-rich contamination of the C-rich AGB population. Thus, we use a modified version of Ashman's $D$ statistic to measure the bimodality of the AGB color distribution \citep{Ashman_1994}. 

The typical Ashman's $D$ statistic measures the distance between the means ($\mu$) of two distributions, scaled by their standard deviations ($\sigma$). However, the distributions of the broadband colors of both C- and O-AGB stars are not well-described by Gaussians due to their long tails.  Thus, for each chemical subtype, we calculate a robust half-width, $\sigma_R$, defined as 
\begin{equation}
    \sigma_R = \frac{P_{84} - P_{16}}{2}, 
\end{equation}
where $P_{84}$ and $P_{16}$ correspond to the 84th and 16th percentile values respectively. This range corresponds in rounded percentages to the standard Gaussian $\pm 1\sigma$ bounds without assuming a Gaussian shape of distribution. While $\sigma_R$ describes the bulk of each distribution better than the standard deviation, which long tails inflate, it still averages the two asymmetric tails of the distribution and describes only the central, Gaussian-equivalent part of each distribution. For a distribution with longer tails than a Gaussian, for example, $(P_{97.5} - P_{2.5})/2$ is larger than 2$\sigma_R$.

We then approximate the degree of bimodality by defining a robust, mixture-separation statistic $D_R$,
\begin{equation}\label{eq:DR_definition}
D_R = \sqrt{2}\frac{\mathrm{med}(C) - \mathrm{med}(O)}{\sqrt{\sigma_{R,C}^2 + \sigma_{R,O}^2}},
\end{equation}
where $\mathrm{med}(C)$ is the median C-AGB color and $\mathrm{med}(O)$ is the median O-AGB color. To maximize physical interpretability and reproducibility, we favor this robust, non-parametric approach over more complex or machine-learning based classifiers.  Because $\sigma_R$ does not describe the tails, $D_R$ should be treated as a relative measure, used to rank colors by how well they separate the two classes, rather than as an absolute measure of contamination. Contamination at the classification boundary depends on the tails, and we measure it directly instead:  the F150W$-$F182M and $I_{\rm F182M}$ classifications agree for 95.5\% of the variability-detected sample (see below), and both agree with the SPHEREx spectroscopic classifications to within $\sim4\%$ (Section~\ref{sec:application_classification}).

Figure \ref{fig:c_sep_plane} shows the combined width of both classes $\sqrt{(\sigma_{R,O}^2+\sigma_{R,C}^2)/2}$ (the denominator of Equation~\ref{eq:DR_definition}), plotted against the separation in median color between C and O stars in every NIRCam filter combination that we obtained (the numerator of Equation~\ref{eq:DR_definition}). Higher $D_R$ indicates that the C- and O-AGB distributions are more easily separated. As a reference, we mark $D_R = 2$, the value above which two Gaussian distributions are cleanly separated \citep{Ashman_1994}. Despite having one of the smaller absolute color separations, F150W$-$F182M is the color with the largest $D_R$ separation statistic in all three groups we tested: the LMC SPHEREx representative subsample ($D_R = 3.13$), the M101 variability-detected population ($D_R = 2.91$), and M101 CMD-only population ($D_R = 2.93$). This strong separation statistic is driven by the narrow width of both the C- and O-AGB color distributions, ranging from $0.05 - 0.17$ mag in all six combinations of chemical class, galaxy, and selection criteria. We also find that F115W$-$F182M was the only other color for which all three samples stayed above the $D_R > 2$ threshold, though its $D_R$ statistic is lower in each case than F150W - F182M, it still performs significantly better than every other color. 

Figure~\ref{fig:windex_vs_color} shows how well separated (by measured $D_R$) four different samples of stars  are when they are classified using both F150W - F182M color and $I_{\rm F182M}$ (See Equation \ref{eq:IF182M}). The two LMC-based samples are the full retrieved SPHEREx sample (not only the representative subset used in the other Results sections of the paper) and a subset of large-amplitude stars built from the full SPHEREx sample by keeping only those with $A_1 > 0.25$ mag ($I$-band amplitude). While not as representative of the color distribution of LMC AGBs, we chose to use the larger sample in order to include more large-amplitude variables. The M101 samples are the CMD-only sample and the variability-detected sample. Based on these results, color separates cleanly in every panel and sample ($D_R = 2.9 - 3.5$), with the SPHEREx spectroscopy serving as the ground truth. However, the water index does not separate the full LMC SPHEREx sample ($D_R = 1.5$), and also has a borderline $D_R = 2.1$ value when applied to the CMD-only sample in M101. However, once the sample is restricted to substantially variable stars like the large-amplitude LMC SPHEREx sample, and the M101 variability-detected sample, this value grows to $2.6$ and $3.2$ respectively. More restrictive cuts of up to $\gtrsim 0.4$ mag for the SPHEREx stars yielded even larger $D_R$ values, up to $\sim3.0$ but also contained only 29 stars --- additional spectra of large-amplitude variables are needed to investigate the trend further. The M101 stars have a minimum detected amplitude of $0.2$ mag in $F160W$, which corresponds to $\gtrsim 0.4$ mag in $I$-band amplitude and can therefore be considered an even larger-amplitude counterpart to the SPHEREx sample. $I_{\rm F182M}$ is thus a classifier that is effective specifically on pulsating stars. The physical implications of this are discussed in Section~\ref{sec:application_classification}. 

The F150W - F182M color and $I_{\rm F182M}$ classification criteria also yield the same chemical type for 95.5\% of the variability-detected sample ($I_{\rm F182M}$ is not applied to the CMD-based sample due to the expectation that the water absorption features it tracks are built by pulsation). The disagreements are dominated by the most-heavily dust-reddened stars. For these stars, circumstellar reddening likely dominates over any detected molecular absorption in F150W - F182M color. On the other hand, the water index continues to track the 1.9~\micron\ absorption, even for these heavily reddened stars. Given the predominance of O-rich AGB stars identified in M101 (C/M $\approx$ 0.1; Section~\ref{sec:methods_metallicity}), and the similarity of its expected metallicity to that of the M31 fields in which \citet{Goldman_2022, Goldman_2026} found the dusty AGB population to be predominantly O-rich, a substantial population of heavily dust-reddened O-rich stars should be present here as well. However, a purely color-based criterion --- even one whose bandpass overlaps a water-absorption feature --- will necessarily assign its reddest members only to the C-rich class. The feature-based index is expected to be less susceptible because it includes a correction for the local continuum slope. The other population showing large disagreements lies at the $I_{\rm F182M}$ boundary itself. This is unsurprising, given that the boundary location is derived empirically, so $I_{\rm F182M}$ values immediately adjacent to it are the least reliable. We will return to these populations and our revised guidelines on classifier choice in Section~\ref{sec:application_classification}. 

\begin{figure*}
\centering
\includegraphics[width=\textwidth]{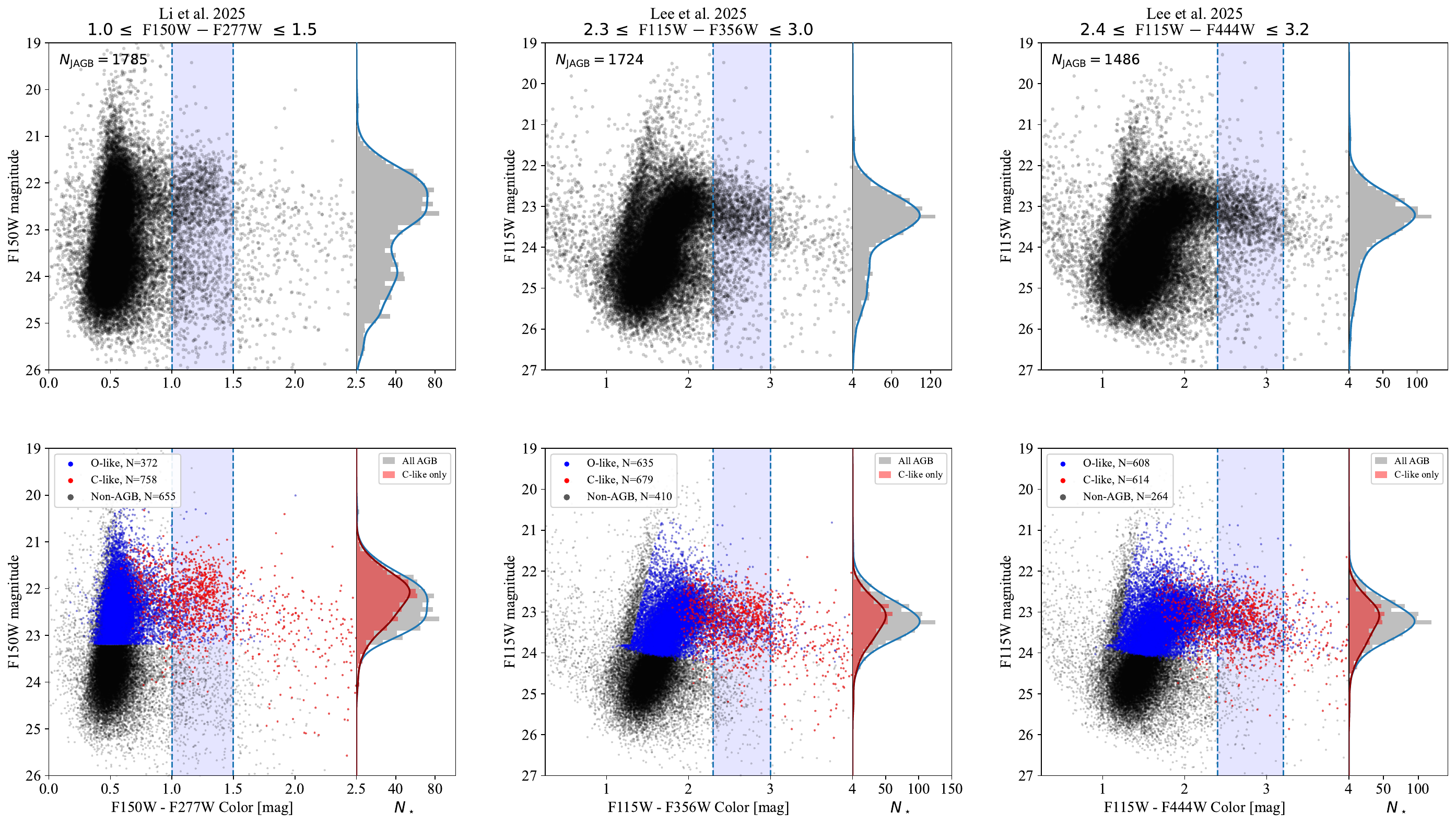}
\caption{J-region color-magnitude diagrams for AGB stars in M101. \textit{Top panels:} The `raw' color-magnitude diagrams in the bands used by the CCHP \citep{Lee_2025_JWSTJAGB} and SH0ES \citep{Li_2025} in their analysis and the luminosity function plotted on the right.  
\textit{Bottom panels:} The same color-magnitude diagrams, but with the stars colored by their likely chemical subtype using the F182M band for classification.
\label{fig:jagb_cmds}}
\end{figure*}

\subsection{Quantifying J-Region Contamination}
\label{sec:JAGB_contamination}

\begin{deluxetable*}{lllrrr}
\tablecaption{Literature JAGB color-selections applied to the LMC and M101 samples \label{tab:jagb_contamination}}
\tablewidth{0pt}
\tablehead{
\colhead{Reference} & \colhead{Window} & \colhead{Sample} &
\colhead{$N_{\rm C}$} & \colhead{$N_{\rm O}$} &
\colhead{O contam.} \\
\colhead{} & \colhead{(mag)} & \colhead{} &
\colhead{in window} & \colhead{in window} &
\colhead{(\%)}
}
\startdata
\cutinhead{\textbf{JWST NIRCam windows}}
\citet{Li_2025} & $1.0 \le \mathrm{F150W} - \mathrm{F277W} \le 1.5$\tablenotemark{a} & LMC -- synth.\ (SRV+Mira) & 17 & 0 & 0.0 \\
 &  & M101 -- variability-detected & 99 & 115 & 53.7 \\
 &  & M101 -- CMD-only & 758 & 372 & 32.9 \\
\hline
\citet{Lee_2025_JWSTJAGB} & $2.3 \le \mathrm{F115W} - \mathrm{F356W} \le 3.0$ & LMC -- synth.\ (SRV+Mira) & 11 & 0 & 0.0 \\
 &  & M101 -- variability-detected & 90 & 180 & 66.7 \\
 &  & M101 -- CMD-only & 679 & 635 & 48.3 \\
\hline
\citet{Lee_2025_JWSTJAGB} & $2.4 \le \mathrm{F115W} - \mathrm{F444W} \le 3.2$ & LMC -- synth.\ (SRV+Mira) & 5 & 0 & 0.0 \\
 &  & M101 -- variability-detected & 91 & 194 & 68.1 \\
 &  & M101 -- CMD-only & 614 & 608 & 49.8 \\
\cutinhead{\textbf{2MASS windows}}
\citet{Madore_and_Freedman_2020} & $1.30\tablenotemark{b} \le J - K_s \le 2.00$ & LMC --\ SPHEREx sample & 25 & 5 & 16.7 \\
 &  & LMC --\ OGLE SRVs+Miras & 4413 & 808 & 15.5 \\
 &  & LMC --\ OGLE all LPVs & 5069 & 1990 & 28.2 \\
\hline
\citet{Nikolaev_and_Weinberg_2000} & $1.40 \le J - K_s \le 2.00$ & LMC --\ SPHEREx sample & 21 & 0 & 0.0 \\
\citet{Ripoche_2020} &  & LMC --\ OGLE SRVs+Miras & 4077 & 189 & 4.4 \\
 &  & LMC --\ OGLE all LPVs & 4431 & 552 & 11.1 \\
\hline
\citet{Freedman_and_Madore_2020} & $1.10 \le J - H \le 1.35$, & LMC --\ SPHEREx sample & 7 & 0 & 0.0 \\
 & $11.9 \le J \le 12.7$ & LMC --\ OGLE SRVs+Miras & 1997 & 14 & 0.7 \\
 &  & LMC --\ OGLE all LPVs & 2024 & 20 & 1.0 \\
\enddata
\tablenotetext{a}{\citet{Li_2025} state that they use $\pm5-6\sigma$ ($\approx 2.3 - 2.8$ mag full width) to identify upper and lower magnitude limits of the J-region in addition to the color window.}
\tablenotetext{b}{This is the operational definition of the blue edge. However, other works in the series specify that the blue edge is an environment-dependent parameter that is adjusted accordingly (e.g., \citealt{Lee_2022} uses 1.5).}
\tablecomments{$N_{\rm C}$ and $N_{\rm O}$ are the classified
C-rich (C-like) and O-rich (O-like) members residing in the given J-region; O contamination is defined as $N_{\rm O}/(N_{\rm C}+N_{\rm O})$. 
LMC colors in the top half of the table are synthetic NIRCam photometry of the SRV+Mira sample (59 O / 34 C); M101 samples are the variability-detected (818 O / 196 C, water-index classification) and CMD-selected (13{,}686 O / 1{,}440 C, $\mathrm{F150W}-\mathrm{F182M} = 0.64$) populations.
The bottom half of the table shows the most commonly referenced 2MASS-based selection windows for the LMC. Here, ``SPHEREx sample'' tallies only the stars with SPHEREx spectroscopy and observed 2MASS colors (83 stars); ``OGLE SRVs+Miras'' is the OGLE SRV+Mira population with observed 2MASS colors and OGLE photometric labels (12,029 of
12,795 with $J$ and $K_s$, 12{,}097 with $J$ and $H$; the labels
are spectroscopically confirmed at 92/93 in the SRV+Mira sample);
``OGLE all LPVs'' adds the OSARGs (91{,}995 stars in total;
85{,}796 with $J$ and $K_s$, 86{,}429 with $J$ and $H$) --- OSARG
photometric labels are the least secure (C-labeled OSARGs are
spectroscopically confirmed at only 1/7), and OSARGs are excluded
from this paper's samples by design.}
\end{deluxetable*}

J-region AGB stars — presumed to be C-AGB stars with low dust reddening — are used as a statistical, CMD-based standard candle. They are identified by their appearance on the CMD as a group of relatively red AGBs that appear to have an approximately constant luminosity over a moderate range in color. Using the JAGB bounds defined in NIRCam filters by both CCHP and SH0ES, we investigate the chemical composition of the AGB population in the J-region in our M101 field. The results of this comparison are shown in Table \ref{tab:jagb_contamination}, along with the citations to the original papers defining the J-regions in these colors. We also compare it to the expected contamination in the equivalent color-magnitude diagram from the LMC using our synthetic colors generated from SPHEREx spectra. 

For the three NIRCam J-region definitions shown in the top half of Table~\ref{tab:jagb_contamination}, no O-rich star from the LMC synthetic sample enters the J-region. The three J-regions contain 17, 11, and 5 AGBs, respectively, and all are C-rich. We are limited by the small size of our SPHEREx sample in the LMC ($N\sim100$ overall, even fewer when considering only J-region C-AGBs), but Section \ref{sec:subsample_rep} shows that these spectra are representative of the color distribution of the larger AGB population in the LMC. Thus, we expect that these LMC J-regions would likely have a near-zero amount of O-AGB contamination for the NIRCam boundaries used in SN Ia hosts. For M101, the picture is quite different. Depending on the filter combination, color range, and method of chemical classification, the contamination ranges from a minimum of $33\%$ to a maximum of $68\%$. Figure \ref{fig:jagb_cmds} shows both the raw CMDs and the CMDs with AGBs classified using atmospheric water absorption overlaid. 

Rather than relying only on the SPHEREx sample, we can also use the full OGLE catalog to determine the contamination in the 2MASS-based J-region windows used for LMC calibrations (bottom half of Table~\ref{tab:jagb_contamination}). Because the blue edge of the J-region is treated as an environment-dependent parameter (see Table~\ref{tab:jagb_contamination}, note $b$), the contamination of any given published calibration depends on the edge adopted. If we use the OGLE photometric classifications of SRVs and Miras, the operational $J-K_s = 1.3$ mag edge results in a population that is 15.5\% O-rich, while the redder $J-K_s = 1.4$ mag edge \citep{Nikolaev_and_Weinberg_2000, Ripoche_2020} limits this contamination to 4.4\% O-rich, and the $J-H$ selection (bounded in both color and magnitude) used by \citet{Freedman_and_Madore_2020} has minimal contamination (0.7\% O-rich). Thus, with suitably drawn boundaries, the assumption of C-rich purity in the J-region is valid for the LMC. However, we argue in Section~\ref{sec:jagb_contamination_discussion} that this purity in the LMC is also likely why the contamination that appears in the J-region of M101 (Table~\ref{tab:jagb_contamination}) evades the method's internal consistency checks. For the more metal-rich Galactic field, \citet{Magnus_2024} had to move the blue edge boundary to $(J-K_s)_0 = 1.5$ mag in order to bring their estimated O-rich contamination rate from $\sim$50\% to below 10\%. The initial results of the spectroscopic CHASE survey \citep{Li_2026}, which selects AGB targets directly from the $1.5 < (J-K_s)_0 < 2.0$ box, suggest that the contamination may be even higher than previously estimated: 76\% of J-region stars are O-rich even when using the redder edge.   

The exact color boundaries vary significantly between different groups and between galaxies, even when identical filters are used. There is some physical motivation for this: the underlying formation of C-AGB stars is inherently tied to local environmental factors such as metallicity and star-formation history. Consequently, this implies that the O-AGB contamination rate can vary significantly not only between galaxies but also between different targeted fields within the same galaxy, as we find when comparing the LMC and M101. The field-to-field variances in JAGB measurements have been noted by both CCHP and SH0ES in the megamaser host galaxy, NGC 4258, which is one of the few JAGB distance-ladder hosts that has observations of multiple fields \citep{Li_2025, Lee_2025_JWSTJAGB}. 

While M101 has JAGB measurements, which are used by both the CCHP and SH0ES teams, the field analyzed in this paper does not appear to be the same as the one used for the JAGB measurements by either group. Thus, these numbers should not be interpreted as a direct measurement of the O-AGB contamination in these previous studies. Without a clean chemical classification applied to the exact field, color bounds, and filter combinations used in a given JAGB measurement, it is not possible to accurately assess the degree of contamination or any systematic effect it may have on the precision of the resulting distance modulus. For a full discussion on how this relates to previous results, see Section \ref{sec:jagb_contamination_discussion}. 

\subsection{Variability's Effect on Color}
\label{sec:var_vs_cmd}

\begin{figure*}
\centering
\includegraphics[width=\textwidth]{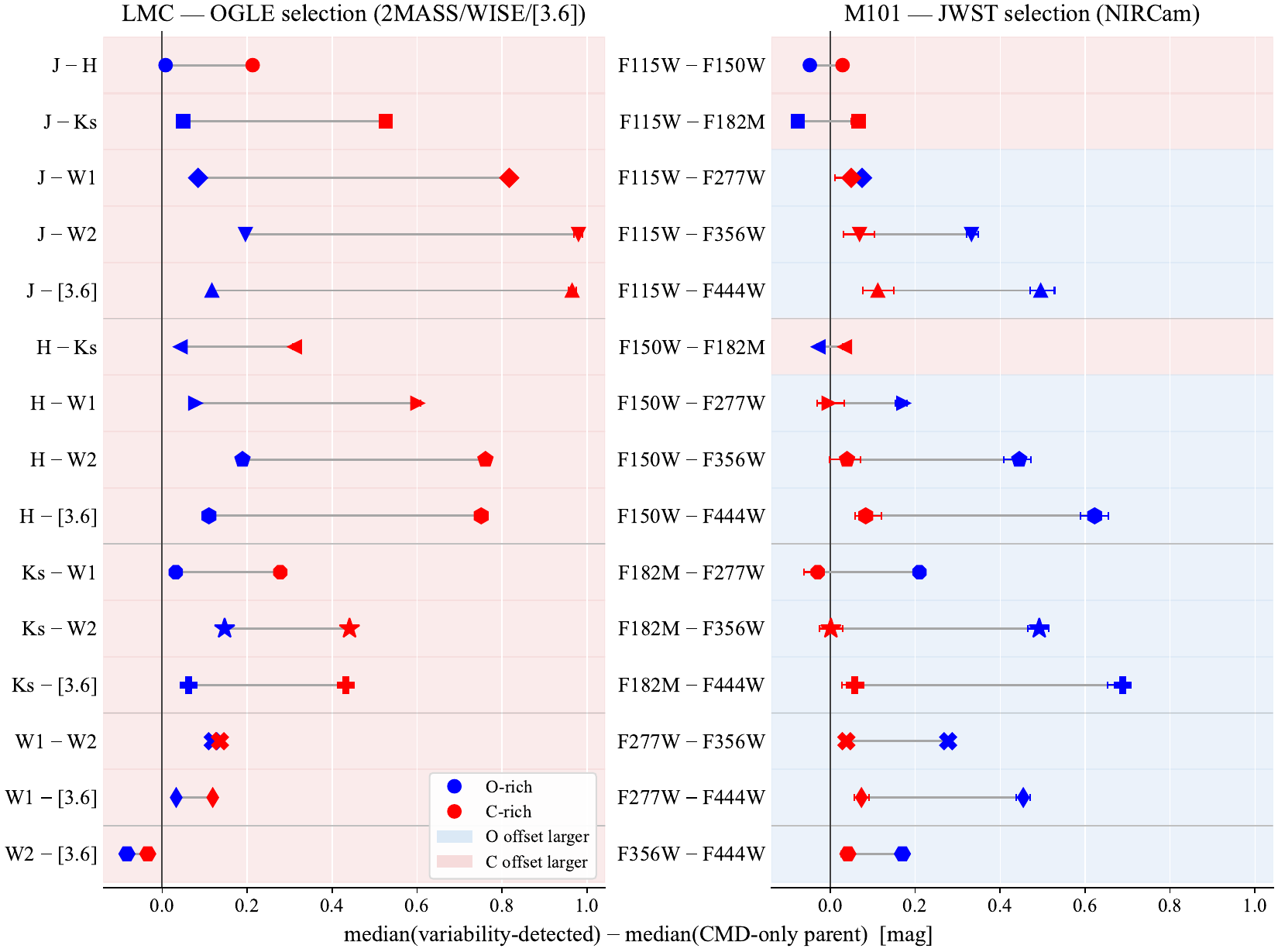}
\caption{Reddening associated with detected variability, separated by color and chemical class, in the LMC (\textit{left}; using observed 2MASS, WISE, and IRAC [3.6] colors) and M101 (\textit{right}; using NIRCam colors). Each point is the median color of the variability-detected sample minus the median color of the low-variability sample (OSARGs in the LMC, CMD-only AGBs in M101). The points are colored blue for O-rich stars and red for C-rich stars. Positive values indicate that the variables are redder. Error bars (often smaller than the symbols) are bootstrapped 68\% intervals. A gray bar is used to connect the two points in each row, highlighting the differences between the O- and C-rich subtypes. The individual rows are shaded by which chemical class (O or C-rich) shows greater reddening with variability. Table \ref{tab:var_amp_2x2} takes these numbers and lists the median offset for each panel, for the C- and O-rich classes. See Section~\ref{sec:var_vs_cmd_reddening} for a full description of the results and Section \ref{sec:var_vs_cmd_sel_eff} for the selection effects that make the M101 values lower bounds. 
\label{fig:var_vs_nonvar}}
\end{figure*}

Within each galaxy, we also examine how the colors of stars identified as variable compare to those that were not identified through variability. 

\subsubsection{Defining the LMC and M101 Samples}
\label{sec:var_vs_cmd_setup}

For the LMC we use the CMD-selected sample described in Section~\ref{sec:lmc_cmd_agb} (19,346 LPVs). We designate the SRVs and Miras as the variability-detected population and the OSARGs as the quasi-constant population. The median $I$-band amplitude of OSARGs in our sample is 0.038 mag, compared to 0.147 and 1.292 mag for SRVs and Miras respectively. It is virtually impossible to detect OSARGs as variable sources in galaxies outside of the Local Group (such as in an SN Ia host) since the photometric uncertainties will be of comparable or greater magnitude than the vast majority of OSARG amplitudes. Thus, the AGB members of the OSARGs are analogous to the AGB stars in more distant galaxies that have no detectable variability and are instead identified via CMD location only. Most SRVs will also be undetectable as variable outside of the Local Group but we retain them in the `variable' AGB population along with Miras. The specific amplitude threshold used to classify a star as an SRV or a Mira is primarily a convention \citep[see][for a more extended discussion]{Trabucchi_and_Mowlavi_2022}. 

Since the majority of LMC AGBs have photometry only, we use the OGLE photometric classifications of C and O spectral types. However, for the 93 stars for which we were able to obtain SPHEREx spectra, we verified in Section \ref{sec:spectral_classification} that they were the same class as their OGLE designations. Only one source, an SRV (LPV-13034, P = 82 d, $A_1$ = 0.054 mag), had a photometric designation that disagreed with the retrieved spectra (it was classified photometrically as an O-rich AGB, but its spectra revealed a distinctive C$_2$H$_2$ + HCN absorption feature). This is discussed further in Section \ref{sec:variable_classification}. Chemical classes for M101 AGB stars and their variable and non-variable designations are described in Section \ref{sec:methods}. 

\subsubsection{Variability Increases Reddening}
\label{sec:var_vs_cmd_reddening}

For each galaxy and chemical type, we compute 15 near-infrared colors (LMC: $J$, $H$, $K_s$, W1, W2, and [3.6]; M101: the six NIRCam bands) and take the difference between the median colors of the variable and low-variability samples (Figure~\ref{fig:var_vs_nonvar}). Table~\ref{tab:var_amp_2x2} lists the median of these 15 differences. Regardless of chemical type or galaxy, the detected variables are redder than their small-amplitude counterparts. The smallest offset, $+0.04$ mag for the C-rich stars in M101, is still not consistent with zero. The offsets for individual colors range from $-0.03$ to $+0.11$ mag, 13 of the 15 are positive, and a typical bootstrap uncertainty for each color is $\sim0.03$ mag. We discuss the biases that may affect the size of these differences in Section~\ref{sec:color_environment}.

Both the LMC and M101 panels of Figure~\ref{fig:var_vs_nonvar} show that the difference is not uniform across colors, and the largest differences are found in wavelength combinations that are most sensitive to the presence of dust, starting at $\sim2-3\mu$m for many AGBs, though this will vary depending on stellar chemistry and the mass-loss rate of the star. F150W-F182M, which we use for classification, is insensitive to whether the star has detected variability. In fact, in M101, the O-rich stars become slightly bluer in this color ($-0.03$ mag) when variable (see Section \ref{sec:color_environment} for interpretation). Pairing a shorter-wavelength band, such as $J$ or $F115W$, where the stellar photosphere largely dominates (and therefore temperature affects color more strongly), with a longer-wavelength band, such as $W_2$ or $F444W$ where dust emission begins to dominate, produces colors that are more sensitive to increased circumstellar dust. In these colors, the O-rich variables can be up to 0.7 mag redder than their CMD-only counterparts. In the LMC these differences are even larger, with the variable C-stars being redder than their OSARG counterparts by almost 1 magnitude in some cases. 

\subsubsection{Selection Effects}
\label{sec:var_vs_cmd_sel_eff}

\begin{deluxetable}{lcccc}
\tablecaption{Reddening as a Function of Variability\label{tab:var_amp_2x2}}
\tablehead{ & \multicolumn{2}{c}{LMC} & \multicolumn{2}{c}{M101} \\
\colhead{Comparison} & \colhead{O-rich} & \colhead{C-rich} &
\colhead{O-rich} & \colhead{C-rich} \\
 & \colhead{(mag)} & \colhead{(mag)} & \colhead{(mag)} & \colhead{(mag)}}
\startdata
Variability-detected $-$ CMD-only & $+0.08$ & $+0.44$ & $+0.26$ & $+0.04$ \\
Top quartile $-$ remaining variables & $+0.08$ & $+0.68$ & $+0.31$ & $+0.41$ \\
\enddata
\tablecomments{Each entry is the median color difference between the AGB samples within each galaxy compared in magnitude space (positive numbers indicate the more variable sample is redder) for the given chemical classification, aggregated across the 15 near-infrared colors examined in Section~\ref{sec:var_vs_cmd}. The corresponding offsets per color are shown in Figure~\ref{fig:var_vs_nonvar}. Row 1 compares the variability-detected and quasi-constant populations of Section~\ref{sec:var_vs_cmd_setup} while row 2 splits the \textit{variable} sample by pulsation amplitude within each galaxy. Numbers given for M101 in row 1 are lower bounds --- detection biases described in Section~\ref{sec:var_vs_cmd_sel_eff} will preferentially redden the CMD-only sample.}
\end{deluxetable} 

In M101, the color difference between the variables and CMD-only sources is likely a lower bound on the true number. There are three main effects that systematically bias the CMD-only subset to be redder, and one that provides a blue bias. Firstly, at a distance of $>6.5$ Mpc, many variables will be undetected due to crowding and photometric noise. Secondly, we also applied strict variability criteria that the source had to be detected in 13 epochs, and have a period that converged in order to minimize contamination from spurious variables. Both of these will relegate many \textit{bona fide} variables to the CMD-only group. Finally, large amounts of circumstellar dust will also dim the star. The reddest and dustiest variables will be dimmed from the sample by their own dust, lowering the redness of the median value. In the other direction, there is potential blue bias on the CMD population resulting from contamination from supergiant stars. These are visible most clearly in the F115W CMDs, as a small branch of stars blueward of the main O-AGB population. However, given that they are not noticeably bluer in the F150W or F182M CMDs, and that supergiant stars are rare, this effect is likely to be smaller than the previous three. 

For the C-stars, the offset of $+0.04$ mag between C-rich variability-detected and CMD-only sources should not be taken as a face-value comparison with the $+0.44$ mag difference in the LMC. The three biases described above that reduce the difference in variable and CMD-only colors affect C-AGBs even more significantly. Additionally, they are not expected to be affected by contamination from supergiants, which are well blueward of the C-AGB distribution. In the LMC, a systematic bias towards increasing the size of the offset arises from the fact that many C-classified OSARGs may not be C-AGBs at all, but rather O-AGBs. We obtained SPHEREx spectra for seven OSARGs originally photometrically classified as C-rich. The spectra revealed that six of these are spectroscopically O-rich (Section \ref{sec:variable_classification}). Moreover, we saw in Section \ref{sec:lmc_cmd_agb} that essentially all stars in the LMC AGB region are long-period variables. Thus, a C-rich star in the LMC without significant variability is more likely to be a misclassified O-rich star than a quasi-constant carbon star. However, similar to M101, the $I$-band amplitude will also miss the most dust-obscured stars, which will work in the other direction to depress the offset. 

Given these competing selection effects, it is also natural to wonder if the redder color is a statement about a threshold of variability rather than any physical reddening caused by pulsation itself. Here we compare colors for each group \textit{within variable samples only} and show that the connection between greater variability and greater reddening is robust to exact thresholds. 

In both galaxies, we compare the variable stars in the top quartile of amplitude with the remaining variables (Table~\ref{tab:var_amp_2x2}, row 2). The most-variable stars are redder in every case, so variability is correlated with increased infrared excess in both galaxies, consistent with earlier studies linking pulsation to circumstellar dust \citep[e.g.,][]{McDonald_Zijlstra_2016, McDonald_and_Trabucchi_2019}. For O-AGBs in the LMC, however, even the most-variable stars are only $+0.08$ mag redder, a modest increase over the small-amplitude stars. In M101, on the other hand, the most-variable O-AGBs show significantly more infrared excess ($+0.31$ mag), much closer to the C-AGBs ($+0.41$ mag). Our results suggest that the infrared excess of O-rich stars depends on environment, consistent with more efficient O-rich dust production at higher metallicity, whereas that of C-rich stars is consistent with being insensitive to the environment. We explore this idea in more detail in Section \ref{sec:color_environment}. 

\subsection{Dust and Variability Asymmetry}
\label{sec:variability_dust}

\begin{figure*}
\centering
\includegraphics[width=\textwidth]{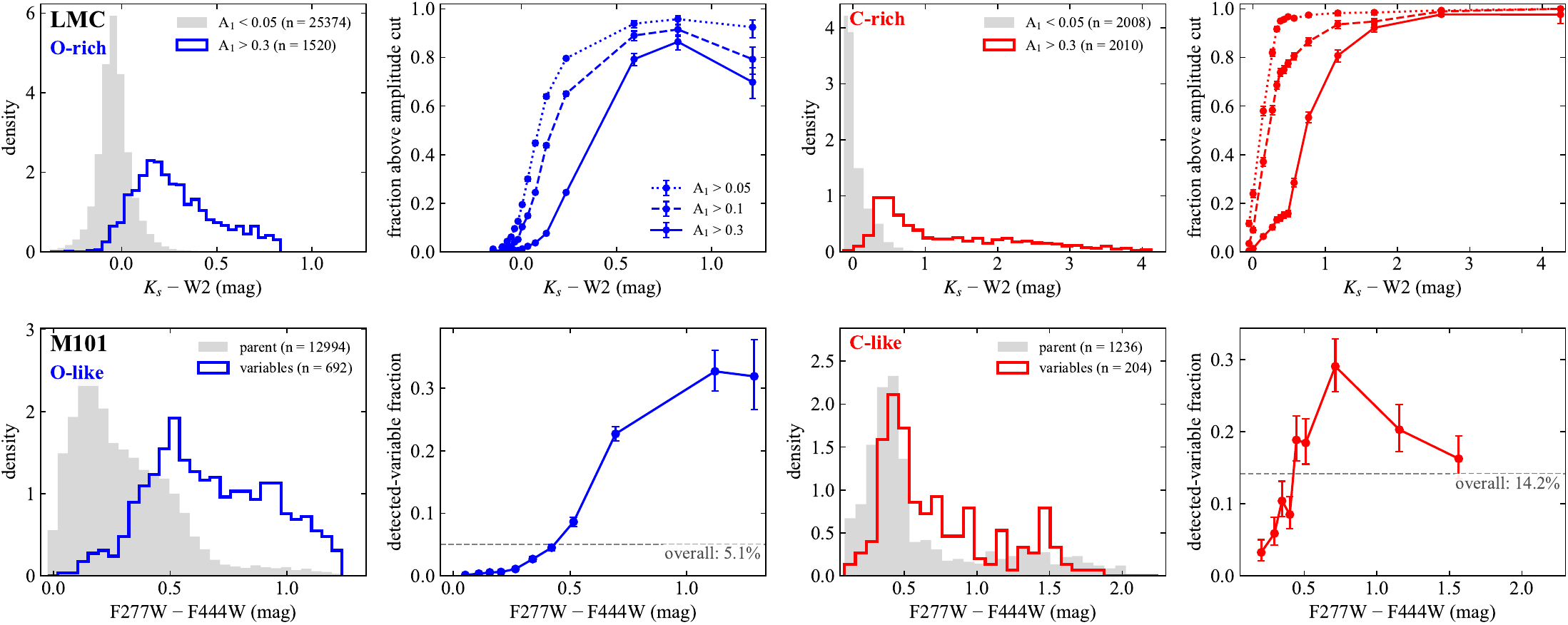}
\caption{\textit{Left panels:} The distribution of O-AGB with low or undetected variability in gray in the LMC (top) and M101 (bottom) as a function of K$_s$ - W2 mag and F277W - F444W mag respectively. The distribution of the O-rich sources with detected (M101) or significant (LMC, $A_1 > 0.3$ mag) variability shown in blue-outlined density histograms. The second column shows the fraction of O-AGB stars detected as variable as a function of color excess. 
\textit{Right panels:} The distributions of C-AGB with low or undetected variability in gray in the LMC (top) and M101 (bottom) as a function of color excess in K$_s$ - W2 mag and F277W-F444W mag respectively. Distributions of C-rich sources with detected or significant variability shown in red-outlined density histograms. The final column shows the fraction of C-AGB stars detected as variable as a function of color excess. 
\label{fig:dust_and_variability}}
\end{figure*}

In the previous section, we conditioned on variability and measured color. Now we invert the question to investigate how variability depends on infrared excess. Even at the class level, we can already see that detected variability is strongly chemistry-dependent. In the LMC, 72\% of C-rich stars in the CMD-selected sample (Section~\ref{sec:lmc_cmd_agb}) are SRVs or Miras, compared to 35\% of O-rich stars. In M101, where the detection threshold for variability is much higher, 14.2\% of C-like stars and 5.1\% of O-like stars classified based on the CMD are also independently detected as variables. In Figure \ref{fig:dust_and_variability} we show the distributions of infrared color of the small-amplitude (LMC) or CMD-only (M101) distributions as gray density-normalized histograms, and the distributions of the variable populations in both galaxies as unfilled, colored histograms depending on chemical type (blue for O-rich and red for C-rich). 

We can see that while the blue and red histograms appear qualitatively similar in both galaxies, the gray histograms of the low-variability stars in M101 are skewed redder. This change is likely the result of the higher threshold of detection for variability that causes many real variables to be thrown out in M101. 

\subsubsection{LMC results}
In the LMC, a star with significant infrared excess is almost always also highly variable. For stars with $K_s - W2 \geq 0.5$, only 5.8\% of O-AGBs and 2.1\% of C-AGBs in the LMC are quasi-constant ($A_1 < 0.05$ mag). This is consistent with all of the dusty stars being variable. If we consider stars in the top decile of reddening, O-rich stars are 6.8 times more likely to be large-amplitude variables ($A_1 > 0.3$ mag; 30.3\%) than the O-rich population as a whole (4.5\%). In contrast, for the half of O-AGBs with below-median reddening, only 0.3\% reach that amplitude. C-rich stars are, on average, much more likely to be highly variable (23.9\%), but they show the same trend as their O-rich counterparts: 95.4\% of the top decile has $A_1 > 0.3$ mag which is 4 times the rate for the entire C-AGB population. 

This relationship does not survive inversion: not all variable stars are dusty (see also Figure \ref{fig:amp_color_sep}, which shows the median colors of LMC AGB stars as a function of $I$-band amplitude). Among the stars with $A_1 > 0.3$ mag, 78\% of the O-rich stars and 26\% of the C-rich stars have $K_s-W2 < 0.5$ mag, and 39\% of the O-rich stars have $K_s-W2 \leq 0.2$, implying little infrared excess. However, variables generally do avoid the truly dust-free half of the population: only 3.8\% of the large-amplitude O-rich stars (and 13.7\% of the C-rich) fall below the median color of their chemical class. Among the quasi-constant ($A_1 < 0.05$ mag) stars only 2.5\% of O-rich and 0.3\% of C-rich stars are found in the top decile of infrared excess.

\subsubsection{M101 results}
M101 reproduces these results quantitatively. Stars we have classified as O-like that are in the top 10\% of F277W - F444W infrared excess are 5.6 times more likely to be variable (28.3\%) than the entire O-AGB/O-like star population. On the other hand, if we examine presumed O-AGB stars that are below the median in infrared excess, only 0.4\% of them are classified as variable. Of the variables that we do detect, only 4.3\% are below the median dust color of the larger, CMD-selected AGB population. 

For presumed C-AGBs in M101, we see a distinctive turnover at the reddest bins, with the percentage of these stars being detected as variable peaking between 50th and 90th percentiles in dustiness at 21.6\%. Once we reach the reddest 10\% of stars, only 12.5\% of them are detected as variable, and of the top 5\% this number drops to 4.2\%. This turnover is the result of detection incompleteness, not physics. This can be seen when we consider only the stars that are above the approximate TRGB location in F150W (thus excluding the most-heavily reddened sources) rather than the full, CMD-selected sample. For this group, when we bin by F277W - F444W color excess, we find that instead there is no turnover; for stars with below-median reddening, 9.1\% are detected as variable, followed by 22.5\% for the 50th to 90th percentile stars, and staying flat (22.2\%) above 90th percentile. When dust reddening is not affecting detectability, dustier stars remain 2.5 times more likely to be detected as variable than less dusty stars. 

In both galaxies, stars with more infrared excess were several times more likely to be variable (consistent with 100\% variable for the reddest stars). However, not all variable stars had significant infrared excess: a substantial fraction of large-amplitude variables in both galaxies have little or no infrared excess. When correcting for selection effects, we find that for increasing values of reddening, the fraction of stars that have detectable variability increases monotonically. We defer the physical interpretation of the asymmetry -- pulsation-driven, dust-enhanced mass-loss -- to Section \ref{sec:pulsation_mass_loss}. 

\section{Discussion} \label{sec:discussion}
Here we consider the physical implications of the analysis presented in the previous section and how these connect with the existing literature. 

\subsection{Evidence for pulsation-driven, dust-enhanced mass-loss}
\label{sec:pulsation_mass_loss}

In Section \ref{sec:variability_dust}, we show that virtually all dusty stars are variable, but not all variable stars are dusty. This asymmetry suggests a causal ordering: variability is a necessary but not (or perhaps not yet) sufficient criterion for dust formation. This result is consistent with our current understanding of mass-loss in AGB stars as being pulsation-driven and dust-enhanced. \cite{Hofner_Olofsson_2018} describe the stellar outflows of AGB stars as arising in two major steps.  First, the shock waves produced by pulsations levitate dense, warm gas out to larger radii ($\sim$ 2-3 R$_*$). Next, at these larger radii, the gas is then able to cool to temperatures low enough ($\sim 1000$K) for molecules to condense into solid dust grains. The dust that forms is still collisionally coupled with the surrounding gas. This allows radiation pressure from the star to accelerate the dust grains, which drag the molecular gas along through collisions, driving a large-scale wind. 

Previous work investigating the relationship between pulsation and mass-loss also supports this conclusion. \cite{McDonald_Zijlstra_2016} examined the question of how period and amplitude correlate with dust production, finding that among variable stars, there is a sharp onset of dust production at a period of $\sim 60$ days. \cite{McDonald_and_Trabucchi_2019} explicitly links the onset of pulsation-driven and dust-enhanced winds to the transition between period-luminosity sequences, consistent with the picture in which stars evolve across sequences toward fundamental-mode large-amplitude pulsation, growing in period and amplitude as they go \citep{Wood_2015, Trabucchi_2017}. Population-level studies such as SAGE and SAGE-VAR \citep{Srinivasan_2009, Riebel_2012, Riebel_2015} have also shown that the mid-infrared excess and fitted mass-loss rates correlate with the period and amplitudes of the LMC AGB population.  

\citet{McDonald_Zijlstra_2016} and \citet{McDonald_and_Trabucchi_2019} detected the onset of dust production through mid-infrared excess ($K_s - [22]$ and $K_s - W3$, respectively). The excesses that we measure ($K_s - W2$ in the LMC and F277W$-$F444W in M101) end at 4.4--4.6~$\mu$m, and we expect that at this near-infrared range a given amount of dust, particularly silicate dust, will produce less excess than in the mid-infrared. Thus, our colors should be less sensitive to the smallest dust columns. A variable star that has little excess in near-infrared color may therefore still be producing dust at a low rate. In the LMC, the criterion that \citet{McDonald_and_Trabucchi_2019} used to define dust-producing stars, $K_s-W3 > 0.3$ mag, corresponds to $K_s-W2 \approx 0$ mag among stars bright enough to be well detected in AllWISE $W3$ ($K_s < 12$ mag; \citealt{Cutri_2013}), close to the colors of our quasi-constant ($A_1 < 0.05$ mag) stars. The value $K_s-W2 = 0.5$ mag used in Section \ref{sec:variability_dust} is an illustrative level for considerably redder stars, not a definition of a dusty star or a lower bound on the dustiness that we can detect. 

Extreme AGB (x-AGB) stars are a photometrically defined class of heavily dust-obscured AGB stars \citep{Blum_2006}, found to be predominantly carbon-rich at Magellanic Cloud metallicities \citep{Boyer_2011, Riebel_2012}. The oxygen-rich counterparts of this heavily obscured phase -- the OH/IR stars -- do exist in the LMC and reach x-AGB-like colors, as shown by OH maser detections \citep{Marshall_2004} with infrared counterparts characterized by \citet{Goldman_2017}, but they are rare at LMC metallicity and rarer still in the SMC \citep{Goldman_2018}. The x-AGB class is therefore predominantly carbon-rich in the LMC environment in which it was first defined, though the chemical composition varies as a function of metallicity. Despite making up $<6\%$ of the AGB population, x-AGB are responsible for producing $>75\%$ of the dust \citep{Riebel_2012}. \cite{Boyer_2015} found with two epochs of observation that in metal-poor dwarf galaxies, at least 33-61\% of dusty extreme AGB stars are variable (a lower limit set by sparse temporal sampling).

Our results build on the existing literature by examining the bidirectional relationship between dust production and variability --- specifically, by evaluating both $P(\rm{Dusty|Variable})$ and $P(\rm{Variable|Dusty})$ --- within two photometrically complete AGB populations. While these studies established that infrared excess scales with pulsation and interpreted the onset as pulsation-driven, measuring the conditional dependence in both directions tests that ordering directly, distinguishing whether dustiness is only correlated with pulsation or whether pulsation is required for dust condensation to occur.  In addition, the 1.9 $\mu$m water index, which draws on the F182M band's sensitivity to molecular water absorption, is remarkably stable under variability selection in both the LMC and M101, changing by $\lesssim 0.04$ mag in either galaxy. This allows this band to function as a control for the broadband colors that are more sensitive to dust, which the same variability selection shifts by $0.5-0.7$ mag. 

The presence of variable stars with little detectable infrared excess in these colors demonstrates that pulsation alone is not enough to guarantee immediate dust production. These stars may never produce significant dust, or may not yet have reached the conditions for efficient dust formation; our single-epoch color observations cannot distinguish between the two. Both possibilities, however, do imply that pulsation is necessary. However, the chemistry-dependent strength of the amplitude-to-dust relation does suggest that time on the AGB may be the cause. Carbon stars show a much tighter correlation between dustiness and amplitude than oxygen-rich stars, partly because carbon dust produces more near-infrared excess per unit mass than silicate dust. Given that carbon stars evolve from O-AGB stars, they are also a more evolutionarily selected population, having already entered the thermally pulsing (TP-AGB) phase by necessity. On the other hand, O-AGBs are an evolutionarily heterogeneous population that ranges from early-AGB stars to highly evolved OH/IR stars. The chemistry dependence of the amplitude-dust relationship also carries important environmental information, which we discuss in more detail in Section \ref{sec:color_environment}. 

\subsection{Environmental dependence of AGB colors: metallicity and dust production}
\label{sec:color_environment}

\begin{figure*}
\centering
\includegraphics[width=\textwidth]{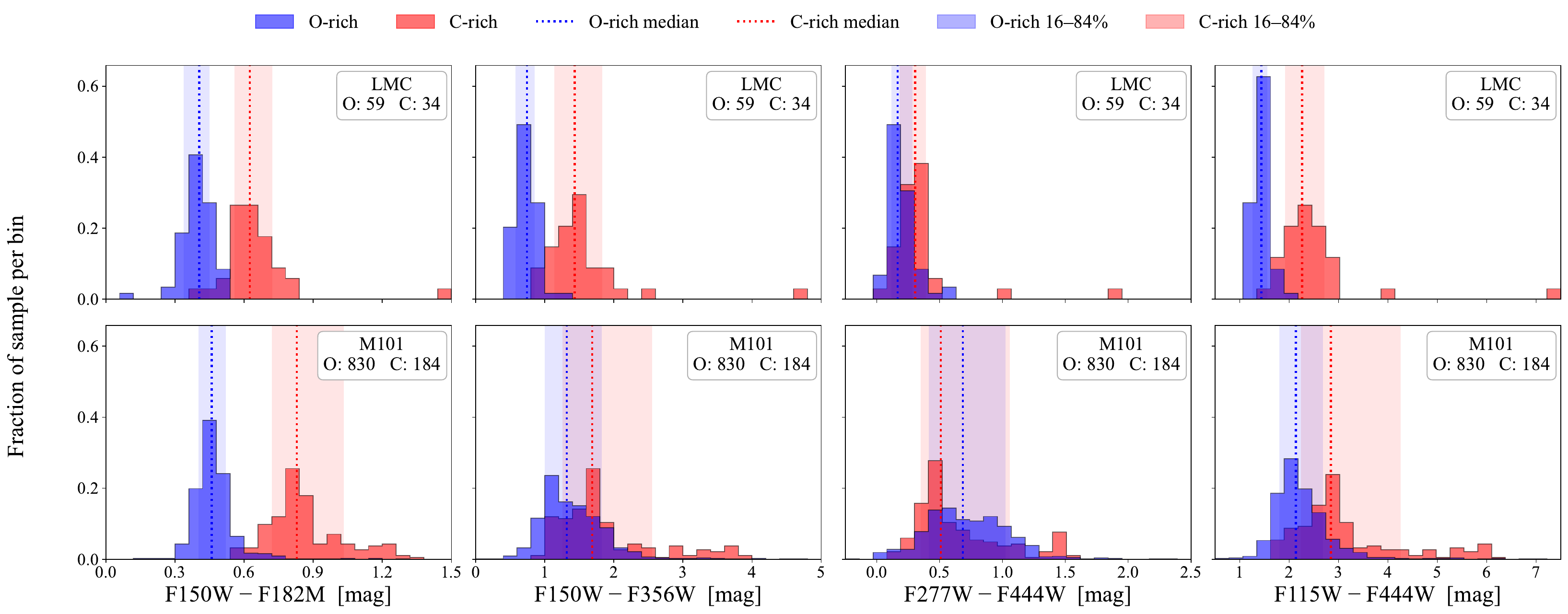}
\caption{\textit{Top row:} Density histograms of synthetic NIRCam colors for LMC stars (O-rich shown in blue, C-rich shown in red). Median color for each chemical type is shown with a dotted line and the 16-84\% range shown as lightly shaded bands. 
\textit{Bottom row:} Density histograms of observed NIRCam colors for M101 stars using the same color scheme as the top. Both rows share x-axis ranges. 
\label{fig:lmc_vs_m101_var_4col}}
\end{figure*} 

In addition to being more evolutionarily selected than O-AGB stars, carbon stars are also theoretically predicted to produce dust with far less dependence on the initial metallicity of their environment. Because carbon stars manufacture their own carbon via the third dredge-up, models predict that they can efficiently produce dust and trigger heavy mass loss regardless of their environment. Models spanning a wide range of initial metallicities consistently find that the carbon dust yield tracks dredged-up carbon excess rather than initial composition \citep{Ferrarotti_Gail_2006, Ventura_2012, Nanni_2013}. This is also supported by spectroscopic observations spanning the Galaxy and the Magellanic Clouds that demonstrate that while the composition of the carbon dust may shift slightly with metallicity, carbon stars with similar pulsational properties produce similar total quantities of dust regardless of their initial environment \citep{Sloan_2008}.

On the other hand, oxygen-rich stars rely on inherited seed elements (e.g., Si, Mg, Fe) to form silicate dust, creating a bottleneck at low metallicities. Observationally, the dust-to-gas ratio in O-rich winds scales roughly linearly with initial metallicity \citep{van_Loon_2000}. At higher metallicities, theoretical models show that silicate dust condenses more efficiently and closer to the stellar photosphere \citep{Nanni_2013}. Extragalactic observations at low metallicity suggest that there may be a dual-trigger mechanism for AGB mass loss \citep{Lagadec_Zijlstra_2008}. 
In carbon stars, the transition to carbon-richness may trigger an AGB superwind (a wind with a high mass-loss rate) at lower luminosities and pulsation amplitudes, whereas oxygen-rich stars at low metallicity must evolve to higher luminosities and larger pulsation amplitudes to compensate for their poor dust opacity. 

Previous studies of high-metallicity, extragalactic AGB populations have demonstrated the necessity of near- and mid-infrared observations for detecting the complete dust-producing stellar population \citep[e.g.,][]{Dalcanton_2012, Boyer_2015}. In addition, previous observations mapping the M31 disk have shown that at near-solar metallicities, carbon star formation is severely suppressed \citep{Boyer_2019} and the global circumstellar mass-loss budget becomes overwhelmingly dominated by oxygen-rich silicates \citep{Goldman_2022}, although the completeness of the carbon-star census at M31's distance remains a caveat to the exact fraction. 

\begin{figure*}
\centering
\includegraphics[width=\textwidth]{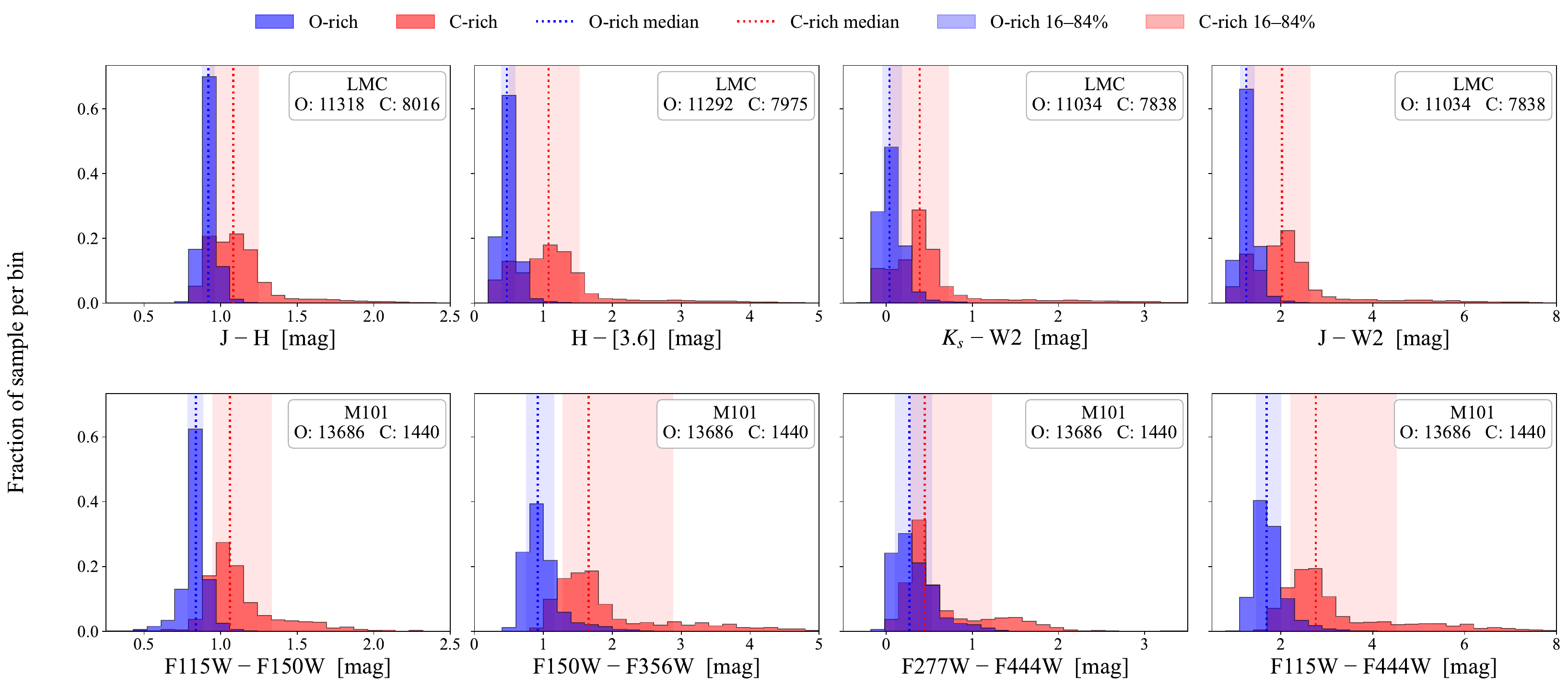}
\caption{\textit{Top row:} Density histograms of observed 2MASS, CatWISE, and IRAC colors for LMC CMD-selected stars (O-rich shown in blue, C-rich shown in red). Median color for each chemical type is shown with a dotted line and the 16-84\% range shown as lightly shaded bands. 
\textit{Bottom row:} Density histograms of observed NIRCam colors for M101 CMD-selected stars using the same color scheme as the top. Both rows share x-axis ranges. 
\label{fig:lmc_vs_m101_cmd_4col}}
\end{figure*} 

Notably, an oxygen-rich-dominated dust budget is not exclusive to metal-rich environments. A recent study by \citet{Nally_2026} on the low-metallicity ($\sim 0.2 Z_{\odot}$) dwarf NGC 6822 finds that O-rich stars supply 60\% of the total dust return, driven largely by a few intermediate-mass stars undergoing hot-bottom burning, with four sources producing half of the galaxy's dust. The integrated budget is therefore sensitive to the star-formation history and can hinge on a handful of extreme sources. At the other extreme, very old populations such as globular clusters also have O-rich-dominated dust production, because their stars are too low in mass to become carbon stars. The quantity we trace here is merely demographic: we examine only the fraction of relatively `ordinary' O-rich AGB stars that have detectable dust production. 

Translating the observed broadband infrared excess into a physical mass-loss rate requires the use of empirical scaling relations, spectral energy distribution (SED) fitting, and assumed fixed gas-to-dust ratios, grain properties, and wind velocities, all of which may vary between environments. Extracting the efficiency of dust condensation from these inferred rates can therefore be complicated by theoretical conversions. Thus, in this work we provide a direct, empirical complement to the previous work by tracing the dust production onset differentially and analyzing only observational reddening rather than inferred mass-loss rates. Throughout, we use infrared excess as a proxy for dust production. However, because the emission per unit dust mass depends on grain composition, we use it to compare stars within the same surface chemistry class rather than to directly compare dust quantities between the C- and O-rich classes. 

As shown in Section \ref{sec:lmc_m101_color_comparison}, the AGB population in M101 is systematically redder than that of the LMC and the offset is strongly wavelength-dependent. The shorter wavelength bands, which are more sensitive to the stellar photosphere and gas, remain nearly stable or show relatively small offsets ($+0.02$ - $0.18$ mag for the O-rich and $+0.14-0.35$ mag for the C-rich stars) while the dust-sensitive colors beyond $\sim 2\mu$m (such as F356W and F444W) exhibit dramatically increased infrared excess, particularly for O-AGBs, which are $0.5-0.7$ mag redder in M101 than in the LMC. 

Figure \ref{fig:lmc_vs_m101_var_4col} shows the distribution of synthetic and observed NIRCam colors for variable AGB stars in the LMC and M101. While the medians shift redward for both C and O stars, the change is more dramatic for O-rich stars. The O-rich distribution is also up to three times as broad in M101 for the same percentile range in the F356W and F444W colors, but has nearly the same width in F150W - F182M. As discussed in Section \ref{sec:var_vs_cmd_sel_eff}, this 0.5-0.7 mag offset represents the difference between the variability-selected populations in both galaxies (i.e., the pulsating stars with the most infrared excess). While the higher detection threshold for variability in M101 means that the exact magnitude of this change cannot be taken at face value as a universal reddening constant, the underlying physical phenomenon is robust. Even when completely removing variability selection and comparing the most conservative CMD-only samples across galaxies, a definitive $\gtrsim 0.1$ mag baseline and greater spread in color persists (shown in Figure \ref{fig:lmc_vs_m101_cmd_4col}). Furthermore, when comparing variables directly with their non-variable CMD-based parent populations in each galaxy, the surge in reddening triggered by variability for oxygen-rich stars is far more dramatic in M101 than in the LMC. 

We interpret this wavelength-dependent difference as most likely reflecting metallicity-dependent dust formation in O-rich stars. O-rich stars show substantially more infrared excess than their LMC counterparts, as expected if dust formation scales with the metallicity of their environment. With these bands, however, we cannot separate the relative quantity of dust from its composition because the near-infrared emission per unit dust mass will also depend on grain properties. While a comparison limited to two galaxies prevents us from conclusively ruling out all other environmental factors, the direction and wavelength structure of the shift are consistent with the theoretical expectation for metallicity-dependent silicate condensation. By directly comparing LMC and M101 stellar populations in the same bandpasses (synthetic photometry for the LMC, observed for M101) we also reduce potential offsets introduced by cross-calibration. 

In contrast, the carbon stars in M101, while still redder than those in the LMC, show a much more stable offset (ranging from $+0.15$ to $+0.20$ mag), consistent with carbon dust production being efficient in both low and high metallicities.  While the \textit{efficiency} of carbon dust production is expected to be negligibly dependent on initial metallicity, our results also demonstrate that this does not necessarily mean the \textit{colors} of carbon stars do not depend on their environment. Regardless of whether we compare CMD-selected or variability-selected samples in the LMC or M101, we consistently find that, overall, C-rich stars are redder in M101 than they are in the LMC. We interpret this $\sim 0.15 - 0.20$ mag difference as reflecting the compositions of the underlying C-star populations in both galaxies. Because the initial oxygen abundance is higher in M101 than in the LMC, a star needs more third dredge-up events to cross the C/O $>1$ threshold and become carbon rich. As a result, the carbon stars that form in M101 are plausibly a more massive and more highly evolved subset than those in the LMC, and such stars are expected to be cooler, dustier, and hence redder. A higher metallicity also lowers the effective temperatures of AGB stars at a given luminosity, which would redden the photospheric colors of both chemical types and may contribute to this offset. 

We note that the variability information used in this study is integral to understanding these differences by providing another axis of information beyond chemical type and infrared excess. While C/O classification describes which channel a star uses for dust formation, it provides asymmetric information about how far a star has progressed towards the heavily mass-losing end of the AGB. Carbon stars have necessarily entered the TP-AGB stage, but O-AGB stars (the AGB ``default'') span everything from early-AGB stars to heavily dust-enshrouded stars near the tip of the AGB. With only the CMD-based classification, the difference in infrared excess for O-AGB stars between the LMC and M101 is only $\sim +0.09$ mag. However, the difference in infrared excess between O-AGB variables in the two galaxies grows to a median value of $+0.36$ mag and up to $+0.74$ mag for dust-sensitive colors. Within galaxies (a comparison that removes cross-calibration and crowding differences between the two galaxies), the variable O-AGBs are redder than the CMD-only O-AGBs by $+0.26$ mag in M101 (a lower bound; see Section \ref{sec:var_vs_cmd_sel_eff}) but only $+0.08$ mag in the LMC (Table \ref{tab:var_amp_2x2}). Variability is thus associated with at least three times more infrared excess in M101.  Variability allows us to compare the infrared excess (and implied dust production) in stars at comparable evolutionary stages across different environments, thus making it possible to measure the turn-on of O-AGB infrared excess between the LMC and M101 directly, without relying on theoretical conversions. 

\subsection{Effect of Contamination on JAGB Measurements}
\label{sec:jagb_contamination_discussion}

Given the results shown in Table \ref{tab:jagb_contamination}, one might wonder if the field examined in this paper is unusually contaminated and would fail to show the self-consistency demonstrated in so many previous works and systematic tests of the JAGB method \citep[e.g.,][]{Zgirski_2021, Parada_2023, Lee_2023, Li_2025}. Here, we demonstrate that insensitivity to the statistical method employed (e.g., mean, median, and mode) in a given field and a near-Gaussian luminosity function (LF) do not guarantee sample purity. In fact, contamination can sometimes increase the apparent robustness of the JAGB fiducial luminosity to various tests. While the photometry is internally consistent, we have not attempted to derive fully calibrated magnitudes and therefore these will likely have a small ($\leq 0.05$ mag) offset from their real apparent magnitudes. Their purpose here is not to provide a JAGB-based distance measurement but to illustrate the effect that population makeup and choice of statistical tool can have on the measured JAGB magnitude. 

Table \ref{tab:jagb_mean_mode} shows the effect of contamination on the mean, median, and modal magnitudes of the JAGB stars originally presented in Figure \ref{fig:jagb_cmds}. Because we had already pre-selected the AGB stars in Section~\ref{sec:m101_cmd_agb}, we do not employ a minimum or maximum magnitude cut on the stars in each JAGB window, and we rely only on color to determine J-region membership. Stars not already identified as AGBs were then excluded from the JAGB measurement to avoid artificially inflating the inconsistencies between the contaminated (mixed C+O-AGB) and uncontaminated (C-AGB only) statistics. For the contaminated `AGB' columns, we use every AGB star in the window for determining the statistics. This replicates the process most commonly used in the JAGB literature of assuming that the J-region is composed of C-AGBs. For the uncontaminated `C' columns, we calculate the same statistics using only the C-classified members ($\mathrm{F150W}-\mathrm{F182M} \ge 0.64$).  We note that the mode is insensitive to the choice of whether non-AGBs (the stars in the color window but not in our AGB sample from Section \ref{sec:m101_cmd_agb}) are included --- varying by only $\leq0.02$ mag --- but mean and median varied by as much as $0.7$ mag. To determine the mode in both cases, we used the peaks of a Gaussian kernel density estimate with a 0.18 mag fiducial kernel. Because this choice matters for a double-peaked luminosity function, we test kernels from $0.10$ to $0.40$ mag below.

\begin{deluxetable*}{lllrrrrrrrrr}
\tablecaption{Mean, median, and mode of J-region stars
\label{tab:jagb_mean_mode}}
\tablewidth{0pt}
\tablehead{
\colhead{Reference} & \colhead{Color} & \colhead{Band} &
\colhead{$\langle m \rangle_{\rm AGB}$} & \colhead{$\langle m \rangle_{\rm C}$} &
\colhead{$\Delta\langle m \rangle$} &
\colhead{$m^{\rm med}_{\rm AGB}$} & \colhead{$m^{\rm med}_{\rm C}$} &
\colhead{$\Delta m^{\rm med}$} &
\colhead{$m^{\rm mode}_{\rm AGB}$} & \colhead{$m^{\rm mode}_{\rm C}$} &
\colhead{$\Delta m^{\rm mode}$} \\
\colhead{} & \colhead{} & \colhead{} &
\colhead{(mag)} & \colhead{(mag)} & \colhead{(mag)} &
\colhead{(mag)} & \colhead{(mag)} & \colhead{(mag)} &
\colhead{(mag)} & \colhead{(mag)} & \colhead{(mag)}
}
\startdata
\citet{Li_2025} & $\mathrm{F150W}-\mathrm{F277W}$ & F150W & 22.32 & 22.24 & $+0.07$ & 22.32 & 22.18 & $+0.14$ & 22.27 & 22.07 & $+0.20$ \\
\citet{Lee_2025_JWSTJAGB} & $\mathrm{F115W}-\mathrm{F356W}$ & F115W & 23.21 & 23.26 & $-0.05$ & 23.21 & 23.19 & $+0.02$ & 23.23 & 23.09 & $+0.14$ \\
\citet{Lee_2025_JWSTJAGB} & $\mathrm{F115W}-\mathrm{F444W}$ & F115W & 23.21 & 23.28 & $-0.07$ & 23.22 & 23.22 & $-0.00$ & 23.23 & 23.09 & $+0.14$ \\
\enddata
\tablecomments{Statistics of the apparent-magnitude distribution of the CMD-selected AGB stars within published JAGB color windows (boundaries are listed in Table~\ref{tab:jagb_contamination}). ``AGB'' uses every AGB star in the window;
``C'' restricts to the C-classified members ($\mathrm{F150W}-\mathrm{F182M} \ge 0.64$). Modes are peaks of a Gaussian-kernel density estimate with a 0.18 mag kernel. %
}
\end{deluxetable*}

Published JAGB analyses adopt different statistics of the luminosity function --- the mean \citep{Madore_and_Freedman_2020, Zgirski_2021}, the median \citep{Ripoche_2020, Parada_2021}, and, in CCHP work since \citet{Lee_2022}, the mode \citep{Lee_2025_JWSTJAGB}. \citet{Li_2024} find that the choice changes the mean SN Ia host distance by up to $\sim$0.19 mag. Contamination affects the mode most (Table~\ref{tab:jagb_mean_mode}). Because the O-AGB stars in the J-region are on average fainter than the C-AGB stars, the contaminated mode is $0.20$ mag fainter than the C-AGB-only mode in the F150W$-$F277W window and $0.14$ mag fainter in both F115W windows. In the F150W window, the two populations make the luminosity function double-peaked, and the all-AGB mode drifts by $0.18$ mag across kernels of $0.10-0.40$ mag; the near-Gaussian F115W luminosity functions give modes stable to $\leq 0.03$ mag over the same range. Yet the contaminated statistics are the more internally consistent ones: the mean, median, and mode of the all-AGB sample span only $0.05$ mag in the F150W window and $0.02$ mag across all six F115W variants (23.21--23.23 mag), compared with $0.17$ and $0.19$ mag for the C-AGB-only samples.

This exercise demonstrates that many of the tests typically used to assess the systematic effects of the JAGB region (e.g., Gaussianity of the LF, consistency between statistical methods) are not reliable indicators of the true sample purity of the J-region. Moreover, in at least some environments and band combinations, the C-AGB LF should not be assumed to be Gaussian, and in fact may be less Gaussian than the O-AGB LF. While we do not measure a C-AGB-based JAGB magnitude --- in part due to the lack of an appropriate absolute distance calibration ---  we also caution that calibration is likely to be challenging and field-dependent.  

Despite having some physical motivation, the JAGB method is first and foremost an \textit{empirical} distance indicator. \citet{Nikolaev_and_Weinberg_2000} proposed the Region J stars as potential standard candles as \textit{long-period variables, calibrated through period--luminosity--color relations}, for the purpose of mapping LMC line-of-sight structure. In this way, period and other variability properties could likewise be used to constrain the masses of the J-region population, essentially through a period--luminosity relation (PLR). 

Though employing a PLR would improve the physical interpretability and robustness of using C-AGBs as a distance indicator, using a PLR instead of CMD-based statistics would also negate the main advantages of the JAGB method: ease of use and reduced observational time requirements. Recent works, such as \cite{Gavetti_2026}, have made significant progress on how our theoretical understanding of C-AGB evolution can be used to inform the creation of a more robust and less environmentally sensitive distance indicator from the J-region without obtaining time-series. Their MavJ (the mean $J$-band magnitude of C-AGBs) is the equivalent of our $\langle m \rangle_{\rm C}$. As shown in Table \ref{tab:jagb_mean_mode}, the difference between the contaminated and C-AGB-only mean magnitudes in our M101 field ($\Delta \langle m \rangle_{\rm C} \leq 0.07$ mag)  is comparable to the 0.05 mag spread in MavJ that \citet{Gavetti_2026} finds across metal-enrichment histories. Our numbers agree with theirs on the stability of the mean, but we attribute this to a different reason. In their models, the mean is robust because the shape of the luminosity function, and hence its peak, tracks the host's enrichment history while the mean does not. They expect that O-rich contamination enters only from massive stars undergoing hot-bottom-burning and therefore should be negligible. In M101, however, we find that the mean is robust to a second, fainter O-rich population inside the J-region. This numerous and fainter population shifts the mode by up to 0.20 mag in our M101 field, but the mean by only $\leq 0.07$ mag. These two effects are independent, but both favor the mean. The median behaves intermediately, matching the C-AGB-only value to $\le 0.02$ mag in the two F115W windows while being dragged towards the faint end by $+0.14$ mag in the F150W window. While this is an $N=1$ SN Ia host study, our results suggest that both mean and median are more robust to contamination than modal magnitude and may merit further comparison. 

Another hypothesis assumes that the entire AGB population within the J-region behaves as a standard candle, regardless of the chemical types of the stars. While there is currently no clear physical motivation to suggest this is the case, the results presented here alone do not conclusively disprove this hypothesis. Comparing the agreement in the JAGB magnitude between geometric anchors at a range of metallicities would be one way to investigate this.

\begin{figure*}
\centering
\includegraphics[width=\textwidth]{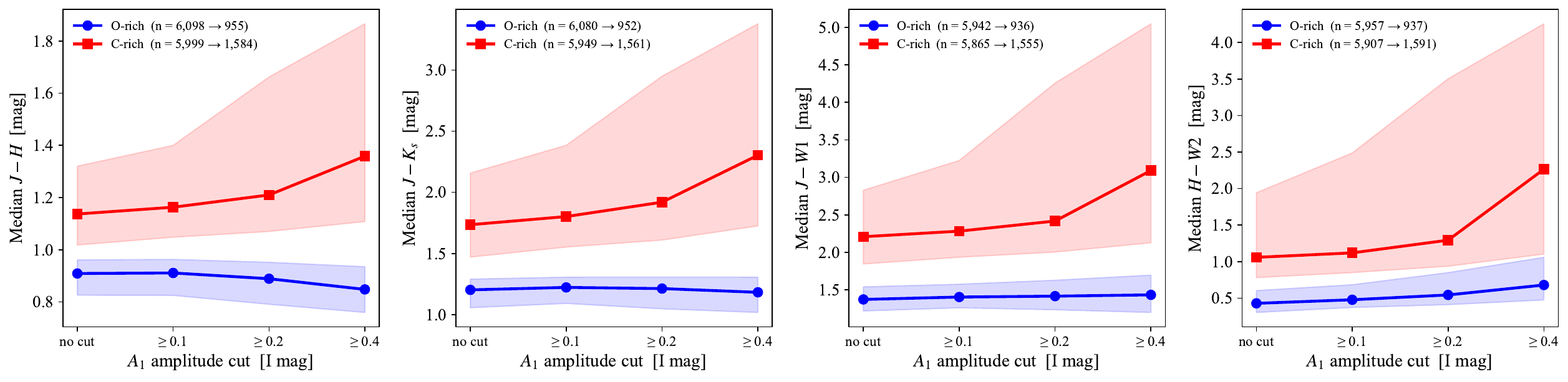}
\caption{The median colors of LMC C (in red squares) and O (in blue circles) AGB stars plotted as a function of their $I$-band amplitude for four color combinations. Shaded regions show the 16-84th percentile regions for both types of star. 
\label{fig:amp_color_sep}}
\end{figure*} 

\subsection{Variability information improves classification}
\label{sec:variable_classification}

The same variability information also improves the chemical classification of AGB stars. In the LMC, the SRV and Mira spectra we obtained for this study confirm 92 out of 93 photometric classifications made by the OGLE team \citep{Soszynski_2009}, suggesting an error rate of less than a few percent. On the other hand, for OSARGs, the same classification scheme is less successful. We retrieved the spectra for a small ($N=7$) number of OSARGs photometrically classified as C-rich; six are spectroscopically O-rich. This is consistent with the OGLE team's interpretation that, while C-AGBs do exist among the OSARG population, they are likely rarer than photometric classifications alone would suggest. 

The classifications of the larger-amplitude variables are likely more robust because C- and O-AGB stars in the LMC differ more strongly in color as their amplitudes increase. Figure \ref{fig:amp_color_sep} shows the median and 16-84th percentile ranges of colors for C- and O-AGB stars in four color combinations. While the spread in color becomes larger at higher amplitudes (see Section \ref{sec:variability_dust}), the separation between the medians of the distributions still grows larger. The change in the distributions is asymmetric. Even at larger amplitudes, the LMC O-AGB distribution's width changes only modestly (e.g., 0.32 mag $\rightarrow$ 0.50 mag in $J-W1$) whereas the distribution width of the C-AGBs is much larger (0.98 mag $\rightarrow$ 2.92 mag in $J-W1$). As discussed in Section \ref{sec:color_environment} this likely reflects the higher near-infrared emissivity of carbon dust and also more efficient dust production of C-AGBs at the metallicity of the LMC compared to O-AGBs at the same metallicity. 

At higher metallicities, dust alone no longer distinguishes the two chemistries, because O-AGB stars also show substantial infrared excess; in Section \ref{sec:co_color_sep}, only the two colors that used F182M stayed bimodal in M101. This is consistent with \cite{Boyer_2024}, who were \textit{``unable to find filter combinations that effectively isolate AGB stellar types from one another, independent of metallicity.''} A direct comparison with the previous NIRCam filters can be found in Appendix \ref{sec:appendix_classification}. The AGB color degeneracy is one-sided (Section~\ref{sec:application_classification}): stars with little dust are almost always O-rich, so a clean O-rich subset is easier to select than a pure C-rich sample, and the most heavily reddened stars are the hardest to classify.

With a greater initial oxygen abundance, O-AGB stars at higher metallicities are also expected to show stronger absorption in the molecules containing oxygen. This effect would make medium-band filters targeting those features more effective. The model grids of \cite{Boyer_2024} indicate that F150W$-$F182M is an effective color at metallicities like the half-solar value of our M101 field and our data also agree. The presumed O-AGB population in the M101 F150W$-$F182M CMD has the same small $\sim 0.04-0.06$ mag distribution width as in the LMC, which does not hold for any of the broadband colors in Figure \ref{fig:lmc_vs_m101_cmd_4col}. Applying the water index to M101, we identify 100 O-AGB candidates among the 246 AGB stars in the heavily reddened tail of the F150W$-$F182M CMD, stars that the color alone would classify as exclusively C-rich. Spectroscopic follow-up will be needed to confirm these classifications, but the direction of the effect is consistent with the asymmetric contamination expected at the metallicity of M101. 

\begin{deluxetable}{ll}
\tabletypesize{\footnotesize}
\tablewidth{0pt}
\tablecaption{Columns of the M101 AGB classification
catalog.\label{tab:catalog_columns}}
\tablehead{\colhead{Column} & \colhead{Description}}
\startdata
ID & Source identifier \\
X, Y & Detector pixel coordinates \\
R.A., Decl. & ICRS coordinates (J2000) \\
F115W \ldots F444W & Vega magnitudes and uncertainties in the six NIRCam bands \\
color & F150W$-$F182M color (mag) \\
$I_{\rm F182M}$ & Water index (Section~\ref{sec:m101_variable_selection}) \\
IR excess & True if F150W$-$F356W $> 1.25$ \\
class$_{\rm cmd}$ & CMD-only class (color cut at 0.64) \\
class$_{0.34}$ & Hybrid class, water-index boundary 0.34 (default) \\
class$_{0.40}$ & Hybrid class, water-index boundary 0.40 \\
\enddata
\tablecomments{Classes are photometric candidates (O/C). The hybrid
columns apply the CMD cut blueward of F150W$-$F356W $= 1.25$ and $I_{\rm F182M}$ redward; see Section~\ref{sec:application_catalog}.}
\end{deluxetable}

\begin{deluxetable*}{rcccccccccccc}
\tabletypesize{\scriptsize}
\tablewidth{0pt}
\tablecaption{The M101 AGB classification
catalog.\label{tab:sample_final_catalog}}
\tablehead{
\colhead{ID} & \colhead{F115W} & \colhead{F150W} &
\colhead{F182M} & \colhead{F277W} & \colhead{F356W} &
\colhead{F444W} & \colhead{color} & \colhead{$I_{\rm F182M}$} &
\colhead{IR excess} & \colhead{class$_{\rm cmd}$} &
\colhead{class$_{0.34}$} & \colhead{class$_{0.40}$}
}
\startdata
 62 & 21.444 & 20.573 & 20.067 & 20.028 & 19.563 & 19.609 & 0.506 & 0.359 & False & O & O & O \\
113 & 22.008 & 20.965 & 20.168 & 19.802 & 19.429 & 19.432 & 0.797 & 0.483 & True  & C & C & C \\
140 & 21.997 & 21.016 & 20.340 & 20.365 & 19.977 & 20.214 & 0.676 & 0.500 & False & C & C & C \\
310 & 22.299 & 21.489 & 20.904 & 20.597 & 20.220 & 20.106 & 0.585 & 0.344 & True  & O & C & O \\
356 & 22.564 & 21.548 & 20.850 & 20.171 & 19.703 & 19.697 & 0.698 & 0.326 & True  & C & O & O \\
\enddata
\tablecomments{Table~\ref{tab:sample_final_catalog} is published in its entirety
in machine-readable format; a portion is shown here for guidance
regarding its form and content. Magnitudes are Vega; the color
column is the F150W$-$F182M color (mag). Photometric uncertainties
and detector coordinates appear only in the machine-readable table.
Column definitions are given in
Table~\ref{tab:catalog_columns}.}
\end{deluxetable*}

\section{AGB Classification Recommendations and the M101 AGB Catalog}
\label{sec:application}

\begin{figure*}
\centering
\includegraphics[width=\textwidth]{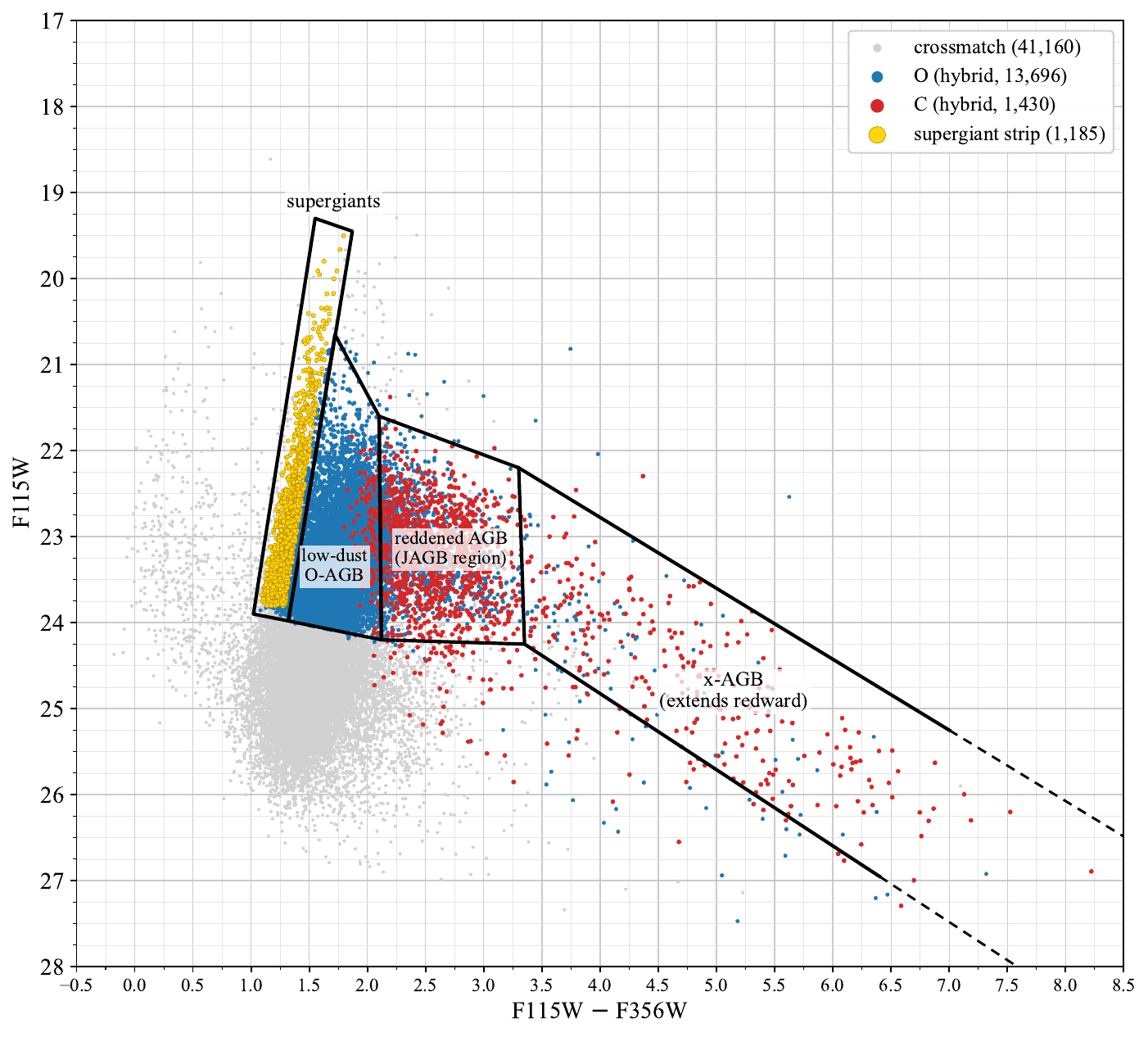}
\caption{M101 F115W - F356W color-magnitude diagram showing the major features in the near-infrared regime. Locations of the relevant features are approximate and meant to be illustrative, but define groups that are found in many resolved stellar populations. 
\label{fig:capstone_cmd}}
\end{figure*} 

\begin{figure}
\centering
\includegraphics[width=0.49\textwidth]{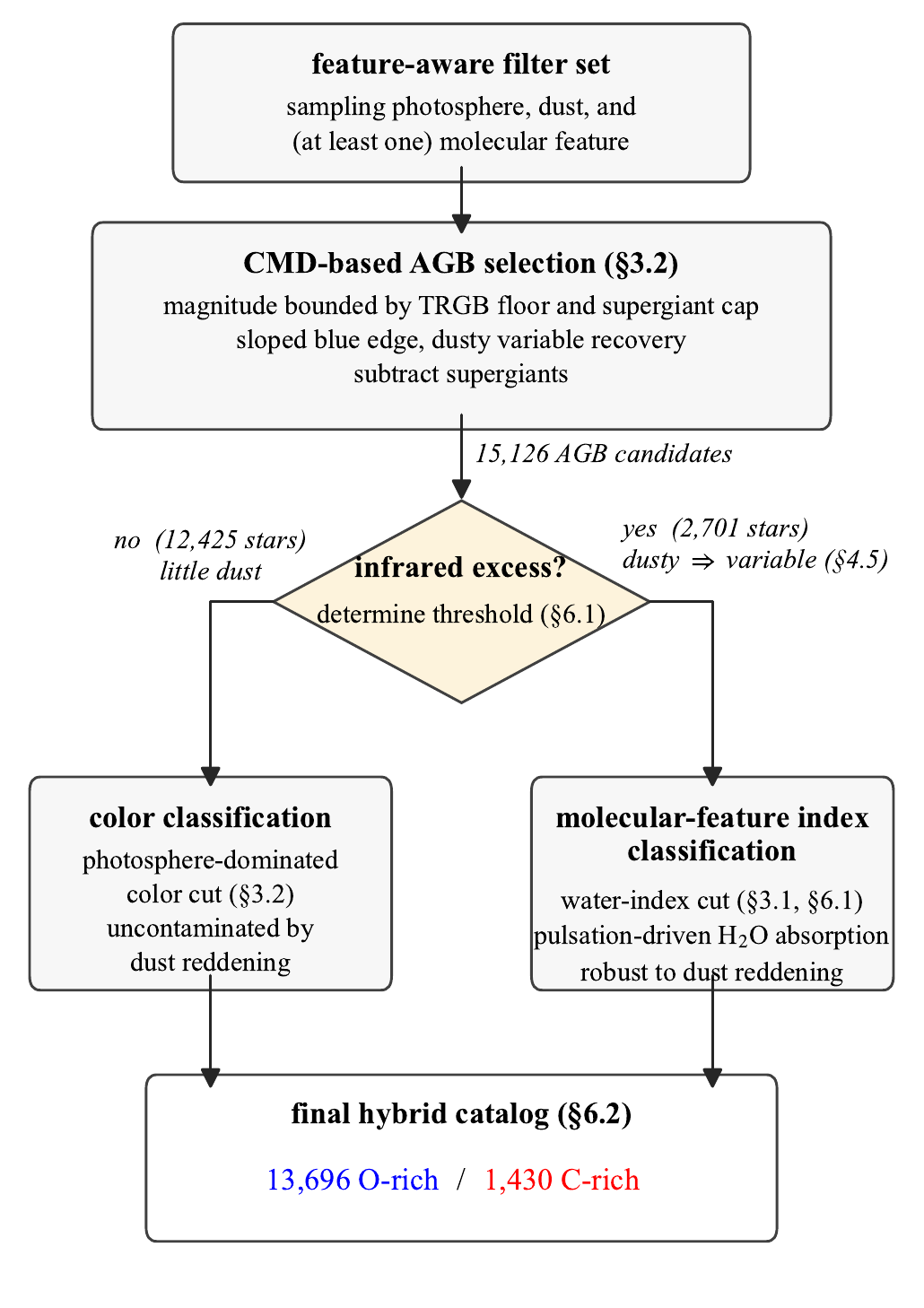}
\caption{Our final decision tree for the chemical classification of O- and C-AGB stars in M101. Specific selection criteria should be rederived \textit{in situ} for each environment, as described in the relevant text Sections labeled in the flowchart. \label{fig:capstone_flowchart}}
\end{figure}

In this section, we distill our results into guidelines for classifying resolved AGB populations with near-infrared photometry and apply them to produce our final M101 AGB catalog. Photometric classifications of AGB stars have generally rested on three working assumptions: that the colors relevant for classification are those of a static photosphere with circumstellar dust added to it, that classification boundaries can be imported from models or a well-studied calibration galaxy such as the LMC, and that the correlation between dust production (as traced by infrared excess) and surface chemistry seen in the LMC holds in other environments. 

Our results depart from these assumptions. We find that the molecular features that most reliably separate C- and O-rich surface chemistries in the near-infrared arise in layers formed by pulsation, which hydrostatic models are not designed to reproduce (Sections \ref{sec:co_color_sep} and \ref{sec:application_scope}). The colors of O-rich stars, and with them the boundaries between the two classes, shift between the LMC and M101 mostly in dust-sensitive colors, while their photospheric colors change little (Section \ref{sec:color_environment}). In population models, the metallicity dependence of AGB colors has been attributed to cooler temperatures, deeper molecular absorption, and the narrower age range of C-stars \citep{Boyer_2024}. While our results do not contradict these explanations, they do indicate that the increased dust production by O-rich stars at higher metallicity, as traced by their infrared excess, dominates in dust-sensitive colors. In our M101 field, dust-reddened O-rich stars are abundant in a region of the color-magnitude diagram that is inhabited almost exclusively by C-rich stars in the LMC (Sections \ref{sec:lmc_m101_color_comparison} and \ref{sec:JAGB_contamination}). We therefore adopt a different approach. 

We treat the classification boundaries as \textit{outputs} measured within each sample, rather than as \textit{inputs} imported from models or from calibrations derived in other environments. We adopt this perspective for three reasons. First, because the colors of O-rich stars shift with environment, a boundary determined in one environment should not be expected to hold in another; Appendix \ref{sec:appendix_classification} gives an explicit example. Second, we demonstrate that a boundary can be measured without prior knowledge of classifications, from the valley in the distribution of a single color or index (single-color bimodality; Section \ref{sec:co_color_sep}), and that its uncertainty can be estimated from the number of stars that change classification across plausible boundary positions. Third, when no bimodality or distinct populations appear, the absence itself contains information that should not be disregarded because it directly shows that the two chemistries cannot be reliably separated in those bands. In the same situation, a boundary carried over from another environment or a model would still produce classifications, but of unknown reliability. 

We prefer this approach to fixed boundaries because it relies only on the physical properties of the stars, and we expect these properties to hold true across environments: the molecular features that distinguish the two chemistries, and the role of pulsation in forming the molecular layers in which those features exist. In a one-dimensional distribution of a color or index, the classification boundary is more appropriately thought of as a threshold. A threshold is not a physical law, but rather an application of one. Its value depends on the population in which it is measured, so importing it into another environment also carries over assumptions about the population, each of which must be verified separately. 

Therefore, we recommend using the same \textit{approach} for classification rather than the same \textit{thresholds} in each environment. We choose the bands based on the physics and molecular features each is expected to sample and then measure the classification thresholds directly from the data. To apply this approach, we recommend filter sets that, at a minimum, sample a molecular feature and measure infrared excess, which identifies stars that must be pulsating (Section \ref{sec:application_bands}). For some effects, we can predict only the direction. For example, we expect the O-rich contamination of the J-region to increase with metallicity, and the classification thresholds to shift with it. However, with only two galaxies, one on either side of the inflection in the C/M--metallicity relation, we cannot determine how large these changes are or the shape of their dependence on metallicity. For such effects, we only give the expected direction and do not fit a relation.

Section \ref{sec:application_classification} applies this approach to our M101 data to derive the classification thresholds, and Section \ref{sec:application_catalog} uses them to produce our M101 AGB catalog. Sections \ref{sec:application_JAGB} and \ref{sec:application_bands} give our recommendations for JAGB measurements and NIRCam filter selection, and Section \ref{sec:application_scope} discusses the connection to model grids.

\subsection{Color as a Proxy for Pulsation}
\label{sec:application_classification}

Figure \ref{fig:capstone_flowchart} summarizes our classification scheme. Here we walk through the logic and derive the central step --- an infrared-excess threshold that decides which classifier should apply to a given star. In Sections \ref{sec:m101_variable_selection} and \ref{sec:m101_cmd_agb}, we present two methods of identifying AGB stars and assigning them provisional chemical classifications. The first, more-traditional CMD-based selection, identifies AGB stars by their brighter-than-TRGB location. We then apply cuts to remove supergiant stars and include selection criteria that recover heavily dust-reddened stars. We use F150W - F182M color to classify the resulting AGBs as C-rich or O-rich (or in our terminology, C-like and O-like). 

The other takes advantage of the time-series baseline from HST WFC3/IR to identify long-period variable stars that meet the basic criteria to be SRV or Mira candidates. The variability cut also removes supergiant stars, which are typically low-amplitude pulsators and keeps only a small number of stars below the TRGB (which are likely early-AGB or RGB stars) while retaining heavily dust-reddened stars. We then develop and use a water index, $I_{\rm F182M}$, to measure the amount of water absorption relative to the expected continuum level at the pivot wavelength of F182M. Stars with detectable water absorption are classified as O-rich candidates while stars without detectable water absorption are classified as C-rich candidates. 

Through a comparison with SPHEREx spectra, we find that, while O-AGB stars, regardless of variability, are well-separated from C-AGB stars in F150W - F182M color, only variable AGB stars are well-separated with $I_{\rm F182M}$. We interpret the cause to be twofold: the water-absorption features themselves grow stronger with pulsation and the colors of the two chemical classes of stars are more distinct at larger pulsational amplitudes. However, where both criteria are applicable, $I_{\rm F182M}$ and F150W - F182M agree to within $\sim4\%$ of each other and with the `ground truth' from SPHEREx spectra. 

The dust-variability asymmetry that we derived in Section \ref{sec:variability_dust}, and interpreted in Section \ref{sec:pulsation_mass_loss} as pulsation preceding dust formation, turns the time-series requirement into a criterion that well-chosen single-epoch photometry can satisfy. Because essentially all dust-reddened AGB stars are variable, a star's location in the dusty regions of the CMD (such as the J-region and x-AGB regions in Figure \ref{fig:capstone_cmd}) inherently implies that it pulsates. Thus, the variability requirement of $I_{\rm F182M}$ is automatically satisfied for exactly those stars whose broadband colors are most degenerate, given that dust-poor AGB stars are rarely C-rich. We note that this proxy only runs in one direction. While a red color guarantees variability, variability alone does not guarantee a red color, as many variables show little dust. Crucially, we are only able to develop this color-based proxy for future single-epoch studies because the time-domain analyses in the LMC and M101 established the necessary ground truth. 

We can also demonstrate this conclusion empirically on the M101 CMD-selected AGB population. The bimodality in $I_{\rm F182M}$, which appears only in variable stars, is present in the red subset of the CMD-selected AGB sample (F150W - F356W $> 1.25$), with a valley at 0.34 mag despite the fact that the bimodality is absent from the broader CMD-selected AGB sample. 

We determined the F150W - F356W boundary by plotting the distribution of $I_{\rm F182M}$ values and F150W - F182M color as a function of F150W - F356W threshold color. At each F150W - F356W threshold, we determined both the location and depth of the valley in the distribution of values from the two classifiers. We found that for F150W - F356W color, at around 1.25 - 1.4 mag the two observables have a comparable valley depth (suggesting roughly equal distinguishing power). However, blueward of this range, F150W - F182M is more effective and has a deeper valley (for the bluest stars $I_{\rm F182M}$ shows virtually no valley), while redward of this range, the valley in $I_{\rm F182M}$ becomes the deeper one. Our adopted threshold of 1.25 sits at the blue edge of this crossover. Because metallicity affects both the dust demographic and the depth of the water absorption, the crossover location is expected to be environment-dependent and should be re-measured in situ when the scheme is applied elsewhere. 

Our final catalog classification and recommendation, therefore, employs a combination of both methods, using the knowledge that dust-reddened stars are universally variable. For our hybrid classification scheme, we recommend employing the CMD-based guidelines for removing supergiant stars, faint, non-AGB contaminants, and recovering heavily dust-reddened stars to create the initial AGB catalog. For stars without noticeable infrared excess, the F150W - F182M color separation is the best option. However, for stars with moderate to heavy infrared excess, which we know must be variable, we apply the water index for classification. We prefer this hybrid method because $I_{\rm F182M}$ allows us to better recover the classifications of x-AGB stars, which are so heavily reddened that they are universally classified as C-rich based solely on F150W - F182M color, despite their water absorption. In the Galaxy, the heavily obscured O-rich stars that occupy the x-AGB region of the CMD are OH/IR stars, whose oxygen-rich chemistry is established not by their broadband colors, but by OH maser emission. The maser serves the same role for these stars that $I_{\rm F182M}$ serves here, discriminating chemistry more precisely in the regime where color cannot. 

This ambiguity is not unique to color-based criteria. \citet{Nally_2026} find that for $\sim10$\% of the stars they classify as C-rich via likelihood-weighted SED fitting, the single best-fitting model is O-rich --- C-rich and dusty O-rich model SEDs fit these sources comparably well, and under the best-fit convention of earlier surveys these stars would have been classified as O-rich --- whereas their O-rich classifications show no such ambiguity. In SED fitting as in colors, the AGB color degeneracy is thus one-sided: C-rich stars generally do not contaminate the low-dust O-AGB region of the CMD, but O-rich stars can contaminate the parts of the CMDs where C-AGBs reside. Feature-based indices like $I_{\rm F182M}$ thus offer a way to independently discriminate between chemical subtypes in the most confusing regions. 

\subsection{M101 AGB Catalog}
\label{sec:application_catalog}

We produce the chemical classifications for our final AGB catalog using the recommendations discussed in the previous section, with the step-by-step process summarized in the flowchart in Figure \ref{fig:capstone_flowchart}. The final catalog consists of the 15,126 AGB candidates first identified in Section~\ref{sec:m101_cmd_agb} alongside updated chemical types. For each star in the catalog, we provide six-band NIRCam photometry, F150W - F182M color, $I_{\rm F182M}$ value, an infrared-excess flag, and the chemical type from three different classification schemes. Table \ref{tab:catalog_columns} contains an explanation for the columns of the final catalog. 

For the class$_{\rm cmd}$ column, classes are the original O-like and C-like classifications derived in Section \ref{sec:m101_cmd_agb} using the F150W - F182M color. The class$_{0.34}$ and class$_{0.40}$ columns give the chemical classifications resulting from the combination of the CMD and $I_{\rm F182M}$ classification scheme described in Section \ref{sec:application_classification}. Using F150W - F356W as a proxy for infrared excess and variability, we used the standard F150W - F182M = 0.64 mag classification (see Figure \ref{fig:windex_vs_color}) for stars blueward of the F150W - F356W = 1.25 mag threshold (see Section \ref{sec:application_classification}). For stars redward of the F150W - F356W = 1.25 mag threshold ($N=$2,701), we instead employed $I_{\rm F182M}$. We used two variants for the $I_{\rm F182M}$ classification --- one based on the valley ($I_{\rm F182M} = 0.40$ mag; Figures \ref{fig:classification} and \ref{fig:windex_vs_color}), determined in the original variability-detected AGB sample from Section \ref{sec:m101_variable_selection}, and one determined from the red population of the F150W - F356W diagram itself ($I_{\rm F182M} = 0.34$ mag). The $I_{\rm F182M} = 0.40$ mag threshold yields 13,921 O-rich candidates and 1,205 C-rich candidates while the 0.34 mag threshold yields 13,696 O-rich candidates and 1,430 C-rich candidates. The CMD-only classification gives a final count of 13,686 O-rich and 1,440 C-rich candidates. Our fiducial result is the $I_{\rm F182M} = 0.34$ mag variant, however, all three agree to within $\sim 200$ stars. 

We note that all of these are photometric classifications --- no spectroscopy exists for these stars. Spectroscopic follow-up would be needed to verify these classifications. The full table will be published in its entirety online in machine-readable format, but a portion of the catalog is shown in Table \ref{tab:sample_final_catalog}. 

\subsection{JAGB Prescriptions}
\label{sec:application_JAGB}

\begin{deluxetable*}{lll}
\tabletypesize{\footnotesize}
\tablewidth{0pt}
\tablecaption{JAGB-region composition and classification behavior by
environment.\label{tab:jagb_environments}}
\tablehead{
\colhead{} & \colhead{LMC ($\approx 1/3\,Z_\odot$)} &
\colhead{M101 field ($\approx 2/3\,Z_\odot$)}
}
\startdata
\textbf{J-region} & Predominantly C-rich: O-contamination of & Mixed: $\approx$33--50\% O-like in CMD-selected \\
\textbf{composition} & the J-region window is 0\% for all three & samples, up to $\approx$68\% for the variability- \\
 & literature windows (\S\ref{sec:JAGB_contamination}, Table~\ref{tab:jagb_contamination}). & detected sample (\S\ref{sec:JAGB_contamination}, Table~\ref{tab:jagb_contamination}). \\
\noalign{\vskip 3pt}\hline\noalign{\vskip 4pt}
\textbf{Why} & O-rich AGB stars show little infrared & O-rich variables redden by $+0.36$ mag \\
 & excess at this metallicity and remain & (median; up to $+0.74$) relative to the LMC \\
 & blueward of the window; C- and O-AGB stars & and cross into the J-region; C- stars are \\
 & are comparably numerous (C/M $= 0.71$; & rare (C/M $= 0.105$; \S\ref{sec:methods_metallicity}), so \\
 & \S\ref{sec:methods_metallicity}). & modest O-rich leakage dominates numerically. \\
\noalign{\vskip 3pt}\hline\noalign{\vskip 4pt}
\textbf{CMD selection} & Broadband color separates the chemical & Degenerate: dust-reddened O-rich stars \\
\textbf{only} & classes cleanly. & overlap C-star colors. \\
\noalign{\vskip 3pt}\hline\noalign{\vskip 4pt}
\textbf{Adding variability} & Confirms the photometric classes (92/93 & Breaks the degeneracy for variables: the water \\
\textbf{$+$ H$_2$O} & spectroscopic confirmations). & index agrees with the CMD cut for 95.5--96.9\% \\
 & & of stars. \\
\noalign{\vskip 3pt}\hline\noalign{\vskip 4pt}
\textbf{Implication for} & J region $\approx$ chemically homogeneous & Mixed population, tracking local environment; \\
\textbf{J-region} & C-star population; single-epoch selection & the single-epoch JAGB magnitude bias can \\
\textbf{chemical purity} & defensible in such low-metallicity & be expected to vary from field to field \\
 & environments. & in a given target galaxy. \\
\noalign{\vskip 3pt}\hline\noalign{\vskip 4pt}
\textbf{C-star progenitor} & \multicolumn{2}{l}{Expected to shift with metallicity, which can potentially affect the J-region} \\
\textbf{mass range} & \multicolumn{2}{l}{luminosity even at perfect chemical purity --- not examined in this work. } \\
 & \multicolumn{2}{l}{Such a test requires independent geometric distances to both galaxies} \\
 & \multicolumn{2}{l}{(available only for the LMC) or period information to obtain masses via a} \\
 & \multicolumn{2}{l}{period--mass relation, deferred to later work.} \\
\enddata
\tablecomments{Rows summarize \S\ref{sec:JAGB_contamination} and
\S\S5.1--5.4; no new measurements. Metallicities:
[Fe/H] $\approx -0.5 \pm 0.1$ for the LMC (stellar tracers) and
[M/H] $\approx -0.2 \pm 0.1$ for the M101 field (C/M-based best
estimate); see Section~\ref{sec:methods_metallicity}. The final row
states a scope limit, not a result: establishing J-region chemical
purity is necessary, but not by itself sufficient, for the JAGB to
serve as a standard candle.}
\end{deluxetable*}

One of the most immediate applications of our classification is to the JAGB method, which selects stars in a fixed color window and assumes that the population inside is uniformly C-rich. Table~\ref{tab:jagb_environments} summarizes how the assumption of C-rich purity in the J-region fares in the two galaxies studied in this paper. At the metallicity of the LMC, expected to be lower than the C/M-metallicity inflection point \citep[see][and Section \ref{sec:methods_metallicity}]{Boyer_2019}, that assumption holds quite well. In our SPHEREx sample --- which is representative of the overall AGB population in the LMC --- every star whose colors fall in the JAGB window, whether defined by 2MASS or NIRCam synthetic colors, is spectroscopically C-rich (see Section \ref{sec:JAGB_contamination} and Table \ref{tab:jagb_contamination}). In the M101 field, which falls in the steep regime of the C/M-metallicity relationship, the same windows are 33--68\% O-rich, dominated by dust-reddened members of its much more numerous O-rich population. Thus, contamination in the J-region is likely environmentally dependent. Because O-rich AGB stars range from dust-poor stars with low infrared excess to heavily obscured OH/IR stars, those that eventually become dust-reddened must --- at some point in their evolution --- transition through the intermediate colors of the J-region. The O-rich fraction of the J-region population at a given time depends on metallicity. At higher metallicities, fewer stars become C-rich and O-rich stars redden more, increasing the O-rich share of the J-region population in two ways at the same time (Section~\ref{sec:color_environment}; Table~\ref{tab:jagb_environments}). 

There is evidence for the environmental dependence of the JAGB in previous studies as well. \citet{Magnus_2024} measured the O-rich contamination of the J-region in the Galaxy using AGB stars chemically classified with Gaia RP spectra and a Gaia--2MASS photometric diagram. They found substantial contamination ($\sim 50\%$) while using standard cuts ($J-K_s$ = 1.3 mag) in the more metal-rich local environment, and an estimated $< 10\%$ contamination when the J-region blue edge boundary is moved to $J-K_s$ = 1.5 mag. While this may initially appear to contradict our results --- for M101, there seems to be no `safe' blue edge boundary in the J-region --- this 10\% contamination is a lower bound by construction, for reasons they identify themselves. The dust-enshrouded O-rich stars that would contaminate the J-region in a higher metallicity environment (such as M101 or our Galaxy) where they are efficiently producing dust are much more difficult for Gaia to detect in the optical and have large parallax uncertainties. As they note, nearby AGB stars tend to be absent as they saturate the 2MASS catalog, removing stars independently of dust production. Among the 258 C-stars from the \citet{Whitelock_2006} catalog, which they cross-checked, only 55 survived the selection, indicating $\sim 20 \%$ completeness. They ultimately conclude that they \textit{``do not confirm the claim in the literature that the absolute calibration of the JAGB method is independent of metallicity up to solar metallicity.''} The more recent results from the CHASE spectral survey, which selects directly from the J-region and retains brighter stars removed by the 2MASS quality cut used by \citet{Magnus_2024}, finds $76\%$ O-rich contamination in their sample \citep{Li_2026}. 

\citet{Li_2025} found that the JAGB mode measurements in two fields from NGC 4258, despite being at the same distance from us, differ by $0.11 \pm 0.02$ mag. They also explicitly state that \textit{``We look for but do not (yet) find a standardizing relation between JAGB LF skew or color dependence and the apparent variation.''} \citet{Lee_2025_JWSTJAGB} find the JAGB magnitude to be on average 0.21 mag brighter in the inner disks of the galaxies in their JWST-based sample, with larger deviations shown in individual galaxies (see their Figure 4). They suggest this may be the result of crowding in inner disk regions, and therefore restrict their measurements to using a galaxy-dependent radial cut. The filters and color windows used by both studies are listed in Table \ref{tab:jagb_contamination}. Radial metallicity gradients would also be a natural additional contributor to this trend: our two-galaxy comparison shows that O-rich contamination of the J-region rises steeply with metallicity, so the population inside the color window should also change with galactocentric radius, independent of any photometric bias due to crowding. 

Several competing effects could contribute to the radial trend identified by \citet{Lee_2025_JWSTJAGB}. Since the net effect of their combination is not expected to be coherent or easily predictable, we list a few contributors but do not attempt to assign a magnitude to each effect. For example, the aforementioned crowding and blending biases measurements towards brighter magnitudes in the more-crowded inner disk of galaxies. Metallicity gradients, on the other hand, would be expected to affect the results in at least two ways --- via chemical composition of the J-region or a possible shift in the C-star progenitor mass range (Table~\ref{tab:jagb_environments}). This work only considers the effect of metallicity on the chemical composition of the J-region because single-epoch photometry cannot separate between these two effects. Convergence of the JAGB mode at large galactocentric radii does establish stability but does not necessarily guarantee purity --- stability by itself does not indicate that the calibrating (such as the LMC) and target fields (in SN Ia hosts) sample equivalent populations. 

While our proposed classifications may allow for the construction of a purely C-rich J-region by breaking the AGB color degeneracy, a pure C-rich J-region is not proof that the JAGB is a standard candle. Although the J-region is selected in near-infrared bands where C-rich stars in the Magellanic Clouds span a narrow range in magnitude over a broad range in color, if the progenitor masses of C-rich AGBs change as a function of metallicity --- as predicted by most AGB evolutionary tracks --- this would be expected to shift the J-region in magnitude, not only in color. However, the single-epoch photometry and variability priors we employ here are not sufficient to determine any changes in the masses of J-region stars. Such a comparison would require a comparison of JAGB luminosities determined in galaxies with precise independent geometric anchors or period information to obtain masses via a period-mass relation, so we defer this to later work. 

In summary, the assumption that the J-region is purely C-rich holds in the LMC but not in our M101 field, and the degree of contamination is expected to vary from field to field within a galaxy. Purity cannot be verified from the J-region luminosity function itself, since a contaminated sample can pass the standard self-consistency checks (Section \ref{sec:jagb_contamination_discussion}), but it can be tested directly with variability and molecular-feature classifications (Section \ref{sec:application_classification}). Establishing purity in this way is necessary, but not by itself sufficient, for the JAGB to serve as a standard candle in its current formulation.

\begin{figure*}
\centering
\includegraphics[width=\textwidth]{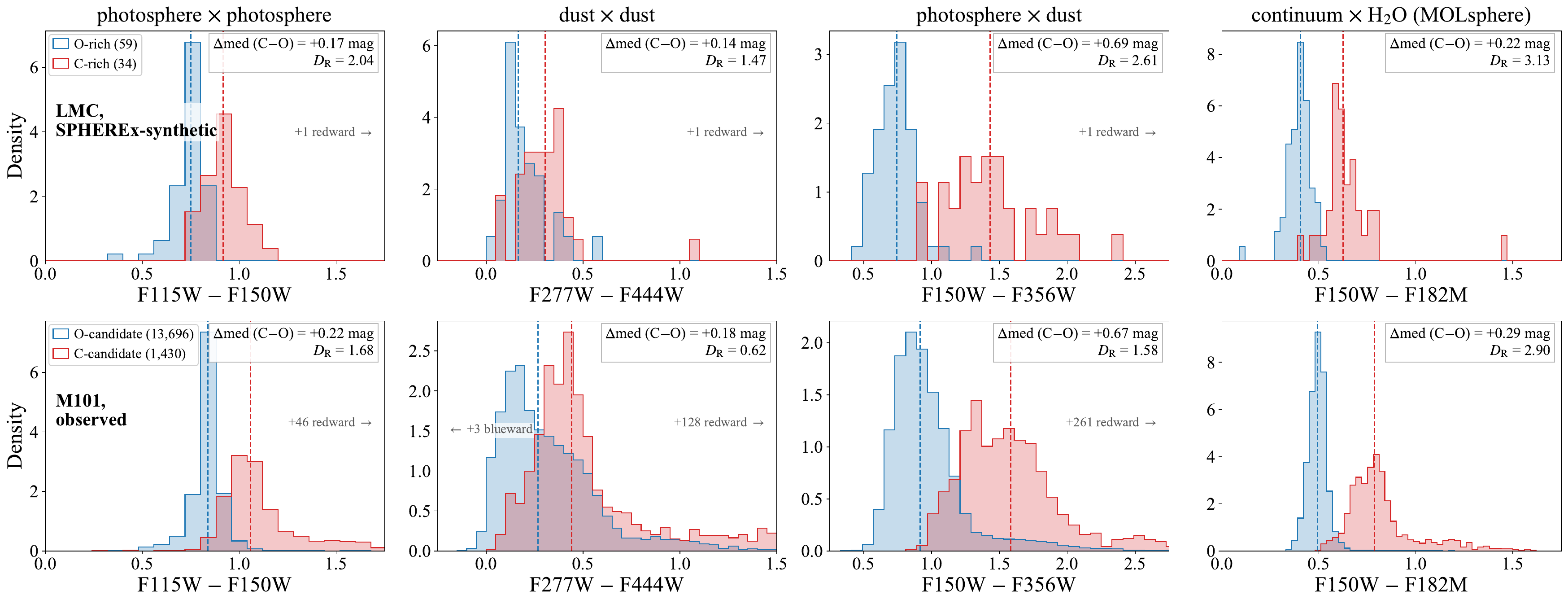}
\caption{Color distributions of O- and C-AGB candidates in four colors, labeled by the atmospheric regime each is most sensitive to (column titles). The top row shows the LMC sample of synthetic NIRCam photometry from SPHEREx spectra. The bottom row shows M101 with the final classifications of Section~\ref{sec:application_catalog}. Histograms are normalized for each chemical class, dashed lines mark median colors for each class, and legends on each panel list the median C-O offset and the robust separation $D_R$ (Section~\ref{sec:co_color_sep}) for the samples shown. Stars outside of the panel's range are tallied on the panel in gray. For the M101 F150W-F182M panel, the separation is partly by construction, since both bands are used in the classification itself. 
\label{fig:capstone_figure}}
\end{figure*} 

\subsection{Guidance for NIRCam Filter Selection}
\label{sec:application_bands}

As detailed in Section \ref{sec:observations_m101}, we selected F182M based on the expected location of water absorption in wavelength space. Here, we briefly outline how our selection could generalize into a prescription for future programs, taking into account the picture from our observations. These observations have shown that a cool AGB star's spectral energy distribution in the 0.75 - 5.0$\mu$m regime carries three separable kinds of information, each primarily dominating different parts of the spectrum: the \textit{photospheric continuum} up to about 2$\mu$m, measured with short-wavelength wide bands, which respond primarily to a combination of temperature and dust opacity; \textit{circumstellar dust}, measured with long-wavelength wide bands, which respond to the onset of the warm dust continuum; and \textit{molecular chemistry}, determined using medium bands placed on individual molecular features. We posit that a filter set that samples all three -- a photospheric band, a dust band, and a molecular band -- will recover more information than any combination drawn from a single regime, regardless of the number of bands it contains. 

Choosing filters without taking these regimes into account results in characteristic failure modes. For example, two well-chosen, short-wavelength, near-infrared wide bands \textit{can} separate C- and O-AGB stars reasonably well, but the narrow wavelength regime compresses the dynamic range. In the $< 2\micron$ regime, broadband colors are sensitive to the differences in dust opacity (silicate dust is much more transparent than carbonaceous dust) and to effective temperature. This allows the difference to be more easily eroded by photometric scatter, as shown in the first column of Figure \ref{fig:capstone_figure}. Two long-wavelength filters that are both sensitive primarily to the presence of dust cannot distinguish between C-rich and O-rich stars at higher metallicities where both classes of AGB stars are efficient dust producers (Section \ref{sec:color_environment}), shown in the second column of Figure \ref{fig:capstone_figure}. As noted by \citet{Gavetti_2026b}, the F277W-F444W combination is a reasonable proxy for the presence of dust. 

On the other hand, because our analysis shows that all dusty stars are expected to be variable, combining a short-wavelength filter with a long-wavelength one gives a larger color separation, as well as information about whether any given star is variable. Classification using molecular bands from molecules formed by pulsation is also more effective on variable stars. This paper focused on one, F182M, but several other NIRCam medium bands have significant wavelength overlap with the locations of other features and are likely to be effective. For O-rich AGB stars, F250M samples H$_2$O and CO (present in both C- and O-AGBs). For C-rich AGB stars, F300M samples C$_2$H$_2$ + HCN and F162M is bracketed by the C$_2$ and CN features. 

Of these filters, only F182M has been tested in both galaxies (on NIRCam photometry in M101 and on synthetic photometry from SPHEREx spectra in the LMC); F250M has been tested only on the LMC synthetic photometry (Appendix \ref{sec:appendix_classification}), and F300M and F162M are motivated by the locations of their features alone. While we note that these filters are promising due to their location relative to the molecular features, we do not provide an explicit ranking because we do not have observations of these bands in resolved stellar populations at higher metallicity, and the hydrostatic model grids are not designed to reproduce molecular layers levitated by pulsation (Section \ref{sec:application_scope}). Doing so would thus suggest a precision that we cannot provide. 

While color-color diagrams can be highly effective at separating C- and O-rich chemistries, our analysis also indicates that single-color bimodality may be more easily translatable across different metallicities. An example is the F090W - F150W vs. F150W - F250M color-color diagram recommended by the Early Release Science team of \citet{Boyer_2024}. Applying their boundaries directly to our LMC SPHEREx sample misclassifies 38\% of our spectroscopically oxygen-rich sample as carbon stars. However, this is expected because, as they note, the boundaries of the WFC3/IR equivalents of these diagrams may depend on metallicity and population age, and thus are typically adjusted for the populations of individual galaxies \citep[e.g., for M31 by][]{Boyer_2019}. The difficulty in using the color-color diagram lies not in the metallicity dependence of its boundaries (as our work shows, metallicity inevitably affects the colors of AGB stars) but that there is no obvious location to draw the boundary in the F090W-F150W vs. F150W - F250M color-color diagram without having prior knowledge of the chemical classifications. 

On the other hand, F150W - F250M, the color on the vertical axis of the ERS diagram, also separates the two chemistries on its own, through single-color bimodality. In our LMC SPHEREx sample, a two-component Gaussian mixture fit to its distribution, made without prior knowledge of the stellar classifications, places the boundary within 0.003 mag of the one that best reproduces the spectral classifications (Appendix \ref{sec:appendix_classification}). The valley in this color is shallower than in F150W - F182M, however, because the longer baseline also responds to circumstellar reddening, which broadens the C-rich distribution. 

At LMC-like and lower metallicities, broadband color and variability information together can classify most AGB stars, since the vast majority of dusty stars are C-rich. For population-level applications such as JAGB distances, broadband colors may therefore be sufficient for chemical classification in such environments. Medium bands become \textit{essential} at higher metallicity, where dust production no longer distinguishes the two chemistries, and for most heavily reddened stars, which broadband colors will assign a C-rich chemistry regardless of their true classification. 

\subsection{Connection to Model Grids}
\label{sec:application_scope}

In the Introduction of this work, we stated our hypothesis that the strongest chemical diagnostics in the 0.75$-$5\micron\, regime might be the features that hydrostatic model grids struggle to reproduce. We also suggested that the strength of these features should correlate with pulsation. The measurements in this paper support this picture: $I_{\rm F182M}$ separates the chemical classes most reliably among stars with detectable variability (Section~\ref{sec:co_color_sep}). 

These results sharpen the usual statement of the validity of model grids. \citet{Aringer_2016} stated, regarding a comparison of synthetic photometry to observed relations and data, \textit{``deviations appear for the coolest giants showing pulsations, mass loss and dust shells, which cannot be described by hydrostatic models.''} Our variability measurements allow us to identify the source of this deviation more precisely, by tying the molecular deviations to atmospheric layers levitated by pulsation \citep{Tsuji_2000, Hofner_Olofsson_2018} and allowing us to predict the relative strength of the departures between models and observations. Model grids predict the photospheric molecular species well, but are inherently not designed to simulate the levitated layers above the photosphere. The historical reliance on these grids for filter selection in water-absorbing bands thus represents a misapplication of the models outside their valid physical regime.  $I_{\rm F182M}$ and similar molecular-feature-based indices are thus only expected to apply where pulsation has built such molecular layers.  

While the behavior of $I_{\rm F182M}$ is coherent in both the LMC and M101, we note that our work has only explored its application within the $\approx 1/3 - 2/3 Z_{\odot}$ regime. This range is, however, an operationally relevant one for distance-ladder applications --- by extending from the LMC, where AGB distance indicators are calibrated, to a SN Ia host galaxy, where they are applied. This metallicity range also brackets both sides of the C/M-metallicity relation inflection point, but we have not substantially tested our selection criteria on more metal-poor or metal-rich environments. In more metal-rich environments, the index is expected to become more effective as the water content of oxygen-rich atmospheres increases. In more metal-poor environments the expectation is less certain. Water absorption has been observed to remain strong in at least one very metal-poor O-rich AGB star \citep{Boyer_2025}, but whether the population-level bimodality needed to set the classification boundary survives at low metallicity is currently untested. 

\section{Conclusions}\label{sec:conclusions}

In this work, we used SPHEREx spectrophotometry of spectroscopically classified LMC AGB stars to synthesize JWST NIRCam photometry, bridging a wide range of archival observations of the LMC with our NIRCam imaging of M101. This creates a baseline that runs from a distance-ladder anchor galaxy to a SN Ia host. Such a direct comparison is otherwise not possible due to the differences in angular size, distance, and apparent brightness between the two systems. Within each galaxy, we selected and classified the AGB population in two independent ways: a traditional CMD-based selection and a variability-based selection paired with a molecular-feature index. We follow both across the two environments, with particular attention to how photometric classification criteria transfer and what the transfer implies for AGB-based distance indicators. We then draw the following conclusions:

\begin{enumerate}
    \item The method used to separate C- and O-AGB stars should be carefully considered based on the goal (e.g., a clean C-AGB population, a clean O-AGB population, or distinguishing dusty red stars) and the host environment of the AGB population.
    \item Variability information improves the photometric classification of AGB stars at both high and low metallicity. Because dust-reddened AGB stars are essentially always variable, infrared excess can be used to identify pulsating stars within a single or few epochs, thus allowing $I_{\rm F182M}$ to be applied where the AGB color degeneracy is most severe, even without time-series data. 
    \item Classification criteria which have boundaries that can be derived within the science sample itself -- e.g., via single-color bimodality, as with $I_{\rm F182M}$ and F150W$-$F182M -- are preferable to criteria requiring externally calibrated boundaries, even at comparable separating power. Metallicity inevitably shifts the boundaries between chemical classes, so a method's transferability into new environments depends on whether the data alone are enough to indicate the boundary locations. 
    \item Dust and variability are linked asymmetrically. We find that essentially all stars with substantial infrared excess are variable while many variables show little dust. By measuring both conditional dependencies -- $P({\rm Dusty}|{\rm Variable})$ and $P({\rm Variable}|{\rm Dusty})$ -- in photometrically complete samples in the LMC and M101, we show that pulsation is a necessary but not sufficient condition for dust production, in both galaxies.
    \item O-rich dust production, as traced by infrared excess, depends on the environment. Variable O-AGB stars in M101 are up to $\sim0.7$ mag redder than their LMC counterparts in dust-sensitive colors while the colors of C-stars change much less between the two galaxies, consistent with O-rich dust condensation tracking metallicity and carbon stars producing their own condensates.
    \item J-region selections that are purely C-rich in the LMC contain 33-68\% O-rich stars in M101, and this contamination evades (and in fact, can strengthen) the standard JAGB self-consistency checks. Variability and molecular-feature-based photometric classifications can be used to classify stars in the J-region, mitigating this contamination.
    \item Where chemical purity of the J-region cannot be established, the mean magnitude is substantially more robust to contamination than the mode in our field. Between the contaminated and pure C-AGB samples, the mean differs by $\leq 0.07$ mag in all three windows, while the mode shifts by up to 0.20 mag. The median is intermediate ($\leq 0.02$ mag in the two F115W windows, but 0.14 mag in the F150W window). This provides an independent, empirical reason to prefer the mean-based indicator that \citet{Gavetti_2026} recommend on theoretical grounds, since the peak of the J-region luminosity function, unlike its mean, depends on the host's metal-enrichment history. The mean merits consideration as the fiducial JAGB magnitude.
    \item A cool AGB star's 0.75--5\,$\mu$m SED carries three separable kinds of information: the photospheric continuum, circumstellar dust, and molecular chemistry. A filter set sampling all three regimes -- a photospheric band, a dust band, and a molecular band -- is expected to recover more classification information than any combination drawn from a single regime, and each regime omitted from a filter set introduces a characteristic failure mode.
    \item SPHEREx spectrophotometry can be used to place resolved Local Group populations and JWST targets on a common photometric system, enabling direct comparisons with no cross-calibration terms -- the bridge this work is built on.
\end{enumerate}

Finally, we emphasize the role of variability information itself. Nearly every conclusion we drew above depended on it --- from the dust--variability asymmetry to the validity condition of the water index to the recovery of the most heavily reddened O-rich stars and the cross-check of photometric classifications. All of these were possible only because pulsation had been measured star by star. Our results indicate that variability is not auxiliary information, but a central axis in extragalactic AGB studies. Pulsation state sequences a star's departure from hydrostatic behavior and precedes dust production, making variability important for studies of AGB evolution in the near- and mid-infrared. 

The prospects for obtaining variability information are also excellent. Roman and Rubin will identify evolved variables in large numbers, and even when precise periods may not be obtainable, detections and amplitudes alone can significantly inform our interpretation of these stars. We hope this work illustrates what variability information can add to studies of evolved stars.

\begin{acknowledgments}
We thank the anonymous referee for their constructive and detailed comments, which improved this paper. We thank Martha Boyer, Steven Goldman, and Elizabeth Tarantino for useful discussions on previous AGB classifications. We thank Jaeyeoung Kim for information regarding GPR C/O classification rates of SPHEREx spectra and Matt Ashby regarding the usage of SPHEREx survey data. We thank Timothy Y. Brooke for his help in retrieving SPHEREx spectrophotometry and Peter Kosec for the spectrum of OGLE-LMC-LPV-00743. Additionally, we thank Warren Hack for his help in designing the observations of HST-GO-17312 and Siddharth Mishra-Sharma for the Claude Max coupon. We thank Nikhil Anand for his support throughout the writing of this paper and suggestion to include a reader's guide. Claude (Anthropic; Fable 5 and Fable 5.1) was used during analysis for code development and note-taking, and during manuscript preparation for fact-checking, verification of quoted values and references, and suggestions for language and structure. 

This material is based on work supported by the National Science Foundation Astronomy \& Astrophysics Postdoctoral Fellowship under Grant No. 2401770. L.B. acknowledges support through the European Space Agency (ESA) Research Fellowship in Space Science. P.A.W. acknowledges support from the South African National Research Foundation (NRF). G.S.A. acknowledges financial support from JWST GO-2875. Support for programs JWST-GO-4087 and HST-GO-17312 was provided by NASA through grants from the Space Telescope Science Institute, which is operated by the Association of Universities for Research in Astronomy, Inc. This research has made use of the SVO Filter Profile Service ``Carlos Rodrigo", funded by MCIN/AEI/10.13039/501100011033/ through grant PID2023-146210NB-I00.

This research has made use of the NASA/IPAC Infrared Science Archive, which is funded by NASA and operated by the California Institute of Technology.
 
\end{acknowledgments}

\vspace{5mm}
\facilities{HST(WFC3/IR), JWST(NIRCam), SPHEREx, Spitzer(IRAC), WISE, NEOWISE, FLWO:2MASS, CTIO:2MASS, OGLE, IRSA, MAST}

\software{Astropy \citep{astropy:2013, astropy:2018, astropy:2022}, Scipy \citep{Virtanen_2020}, DOLPHOT \citep{Dolphin_2016, Weisz_2024}, DAOPHOT/ALLSTAR/ALLFRAME \citep{Stetson_1987, Stetson_1994}}

\appendix

\section{Additional Dataset Details}
\label{sec:appendix_datasets}

Here we provide details on the HST and JWST observations used in this paper. Table \ref{tab:jwst_observations} lists the JWST observations used in this paper while Table \ref{tab:hst_observations} lists the updated HST time-series baseline used. 

After retrieval from MAST, the Stage 2 and Stage 3 products were processed with calibration pipeline v1.12.5 under CRDS context \textit{jwst\_1202.pmap}, which includes the updated NIRCam photometric zeropoints. Photometry used \textsc{DOLPHOT} 2.0 with its NIRCam module \citep{Dolphin_2016, Weisz_2024} with the F150W mosaic used as the astrometric reference and 12 input frames per run (three filters $\times$ four dithers). We adopted the ERS parameters: \texttt{FitSky}=2, \texttt{RCombine}=1.5,
\texttt{UseWCS}=2, \texttt{SecondPass}=5, \texttt{ApCor}=1,
\texttt{PSFPhot}=1, \texttt{InterpPSFlib}=1, with
\texttt{img\_RAper}=2 and \texttt{img\_RChi}=1.5 for the short-wavelength chips and 3 and 2.0 for the long-wavelength detector. We then merged the short- and long-wavelength catalogs as described in Section \ref{sec:hst_jwst_crossmatch}. 41,160 Module~B sources were detected in all six bands. While Module~A was photometered, it is not used here. 
 
\begin{deluxetable}{lllrcc}
\tabletypesize{\scriptsize}
\tablecaption{Summary of JWST/NIRCam Observations}
\tablewidth{0pt}
\tablehead{\colhead{Date} & \colhead{Mod.} & \colhead{Filter} & \colhead{Exp.\ (s)} & \colhead{PID} & \colhead{Obs.\ ID}
}
\startdata
    2024-02-23 & A, B & F115W & 1674.94 & 4087 & o001 \\
    2024-02-23 & A, B & F150W & 1674.94 & 4087 & o001 \\
    2024-02-23 & A, B & F182M & 1674.94 & 4087 & o001 \\
    2024-02-23 & A, B & F277W & 1674.94 & 4087 & o001 \\
    2024-02-23 & A, B & F356W & 1674.94 & 4087 & o001 \\
    2024-02-23 & A, B & F444W & 1674.94 & 4087 & o001
\enddata
\tablecomments{All JWST NIRCam observations analyzed in this paper. Full data set can be accessed at \dataset[doi:10.17909/168e-1683]{https://doi.org/10.17909/168e-1683}.
}
\label{tab:jwst_observations}
\end{deluxetable}
\onecolumngrid

\begin{deluxetable*}{lllrccc}
\tabletypesize{\scriptsize}
\tablecaption{Summary of HST Observations}
\tablewidth{0pt}
\tablehead{\colhead{Epoch} & \colhead{Date} & \colhead{Filter} & \colhead{Exposure (s)} & \colhead{No. Dithers} & \colhead{Proposal ID} & \colhead{Dataset}
}
\startdata
    1 & 2014-10-09 & F160W& 1398.46 & 2 & 13737 & ICOY01 \\
    1 & 2014-10-09 & F110W & 1598.47 & 2 & 13737 & ICOY01 \\
    2 & 2015-01-24 & F160W& 827.72 & 3 & 13824 & ICL0A1 \\
    3 & 2015-04-06 & F160W& 1398.46 & 2 & 13737 & ICOY02 \\
    3 & 2015-04-06 & F110W & 1598.47 & 2 & 13737 &  ICOY02 \\
    4 & 2015-07-14 & F160W& 1398.47 & 2 & 13737 & ICOY04 \\
    4 & 2015-07-14 & F110W & 1598.47 & 2 & 13737 & ICOY04 \\
    5 & 2015-09-10 & F160W& 1655.45 & 6 & 13824 & ICL0A2 \\
    6 & 2016-02-19 & F160W& 1398.46 & 2 & 14166 & ICZJ01 \\
    6 & 2016-02-19 & F110W & 1598.47 & 2 & 14166 & ICZJ01 \\
    7 & 2016-09-27 & F160W& 8795.40 & 6 & 14166 & ICZJ03 \\
    7 & 2016-09-27 & F110W & 5195.40 & 6 & 14166 & ICZJ03 \\
    8 & 2017-06-17 & F160W& 1911.74 & 4 & 14678 & ID6V02 \\
    8 & 2017-06-17 & F110W & 4011.74 & 4 & 14678 & ID6V02\\
    9 & 2021-12-03 & F160W& 1311.76 & 4 & 16744 & IEQ601 \\
    9 & 2021-12-03 & F110W & 1211.74 & 4 & 16744 & IEQ601 \\
    10 & 2021-12-24 & F160W& 1311.76 & 4 & 16744 & IEQ602 \\
    10 & 2021-12-24 & F110W & 1211.74 & 4 & 16744 & IEQ602 \\
    11 & 2022-02-10 & F160W& 1311.76 & 4 & 16744 & IEQ603 \\
    11 & 2022-02-10 & F110W & 1211.74 & 4 & 16744 & IEQ603 \\
    12 & 2022-04-16 & F160W& 1311.76 & 4 & 16744 & IEQ604 \\
    12 & 2022-04-16 & F110W & 1211.74 & 4 & 16744 & IEQ604 \\
    13 & 2022-08-19 & F160W& 1311.76 & 4 & 16744 & IEQ605 \\
    13 & 2022-08-19 & F110W & 1211.74 & 4 & 16744 & IEQ605 \\
    14 & 2023-07-19 & F160W& 2661.75 & 4 & 17312 & IF4C01 \\
    15 & 2024-08-13 & F160W& 2404.34 & 4 & 17312 & IF4C02 \\
    16 & 2024-08-16 & F160W& 2404.34 & 4 & 17312 & IF4C03
\enddata
\tablecomments{All HST observations analyzed in this paper. Full data set can be accessed at \dataset[doi:10.17909/168e-1683]{https://doi.org/10.17909/168e-1683}
}
\label{tab:hst_observations}
\end{deluxetable*}
\onecolumngrid

\section{Computation of Synthetic Magnitudes}
\label{sec:appendix_synth_mag}

Using the SPHEREx spectra, we computed synthetic Vega magnitudes in 2MASS, WISE, JWST NIRCam, HST WFC3/IR, Roman WFI, and Spitzer IRAC bands. Filter-transmission functions were retrieved from the SVO Filter Profile Service \citep{Rodrigo_2012, Rodrigo_2020, Rodrigo_2024}. 

After processing, the mean spectra are in $\mu$Jy (normalized to an arbitrary phase in the pulsational cycle, as described in Section \ref{sec:mean_spec}). We first convert this to $F_{\lambda}$. Then we find the wavelength overlap for each filter with the spectra and build a common fine grid and linearly interpolate the filter transmission and the spectrum onto the grid. We then integrate the bandpass-averaged flux. Here we use the `Detector Type' classifications given by SVO to determine how to use the transmission curve for the synthetic magnitude calculations. All filters were tagged as either photon or energy counters depending on the format of the response function. For energy counters the average flux was given by, 
\begin{equation}
f(\lambda_{\text{eff}}) = \frac{\int T(\lambda)f(\lambda)d\lambda}{\int T(\lambda)d\lambda}, 
\end{equation}
while for photon counters,
\begin{equation}
f(\lambda_{\text{eff}}) = \frac{\int T(\lambda)f(\lambda)\lambda d\lambda}{\int T(\lambda)\lambda d\lambda}. 
\end{equation}
where $T(\lambda)$ is the filter transmission as a function of wavelength, and $f(\lambda)$ is the flux of the spectrum as a function of wavelength. 

We then compute Vega magnitudes by integrating a reference spectrum with the same filter and the same convention. The reference spectrum is a 9600 K blackbody approximation normalized to Vega's flux at 0.5556$\mu$m. We note that while colors are insensitive to this choice, a Vega model file would be more appropriate for the zero points of individual magnitudes.

\section{Comparison with Literature Color-Color Diagrams}
\label{sec:appendix_classification}

\begin{figure}
\centering
\includegraphics[width=0.49\textwidth]{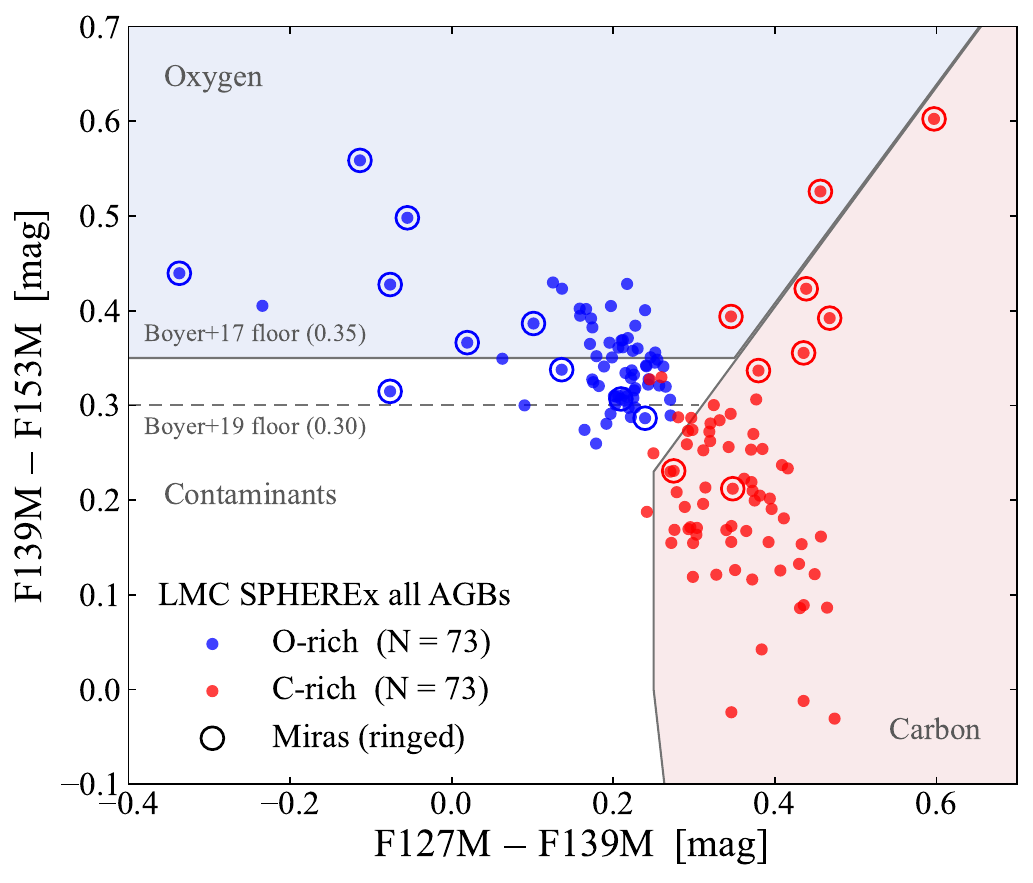}
\includegraphics[width=0.49\textwidth]{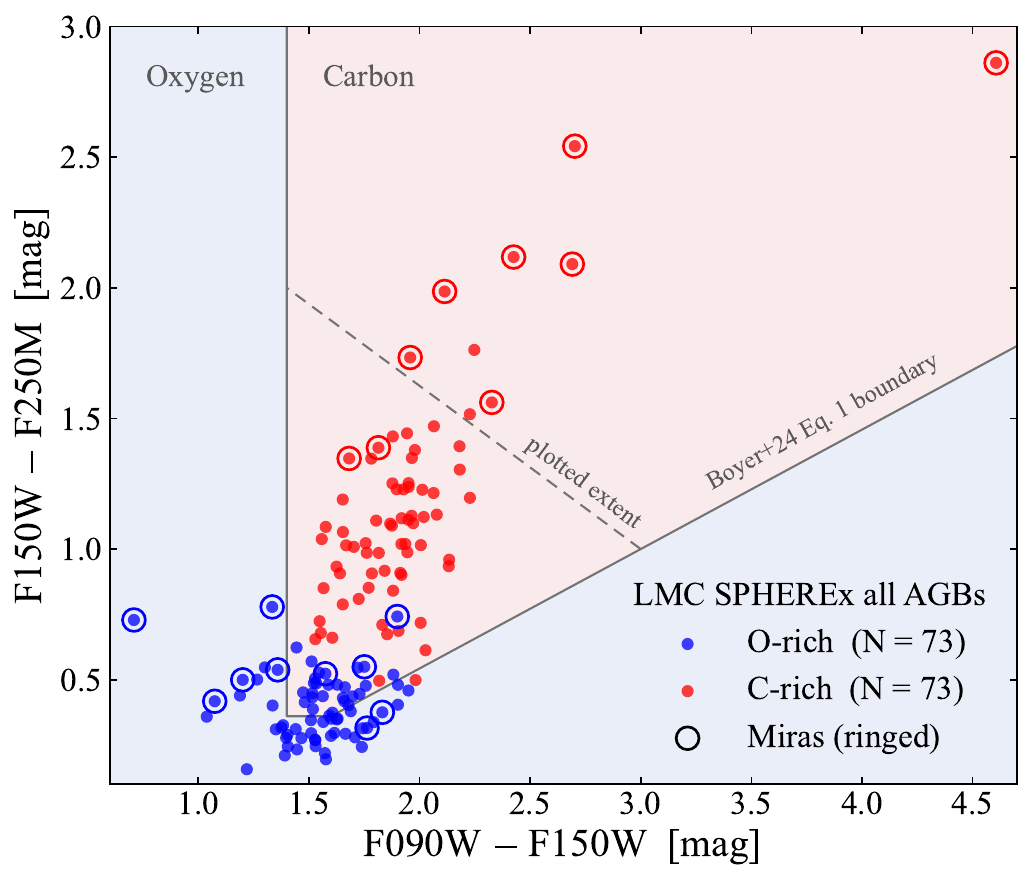}
\caption{\textit{Left:} The HST WFC3/IR medium-band color-color diagram of \citet{Boyer_2017}, applied to synthetic photometry of SPHEREx mean spectra of all LMC SPHEREx AGBs (full retrieved sample). Blue points are stars spectroscopically classified as O-rich and red points as C-rich. Shaded regions are the O-rich (blue) and C-rich (red) classification regions of \citet{Boyer_2017}. The unshaded region is designated as contaminants. The solid horizontal line represents the \citet{Boyer_2017} floor of F139M $-$ F153M $=0.35$ mag and the dashed line marks the \citet{Boyer_2019} floor at 0.30. \textit{Right:} The JWST NIRCam F090W$-$F150W vs. F150W$-$F250M diagram of \citet{Boyer_2024} with boundaries drawn from their Equation 1 applied to the same stars. The gray solid lines mark their boundary between C-rich (red shading) and O-rich (blue shading) AGB stars, calibrated using HST-classified stars in WLM. The C-rich side of the polygon is open toward the redder end and the dashed segment marks the extent of the polygon as it was plotted in their paper, rather than a classification boundary. This scheme has no contaminant region because it is applied to stars already classified as TP-AGB stars. Miras are represented as ringed points in both panels.  \label{fig:HST_medium_color_color}}
\end{figure}

Here we test two literature medium-band classifications using HST WFC3/IR and JWST NIRCam against our spectroscopically classified LMC sample from SPHEREx. In particular, we examine the HST WFC3/IR F139M $-$ F153M vs. \ F127M $-$ F139M diagram \citep{Boyer_2013, Boyer_2017}, which has been applied in external galaxies across a range of metallicities \citep[e.g.,][]{Boyer_2017, Goldman_2022} and the more recent JWST NIRCam F150W $-$ F250M vs. F090W $-$ F150W color-color diagram recommended by JWST Resolved Stellar Populations Early Release Science program team. 

The left panel of Figure \ref{fig:HST_medium_color_color} shows the LMC SPHEREx AGBs (full retrieved sample; 73 O-rich, 73 C-rich) in the WFC3/IR medium-band diagram. Literature studies adjust the location of the contaminant floor \textit{in situ} in new environments. Comparing the classifications to the F139M $-$ F153M = 0.30 mag floor used by \citet{Boyer_2019}, we find that 64/73 O-rich stars in our sample fall in the O-rich region, nine fall in the contaminant region, and none fall in the C-rich region. Of the 73 C-rich stars, 65 fall in the C-rich region, and four of the remaining eight fall in the O-rich region, and the other four among the contaminants. The C/O boundary in Figure \ref{fig:HST_medium_color_color} is quite robust, and adjusting the contaminant floor is well-motivated in order to recover more O-rich stars (using the 0.35 mag boundary, only 31 out of 73 stars would reside in the O-rich region). We note that the full set of retrieved spectra from SPHEREx are not a representative distribution in color space of the larger LMC AGB population. This population has a larger spread in color than the representative sample, due to additional random retrievals of Miras, which are highlighted in the diagram with an additional ring. 

The F127M band is approximately a continuum band for AGB stars and, at a shorter wavelength, is more sensitive to photospheric emission. F153M sits on the 1.53\micron\ C$_2$H$_2$+HCN feature in C-rich stars, while F139M is located almost directly on top of an H$_2$O absorption feature. The combination of continuum-band + molecular-feature sampling closely matches the approach we recommend here. Furthermore, we note that, among the C-rich population, the Miras appear to have differences in color relative to the lower-amplitude C-rich pulsators. This is because the depth of the C$_2$H$_2$ + HCN feature is likely enhanced by the presence of pulsation, allowing more of these more fragile molecules to form. 

The right panel of Figure \ref{fig:HST_medium_color_color} shows the JWST color-color diagram and hand-drawn boundaries from \citet{Boyer_2024}. Using observations of the star-forming galaxy WLM, which was previously observed in the HST WFC3/IR medium-band filters described above, they drew the black boundaries in the right panel to mark the best separation between C- and O-rich AGB subtypes in their data. Applying these boundaries without new, \textit{in situ} correction to our LMC-based SPHEREx sample results in 28 out of 73 O-rich stars falling within the C-rich boundary. The LMC is significantly more metal-rich than WLM, the environment in which these boundaries were validated, which likely explains the difference in color. We expect more dusty and water-reddened M-type (O-rich) stars in the LMC than in the lower-metallicity WLM, so the direction of the misclassification is coherent. Unlike the HST medium-band classifications --- which translate well because they measure the strength of molecular features in both O- and C-rich stars --- the only medium band used in the JWST NIRCam color-color diagram is F250M. 

\begin{figure}
\centering
\includegraphics[width=0.49\textwidth]{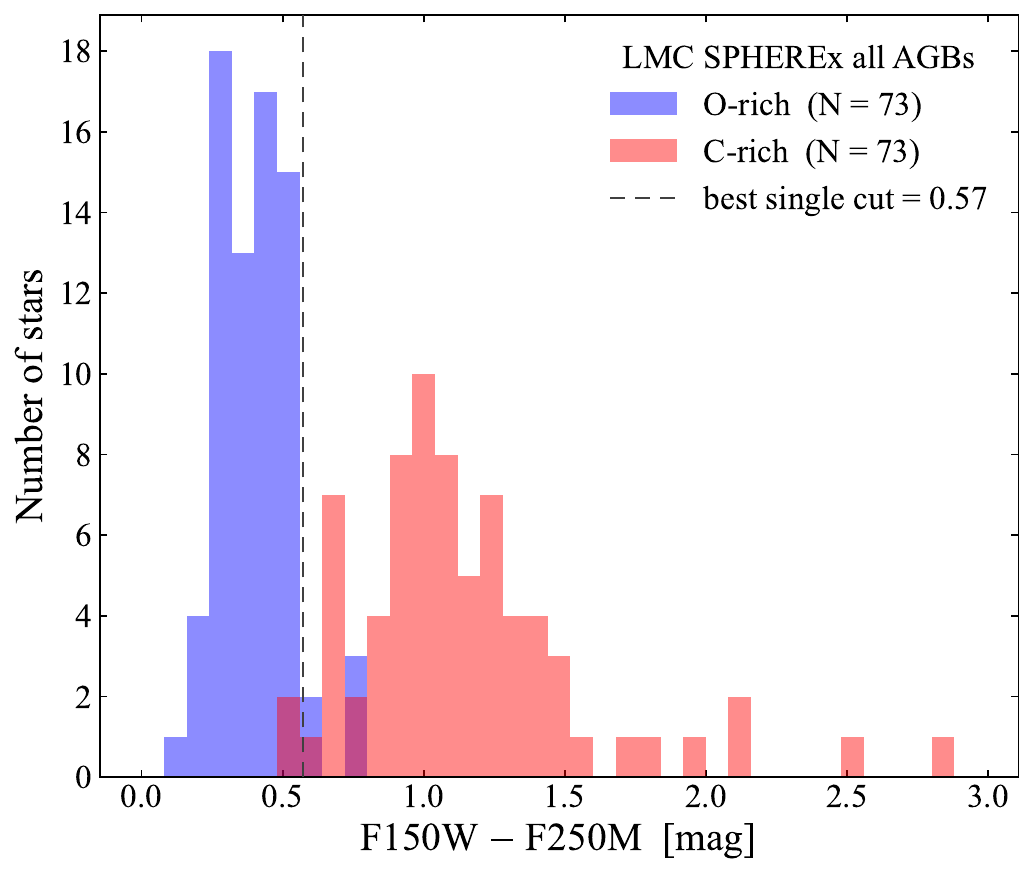}
\includegraphics[width=0.49\textwidth]{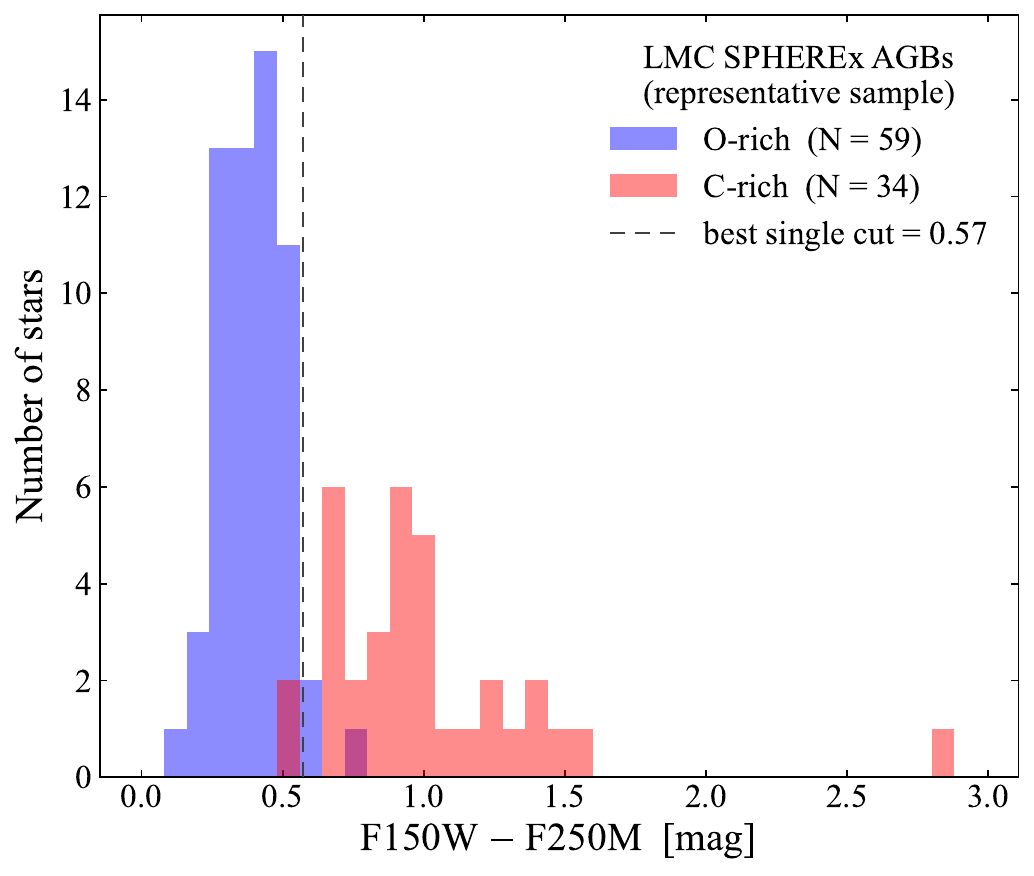}
\caption{Distributions of F150W$-$F250M color for the spectroscopically classified LMC SPHEREx AGBs, with O-rich in blue and C-rich in red. \textit{Left:} The full retrieved sample (73 O-rich, 73 C-rich). \textit{Right:} The representative subsample (59 O-rich, 34 C-rich). The dashed line in both panels is the 0.57 mag threshold (denoting C-rich to the red and O-rich to the blue) set using the full sample. 71 out of 73 C-rich and four out of 73 O-rich stars lie to the red of this line in the left panel while 32 of 34 C-rich and two out of 59 O-rich stars lie redward of this line in the right panel. \label{fig:JWST_1d_classifier}}
\end{figure}

We also test the color-color diagram method against single-color bimodality. To make a direct comparison, we use the F150W$-$F250M color, which is the color from \citet{Boyer_2024} that is most aligned to our method. Applying a color-only threshold to the SPHEREx sample recovers two well-separated populations without importing boundaries from a previous classification. However, for this color combination, the separation was less consistent across the methods employed. A two-component Gaussian mixture fit to the F150W$-$F250M distribution of all 146 stars yields an equal-posterior boundary of $0.575$ mag. This value is within $0.003$ mag of the boundary that best reproduces the spectroscopic classifications ($0.572$ mag). Both thresholds assign 71 of 73 C-rich stars to the C-rich side of the diagram and admit four out of 73 O-rich stars. However, the kernel-density minimum strategy we use to check the F150W$-$F182M color threshold (Section~\ref{sec:m101_cmd_agb}) does not recover this boundary because the broad C-rich distribution washes out the valley between the two chemical subtypes. The full SPHEREx sample yields a bimodal distribution for kernels up to 0.18 mag, but the representative subsample is unimodal if we use kernels wider than 0.10 mag. For the full SPHEREx sample, we find that the two groups are separated at $D_R =2.97$ while the representative subset gives $D_R = 2.29$. The one-dimensional distribution in color is shown in Figure \ref{fig:JWST_1d_classifier} for both the full SPHEREx sample and the representative sample. 

This difference in the behavior of this sample can be attributed to the width of the C-rich distribution in this color. The half-width of the C-AGB population in F150W$-$F182M is 0.09 mag, but grows to 0.31 mag in F150W$-$F250M, while the separation of the medians for the two subtypes only increases from 0.26 to 0.69 mag. This leads to worse separability, even at LMC metallicities and is reflected in the lower $D_R$ value relative to the F150W$-$F182M sample ($D_R = 3.48$ and $3.13$ for the full and representative subset, respectively). While both bands sample H$_2$O absorption in O-rich stars, the longer baseline from F150W to F250M also mixes the effect of molecular absorption with circumstellar reddening, which broadens and skews the C-rich distribution redward (see Figure~\ref{fig:JWST_1d_classifier}, ringed Miras). 

This exercise demonstrates that the one-dimensional approach transfers to a color that it was not designed for. In F150W$-$F250M, a threshold placed without reference to the underlying spectroscopic classifications recovers the correct chemistry for 140 of 146 stars, whereas the same bands in the color-color diagram admit 28 O-rich stars into the C-rich region (117 of 146 correct) when transferred outside of the calibration galaxy. While shifting color-color boundaries are expected and unavoidable, it is less clear how to draw new color-color boundaries in a different environment without prior classification. A detailed spectroscopic comparison of all molecular features in this wavelength regime against hydrostatic model grids is beyond the scope of this population-level study and will be explored in future work.

\bibliography{multibandcmds}{}
\bibliographystyle{aasjournalv7}

\end{document}